\documentclass[]{article}

\usepackage{enumerate}
\usepackage[a4paper]{geometry}
\usepackage{natbib}
\usepackage{graphicx}
\usepackage{multirow}
\usepackage{changepage}
\usepackage{amsfonts,amsmath,amssymb}
\usepackage[dvipsnames,table]{xcolor}
\usepackage{url}
\usepackage{hyperref}
\usepackage{subfigure}

\usepackage[utf8]{inputenc}
\usepackage{xr}
\newtheorem{theorem}{Theorem}

\begin{document}

\title{Deconvolution density estimation with penalised MLE}

  \author{Yun Cai \and   Hong Gu\thanks{The authors gratefully acknowledge funding from NSERC} \and  Toby Kenney\thanks{The authors gratefully acknowledge funding from NSERC}\\
Department of Mathematics and Statistics, Dalhousie University}

  \maketitle

\begin{abstract}
Deconvolution is the important problem of estimating the distribution
of a quantity of interest from a sample with additive measurement
error. Nearly all methods in the literature are based on Fourier
transformation because it is mathematically a very neat
solution. However, in practice these methods are unstable, and produce
bad estimates when signal-noise ratio or sample size are low. In this
paper, we develop a new deconvolution method based on maximum
likelihood with a smoothness penalty. We show that our new method has
much better performance than existing methods, particularly for small
sample size or signal-noise ratio.

\end{abstract}

\noindent%
         
\section{Introduction}

Measurement error is a common problem with data. It occurs when the
apparatus measuring a variable is not perfect, and the value it
returns is a random variable, based on the true value. A simple
example is additive measurement error, where the recorded value is the
true value (which is itself a random variable) plus a random
error. More formally, the observed value is given by $Y=X+E$, where
$X$ is the true value and $E$ is a random measurement error. This can
happen with a lot of measurement apparatus. It can also happen when $Y$
is an estimated quantity (such as an MLE estimate from a particular
model on some data) for which there is no analytic solution. In this
case, the estimates $Y$ are subject to convergence error, which
behaves like measurement error.

\begin{figure}
\centering
\makebox{\includegraphics[width=8cm]{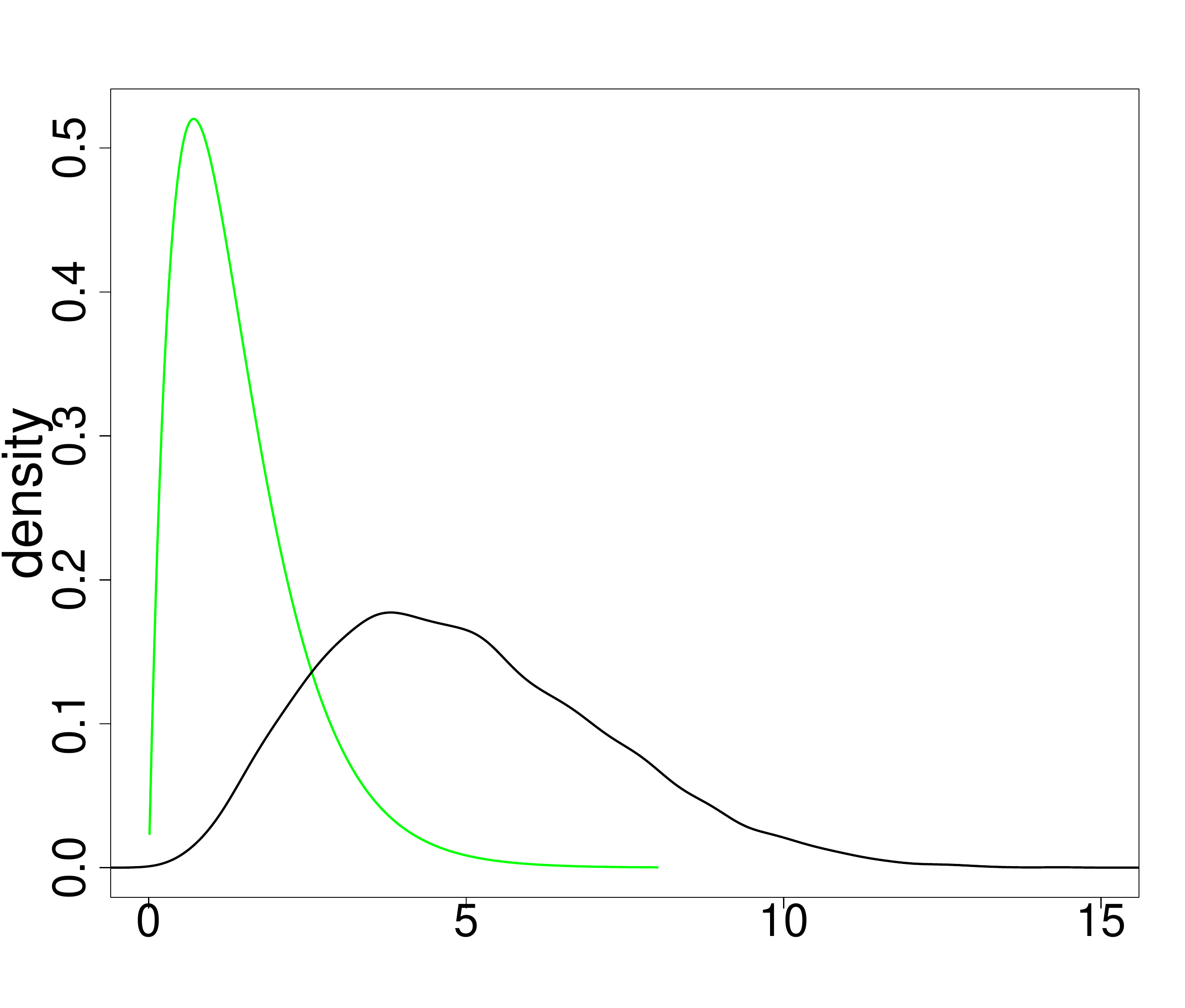}}
\caption{The black curve is the density of the contaminated data. The green curve is the density of the underlying truth.}
\label{example}
\end{figure}

In this paper, we look at the problem of estimating the density of the
underlying variable $X$ from a sample of observations with measurement
error. This is referred to as deconvolution. Figure~\ref{example}
shows an example of this problem. The green curve is the density
function of interest: it follows a scaled chi-squared distribution
with 4 degrees of freedom. The black curve is the density of variable $Y$ with
measurement error following a scaled beta distribution. We see that
the distribution with measurement error is very different from the
original distribution, so some method is needed to correct for this
difference.

Formally, we have a sample of observations $Y_1,\ldots, Y_n$ given by
the additive error model $Y_j=X_j+\epsilon_j$, where $\epsilon_j$ are
i.i.d. The latent variables $X_j$ are i.i.d. with density $f_x$, but
are not observed. We are interested in estimating the density $f_x$
from this data, and either a known error distribution for
$\epsilon_j$, or a separate sample from the error distribution (for
example obtained by repeated measurements of one observation).

There have been a number of methods developed for this problem, almost
all based on the characteristic function $\chi_X(t)={\mathbb
  E}{e^{iX}}$. The key to the methods is that the convolution that
gives the distribution of $Y$ becomes pointwise multiplication of
characteristic functions. That is, if $Y$ has distribution function
given by the convolution $F_Y(y)=\int_{-\infty}^\infty
f_x(x)F_\epsilon(y-x)\,dx$, then the characteristic function is given
by elementwise multiplication
$\chi_Y(t)=\chi_X(t)\chi_\epsilon(t)$. Elementwise multiplication is
easily inverted by elementwise division,
i.e. $\chi_X(t)=\frac{\chi_Y(t)}{\chi_\epsilon(t)}$, so if we know the
characteristic functions of $Y$ and $\epsilon$, then we can calculate
the characteristic function of $X$. Different deconvolution methods in
the literature are based on different estimators for the
characteristic function of $Y$ (and sometimes $\epsilon$), and
different regularisation and correction (the estimated $\chi_X$ is not
guaranteed to be the characteristic function of a distribution).  This
formulation in terms of the characteristic function also highlights
the difficult cases --- when $\chi_\epsilon(t)$ is very small, the
quotient becomes much larger, so estimation errors in the
characteristic function are magnified. In cases where
$\chi_\epsilon(t)$ converges quickly to zero as $t\rightarrow\infty$,
the error distribution is called supersmooth, and the deconvolution
problem is particularly challenging.

One widely used method is by \citet{liu1989consistent}. They use a
kernel density estimator for the characteristic function $\chi_Y$, a
known error distribution, and a method based on minimising mean
squared error (MSE) to select the boundary and bandwidth. They prove
the method is consistent in cases where the error distribution and
kernel function are symmetric.

Assuming the error distribution as known is unrealistic. A more
reasonable approach is to model the error distribution parametrically
or nonparametrically.  \citet{delaigle2004bootstrap} used the moment
estimators for the parametric error distribution parameters in their
example, with the main contribution of the paper being a bootstrap
bandwidth selection method for the deconvolution kernel density
estimation of $f_x$. The R package \texttt{decon}~\citep{wang2011deconvolution} implements their method for Gaussian or
Laplace error, using either direct computation or a fast Fourier
transform (FFT).

The error distribution can be nonparametrically estimated from repeated observations of the contaminated variable $Y$, see for example,  \citet{delaigle2008deconvolution}
and \citet{comte2014deconvolution}.  Alternatively it can be estimated from a pure error sample (which can be obtained through repeated measurements of
the same quantity which is independent of the observed sample of $Y$). For example, \citet{kerkyacharian2011localized} used the empirical characteristic
function from a pure error sample for estimating $\chi_\epsilon$. 
There are a few other approaches to deconvolution with a pure error
sample, mainly differing in details such as bandwidth selection. 

The R package \texttt{deamer}~\citep{stirnemann1deconvolution}
implements several deconvolution methods based on the FFT algorithm,
including situations for known error density; for unknown error
density with an auxiliary sample of i.i.d. pure errors (method by
\citet{comte2011data}) which is only proven consistent under the
assumption that $f_x$ is ordinary smooth or supersmooth; and for
unknown error density with replicate observations for variable $Y$
with the assumption that the error distribution is symmetric around
zero (methods by \citet{delaigle2008deconvolution} and
\citet{comte2014deconvolution}).

While the Fourier-based methods are mathematically elegant,
estimation of the characteristic function is much more challenging
than estimation of other distributional quantities, and because of the
division by $\chi_\epsilon(t)$, values where $\chi_\epsilon(t)$ is
small can cause instability in the estimates. The methods in the
literature get around this by limiting the range of $t$. This hard
limitation can result in poor estimation.  Another source of
significant errors comes in the correction stage where the estimate
$\hat{\chi_X}(t)=\frac{\hat{\chi_Y}(t)}{\hat{\chi_\epsilon}(t)}$ is
adjusted to become the characteristic function of a distribution.
Because of these difficulties, existing methods perform very poorly
unless the sample size and the signal-noise ratio (SNR) are both
large.

In this paper, we develop a completely new method based on maximising the 
log-likelihood of the data plus a smoothness penalty on $f_x$,
similar to the penalty used in smoothing splines. This method can
overcome the difficulties with the Fourier-based methods, and produce
better estimates, particularly when the sample size is small, or the
SNR is low.

The outline of our paper is as follows. In
Section~\ref{MethodSection}, we introduce our method and define the
estimator. In Section~\ref{ComputationSection}, we deal with the
difficult computational challenges of the estimator. We discuss the
theoretical convergence properties of our method in
Section~\ref{TheorySection}, with proofs in the appendix. In
Section~\ref{SimulationSection}, we compare penalised MLE with \texttt{decon}
and \texttt{deamer} on simulated datasets. In Section~\ref{RealDataSection}, we
apply our method to real data and compare the performance with \texttt{decon}
and \texttt{deamer}. The paper finishes with the conclusions in
Section~\ref{ConclusionSection}.

\section{Deconvolution based on penalised log-likelihood}\label{MethodSection}

We want to estimate the density function $f_x(x)$ of the continuous random
variable $X$ from a sample $\{y_1, y_2, \ldots, y_n\}$ of the random
variable $Y=X+\epsilon$, where $\epsilon$ is another random
variable. For simplicity, we will start with the case where the
distribution of $\epsilon$ is known. If the distribution of $\epsilon$
is unknown, but we have a pure error sample, $\{e_1, e_2, \cdots,
e_M\}$, then we may apply our method with the empirical distribution
of $\epsilon$ from this sample. We will also assume that $X$ has
finite support $[l,u]$. In theory, we could set the support to be
$(-\infty,\infty)$, but this causes practical challenges with the
optimisation.

The density function $f_y$ is obtained via the convolution
$$f_y(y)={\mathbb E}_{\epsilon}(f_x(y-\epsilon))$$
so the log-likelihood of our data for a particular density function
$f_x$ is
$$\sum_{i=1}^{n}\log f_y(y_i)=\sum_{i=1}^{n}\log \left(\int_l^u f_x(x) d\mu_\epsilon(y_i-x)\right)$$
where $\mu_\epsilon$ is the probability measure of the error
distribution. (We need to assume that $X$ has a continuous
distribution, but our method has no problems with $\epsilon$ having a
discrete or mixed distribution.)

Our method is to minimise the negative
log-likelihood function of $\{y_1, y_2, \cdots, y_n\}$ plus a penalty
term on the smoothness of function $f_x$. The smoothness penalty
$\psi(f_x)$ is given by
$$\psi(f_x)=\langle f_x'',f_x''\rangle=\int_{l}^{u}f''(x)^2dx$$
where $\langle.,.\rangle$ denotes the inner product on the Hilbert
space $L^2[l,u]$.
This
penalty is widely used in the smoothing splines method
\citep{green1993nonparametric}. This gives us the penalised negative
log-likelihood function:

\begin{equation}
J=-\sum_{i=1}^{n}\log\left(\int_l^u f_x(x) d\mu_\epsilon(y_i-x)\right)+\lambda_n \langle f_x'',f_x''\rangle
\label{obj}
\end{equation}
where $\lambda_n$ is the smoothness penalty tuning parameter used to
control the smoothness of $f_x$. To minimise $J$, we will
rewrite the log-likelihood term by integrating by parts twice, to
express the log-likelihood as a function of $f_x''$. We can then solve
for the optimal function $f_x''$ and obtain $f_x$  by
integrating twice.

Using integration by parts twice to compute the integral defining $f_y(y_i)$, we get
\begin{align}
f_y(y_i)=&   \int_l^u f_x(x)d\mu_\epsilon( y_i-x) \nonumber\\
=&[-f_x(x)F_\epsilon( y_i-x)]_l^u+\int_l^u F_\epsilon( y_i-x)f'_x(x) dx\nonumber\\
=&[-f_x(x)F_\epsilon( y_i-x)]_l^u-[f_x'(x)H( y_i-x)]_l^u+\int_l^u f_x''(x)H( y_i-x) dx  \label{eq3}
\end{align}
where $F_\epsilon$ denotes the cumulative distribution function of $f_\epsilon$ and
$H(v)=\int_{-\infty}^v  F_\epsilon(u) du$.

We assume that $f_x$ is a density function and $[l,u]$ contains the
positive support of $f_x$, so without loss of generality we let
$f_x(l)=f_x'(l)=0$ and $f_x(u)=f_x'(u)=0$. The first two terms in~\eqref{eq3} vanish, giving
$$f_y(y_i)=\int_l^u f_x''(x)H( y_i-x) dx=\langle f_x'', h_i(x) \rangle$$
where $h_i(x)=H( y_i-x)$.

Using this inner product, our objective function (\ref{obj}) becomes
\begin{equation}
J=-\sum_{i=1}^{n}\log(\langle f_x'',h_i\rangle)+\lambda_n\langle f_x'',f_x''\rangle \label{obj2}
\end{equation}
We want to minimise this $J$ subject to the constraints:
\begin{align}
f_x(l) =f_x'(l)&=0, \label{cons1}\\
f_x(u) =f_x'(u)&=0,  \label{cons2} \\
\forall  x\in (l,u), f_x(x) & \geqslant 0,   \label{cons3} \\
 \int_l^u f_x(x)dx &=1.  \label{cons4}
\end{align}

Constraint~\eqref{cons1} gives a unique solution for $f_x$ given $f_x''$. The other conditions can be rewritten as conditions on $f_x''$.

From~\eqref{cons1} and~\eqref{cons2}, we get $\langle
f_x''(x),x\rangle$=0 and $\langle f_x''(x),1\rangle$=0.  From
Constraints~\eqref{cons1},~\eqref{cons2} and~\eqref{cons4} we get $\langle
f_x''(x),x^2\rangle=2$.  For the non-negativity constraint,~\eqref{cons3},
we integrate $f_x'(x)$ by parts to get the following inner products.

\begin{align}
f_x(x)&=\int_l^xf_x'(r)dr
=[f_x'(r)(r-x)]_l^x-\int_l^xf_x''(r)(r-x)dr
=\langle f_x''(r),(x-r)_{+}\rangle\label{evaluateleft}\\
f_x(x)&=-\int_x^uf_x'(r)dr
=-[f_x'(r)(r-x)]_x^u+\int_x^uf_x''(r)(r-x)dr
=\langle f_x''(r),(r-x)_{+}\rangle\label{evaluateright}
\end{align}
We can also take any affine combination of
Equations~\eqref{evaluateleft} and~\eqref{evaluateright} to evaluate
$f_x(x)$. That is $f_x(x)=\lambda\langle
f_x'',(x-r)_+\rangle+(1-\lambda)\langle f_x'',(r-x)_+\rangle$ for any
$\lambda$. In particular, it is convenient to set
$b_x(r)=\frac{(u-x)^2((u-x)+3(x-l))}{(u-l)^3}(x-r)_++\frac{(x-l)^2((x-l)+3(u-x))}{(u-l)^3}(r-x)_+-2\frac{(u-x)^2(x-l)^2}{(u-l)^3}$, so that
\begin{adjustwidth}{-4cm}{-4cm}
\begin{align*}
\langle b_x,1\rangle&=\frac{(u-x)^2((u-x)+3(x-l))}{(u-l)^3}\int_l^x (x-r)\,dr+\frac{(x-l)^2((x-l)+3(u-x))}{(u-l)^3}\int_x^u
(r-x)\,dr
\\
&\qquad\qquad\qquad\qquad\qquad\qquad\qquad\qquad
-2\frac{(u-x)^2(x-l)^2}{(u-l)^3}\int_l^u1\,dr\\
&=
\frac{1}{2}\left(\frac{(u-x)^2((u-x)+3(x-l))}{(u-l)^3}(x-l)^2+\frac{(x-l)^2((x-l)+3(u-x))}{(u-l)^3}(u-x)^2\right)
-2\frac{(u-x)^2(x-l)^2}{(u-l)^2}\\
&=0\\
\langle b_x,r\rangle&=\langle b_x,r-x\rangle\\
&=\frac{(u-x)^2((u-x)+3(x-l))}{(u-l)^3}\int_l^x (r-x)(x-r)\,dr+
\frac{(x-l)^2((x-l)+3(u-x))}{(u-l)^3}\int_x^u(r-x)(r-x)\,dr\\
&\qquad\qquad\qquad\qquad\qquad\qquad\qquad\qquad
-2\frac{(u-x)^2(x-l)^2}{(u-l)^3}\int_l^u(r-x)\,dr\\
&= \frac{1}{3}\left(-\frac{(u-x)^2((u-x)+3(x-l))}{(u-l)^3}(x-l)^3+
\frac{(x-l)^2((x-l)+3(u-x))}{(u-l)^3}(u-x)^3\right)\\
&\qquad\qquad\qquad\qquad\qquad\qquad\qquad\qquad
-\frac{(u-x)^2(x-l)^2}{(u-l)^3}
((u-x)^2-(x-l)^2)\\
&=\frac{(u-x)^2(x-l)^2}{(u-l)^3}\left(\frac{1}{3}\left(3(u-x)^2-3(x-l)^2\right)-((u-x)^2-(x-l)^2))\right)\\
&=0
\end{align*}
\end{adjustwidth}

And the non-negativity constraint is $\langle
f_x'',b_x\rangle\geqslant 0$ for all $x\in(l,u)$.

Thus we have converted the constraints in the minimisation problem
from~\eqref{obj2} to Hilbert Space inner product conditions on $f_x''$:

\begin{align}
J&=-\sum_{i=1}^{n}\log(\langle f_x'',h_i\rangle)+\lambda_n\langle
f_x'',f_x''\rangle\label{obj3}\\
\langle f_x'',1\rangle &=0 \\
\langle f_x'',x\rangle &=0 \\
\langle f_x'',x^2\rangle &=2 \\
\langle f_x'',b_x\rangle &\geqslant 0,   \forall  x\in (l,u)\label{lastconstraint1}
\end{align}

The advantage of writing the optimisation problem in this way is that
in the Hilbert space $L^2$, we can decompose $f_x''$ as a linear
combination of $\{h_i|i=1,\ldots,n\}, 1, x, x^2$, and
$\{b_x|x\in(l,u)\}$ plus a function orthogonal to all these
elements. If we let $f_0$ be the linear combination of these elements, and
let $f_\perp$ be the orthogonal part, we see that $f_\perp$ only
affects the penalty term $\langle f_x'',f_x''\rangle$, and because of
the orthogonality, $\langle f_x'',f_x''\rangle=\langle
f_0,f_0\rangle+\langle f_\perp,f_\perp\rangle$ is clearly minimised
when $f_\perp=0$, so we have shown that the optimal solution is a
linear combination of $\{h_i|i=1,\ldots,n\}, 1, x, x^2$, and
$\{b_x|x\in(l,u)\}$. Thus, we have reduced the constrained
optimisation over the whole Hilbert space $L^2((l,u))$ to a
constrained optimisation of the coefficients in this basis. This is
still infinite dimensional. However, we can get a good approximate
solution to the problem by only requiring a finite subset of the
non-negativity constraints. After we restrict to this condition, the
problem has become a standard finite-dimensional constrained
optimisation problem.

More precisely, suppose we choose $k-3$ basis functions of $\{b_x|x\in(l,u)\}$, corresponding to $k-3$ values for $x$, denoted by  
$x_{n+4},\ldots,x_{n+k}$, as constraint values, and want to minimise
the objective function $J$ subject to
$f_x(l)=f_x'(l)=0$,  conditions (11)-(13) and condition (14) for $i=n+4,\ldots,n+k$.
Let the basis functions be given by
\begin{equation}
  k_i(r)=\left\{\begin{array}{ll}
  h_i(r)&\textrm{if }i=1,\ldots,n\\
  1&\textrm{if }i=n+1\\
  r&\textrm{if }i=n+2\\
  r^2&\textrm{if }i=n+3\\
  b_{x_i}(r)&\textrm{if }i=n+4,\ldots,n+k
  \end{array}\right.
\end{equation}
From the above argument,
we know that the solution to the optimisation problem is given by
\begin{equation}
 f_x''(r)=\sum_{i=1}^{n+k}\alpha_i k_i(r)
\end{equation}
for some coefficients $\alpha_1,\ldots,\alpha_{n+k}$. If we form a
matrix of inner products of these basis terms by
\begin{equation}
A_{ij}=\langle k_i,k_j\rangle
\end{equation}
then the optimisation problem from
Equations~\eqref{obj3}--\eqref{lastconstraint1} can be rewritten as:
Minimise
\begin{equation}
J=-\sum_{i=1}^n \log\left(\sum_{j=1}^{n+k}A_{ij}\alpha_j\right)+\lambda_n\sum_{i=1}^{n+k}\sum_{j=1}^{n+k}A_{ij}\alpha_i\alpha_j\label{obj4}
\end{equation}
subject to
\begin{align}
\sum_{j=1}^{n+3}A_{(n+1)j}\alpha_j&=0\\
\sum_{j=1}^{n+3}A_{(n+2)j}\alpha_j&=0\\
\sum_{j=1}^{n+k}A_{(n+3)j}\alpha_j&=2\\
(\forall
i\in\{n+4,\ldots,n+k\})\sum_{j=1}^{n+k}A_{ij}\alpha_j&\geqslant 0\label{lastcondition2}
\end{align}

\section{Practical Optimisation Issues}\label{ComputationSection}

While the optimisation problem in
Equations~\eqref{obj4}--\eqref{lastcondition2} looks like a fairly
standard multidimensional optimisation problem, it is not completely
straightforward. There are a number of choices that need to be made
for the optimisation. Our implementation of the method uses the
Nelder-Mead method~\citep{nelder1965mead} in the \texttt{optim}
function from the \texttt{stats} package in \texttt{R}. This is a
simplex-based optimisation method. It has the advantage of being
robust. This is particularly useful for our problem because the
log-likelihood function is only defined for values of the parameters
such that the convolved density is positive, so less robust methods
sometimes produce invalid parameter values.

In addition to choice of optimisation method, there are a number of
particular challenges present in this problem. In this section, we
discuss the approach taken to deal with the following challenges: what
values to set for $l$ and $u$; which non-negativity constraints to
impose; initial values for parameters; computational singularity; and
selection of tuning parameter.

\subsection{Choosing $l$ and $u$}

In theory, as $l$ and $u$ tend to $-\infty$ and $\infty$ respectively,
we should expect the solution to converge to the solution for support
$(-\infty,\infty)$. However, in practice, ensuring the non-negativity
constraints all hold becomes more difficult when the support is large
and far from the observed data. Also, setting a narrower support
reduces the computational difficulty of calculating the necessary
inner products $A_{ij}$.

We initially set the support by first taking the empirical support
$l_Y,u_Y$ of the observed data points, and the empirical support
$l_e,u_e$ of the pure error sample. If the support of $X$ is $(l,u)$
and the support of $\epsilon$ is $(l_e,u_e)$, then the support of $Y$
is $(l_Y,u_Y)=(l+l_e,u+u_e)$. Therefore, the support of $X$ is given
by $l=l_Y-l_e$ and $u=u_Y-u_e$. However, because it is easier to
adjust the boundaries inward, and because we only have estimated
values of $l_Y$, $u_Y$, $l_e$ and $u_e$, we start with a wider
interval than this.

When we fit the P-MLE for the initial support, it often happens that
$\hat{f}_x$ is negative near the boundaries and positive away from the
boundaries. We then use an adaptive method to shrink the boundaries so
that $\hat{f}_x$ is positive on $(l,u)$. Suppose our current estimate
for the support is $(l_k,u_k)$, and $\hat{f}_x$ is negative on the
interval $(l_k,w_k)$ with a minimum at $m_k$, and on the interval
$(v_k,u_k)$ with a minimum at $n_k$. Then our next estimate for the
support is $\left(\frac{l_k+m_k}{2},\frac{n_k+u_k}{2}\right)$.

\subsection{Non-negativity constraints}

To ensure the non-negativity of density function, a large number of 
constraints will be involved in the optimisation problem. This may cause numerical singularity of matrix $A_{ij}$ and make the computational issues nontrivial. 
So for simplicity, we 
choose 30 evenly spaced points to cover the full range on the support 
of $f_x$. With the smoothness penalty, the estimated function will often be non-negative on the whole support. 

\subsection{Initial values}

Random starting points are commonly used in many optimisation
problems.  However, in our case, random coefficients often do not
produce valid density functions to satisfy the non-negative
constraints. When the estimated density is negative somewhere, the
log-likelihood cannot be computed. As a result, we need a better
method to find the appropriate starting points. We take a kernel
density estimate for $f_Y$ as our starting point. We use a Gaussian
kernel, with bandwidth estimated using Silverman's
rule-of-thumb~\citep{Silverman1986}, in our implementation. We then
project this estimated kernel density of $f_Y$ orthogonally into the
space spanned by our basis functions to get the initial
coefficients. This can be easily achieved by evaluating the functions
of $f_Y$ and all basis functions at a large number of values over the
support and solving a regression problem.

\subsection{Computational Singularity}

A large number of basis functions are translations of the function
$H$. When $H$ is a relatively smooth function, and when there are a
lot of observed points, these translations can become close to be
linearly dependent, which can cause numerically unstable solutions, or
convergence problems and even sometimes incorrect results. We
alleviate this problem using a subsampling approach on the basis. We
choose a subsample of size $S$ such that for a sample of $S$
observation points, the translations $h_i$ are not computationally
singular. We then solve the optimisation problem for this subsample of
basis elements (note that we use the whole data set to compute
log-likelihood). In our experience, $S=30$ usually works well. We
repeat this for a number of subsamples, and average the results. To
improve the reliability, the subsamples are stratified by dividing the
support of $Y$ into $S$ intervals and selecting one sample point from
each interval. This ensures that the subsample points are well
spread-out, reducing the computational instability. We take enough
subsamples to ensure that with high probability, the majority of basis
elements have been used in at least one subsample. Even with this
subsampling, it occasionally still happens that the optimisation fails
for some subsamples. In these cases, a replacement subsample is drawn.

\subsection{Selecting $\lambda_n$}\label{TuningSubsection}

The smoothness parameter $\lambda_n$ needs to be selected. This is a
tuning parameter, so the usual approach is via cross-validation. We
divide the data into a training set and a validation set, use the
training set to fit $f_x$, then evaluate the likelihood on the validation
data, where the log-likelihood is $\sum_j\log((\hat f_x*\hat f_e)(y_j))$. 
We then choose
$\lambda_n$ to maximise this cross-validated log-likelihood.

In cases where computation time is limited, and we need to fit our
method more quickly (for example in the simulations in
Section~\ref{SimulationSection}), we have used the following heuristic
approach to quickly select $\lambda_n$.

The role of $\lambda_n$ is to find a reasonable balance between the
log-likelihood and the smoothness penalty. For a given sample size,
the right balance should be based on controlling the relative size of
the two terms. Thus, we take the initial density estimate (before
optimisation, since we need to choose $\lambda_n$ before we can
optimise) and compute the derivatives of the log-likelihood and
derivatives of the smoothness penalty terms with respect to all the
coefficients. We then set 
$\lambda_n=\frac{1}{R}\frac{\sum_{i=1}^{n+k} \left|\frac{\partial
    l(f_x;y)}{\partial \alpha_i}\right|}{\sum_{i=1}^{n+k}
  \left|\frac{\partial \langle f'',f''\rangle}{\partial
    \alpha_i}\right|}$
for some constant $R$.
From experience, we found that the following values for $R$ work
well.

\hfil\begin{tabular}{rc}
Sample size & $R$\\ 
\hline
  30      & $10^4$ \\
  100     & $10^5$ \\
  300     & $10^6$
\end{tabular}

\section{Theory}\label{TheorySection}

In this section, we discuss the theoretical large-sample performance
of our method. For typical deconvolution problems, theoretical
performance is controlled by the identifiability of the deconvolution
problem. In particular, if one of the Fourier coefficients in the
error distribution is zero, then the deconvolution problem is not
identifiable, since adding the appropriate Fourier term to a ``true''
distribution does not change its convolution. In practice,
distributions rarely have zero Fourier coefficients, so the method is
not completely unidentifiable, but when the error distribution is
super-smooth, then its Fourier coefficients quickly converge to zero,
which makes the problem practically unidentifiable. We resolve this
issue by studying the error in the convolved side, rather than the
deconvolved side. That is, we can show that $\hat{f}_x*f_e\rightarrow
f_y$ under general circumstances. This allows us to avoid worrying
about ordinary smooth and super-smooth error distributions. We are
able to show that P-MLE consistently produces a valid solution to the
deconvolution problem. If there are multiple valid solutions, then the
choice made by P-MLE may or may not be the truth. We are able to show
that

\begin{theorem}
  Let the true density $f_x$ be twice continuously differentiable, and
  let the convolved density be $f_y=f_x*f_e$. Let $\hat{f}_x$ be the
  P-MLE estimate for $f_x$ and $\hat{f}_y=\hat{f}_x*f_e$.  If
  $\lambda_n=C_1 n^{\frac{7}{8}}\log(n)^{\frac{1}{8}}\sqrt{|u-l|}$,
  for a certain constant $C_1$, then almost surely, for all
  sufficiently large $n$, $\lVert
  \hat{f}_y-f_y\rVert_\infty<C_2n^{-\frac{1}{32}}|u-l|^{\frac{1}{8}}\log(n)^{\frac{1}{32}}$
  for some constant $C_2$.  (The constants $C_1$ and $C_2$ depend on
  the smoothness $\psi(f_x)$ and $\psi(f_y)$ of the true distribution,
  but not on $n$, $l$, or $u$.)
\end{theorem}

The details of this proof are in the supplementary appendix. Here, we
briefly outline our approach.

\begin{enumerate}[1.]

\item By the
  Dvoretzky-Kiefer-Wolfowitz
  inequality~\citep{DvoretzkyKieferWolfowitz},
letting $G$ denote the cumulative distribution function
$G(t)=\int_{-\infty}^t f_y(s)\,ds$ and $G_n$ denote the empirical distribution
function of the sample $y_1,\ldots,y_n$,
  the inequality
\begin{equation}
  \lVert G_n-G\rVert_\infty\leqslant\sqrt{\frac{\log(n)}{n}}\label{DKW}
  \end{equation}
  almost surely holds for all sufficiently  large $n$.

\item We construct a sequence of estimators $f^*_n$ such that
  $\psi(f^*_n)<2\psi(f_x)$, but the average likelihood
  $\frac{1}{n}\sum_{i=1}^n (f^*_n*f_e)(y_i)$ of the convolved
  density is bounded below by the negative entropy of the true data
  distribution minus a $O(n^{-\frac{1}{2}}\log(n)^{\frac{3}{2}})$ term.

\item We construct a sequence of estimators $\tilde{g}_n$ from the
  data such that $\lVert
  \tilde{g}_n-f_y\rVert_\infty=O\left(\left(\frac{\log(n)}{n}\right)^{\frac{1}{4}}\right)$
  and for any Lipschitz density $g_L$, the log-likelihood
  $\frac{1}{n}\sum_{i=1}^n g_L(y_i)$ is accurately approximated by
  $\int_{l}^u\tilde{g}_n(x)\log(g_L(x))\,dx$.

\item Using the sequence $\tilde{g}_n$, we show that for any density
  function $f$, if $\lVert f*f_e-f_y\rVert_\infty>\rho_n$, for a
  suitably chosen $\rho_n\rightarrow 0$ then the penalised average
  log-likelihood of the data is smaller than the penalised average
  log-likelihood for the estimator $\hat{f}_n$.
  From this, we deduce that the P-MLE, $\hat{f}_n$ must almost surely
  satisfy $\lVert \hat{f}_n*f_e-f_y\rVert_\infty\leqslant \rho_n$ for
  all sufficiently large $n$.

\end{enumerate}

\section{Simulations}\label{SimulationSection}

In this section, we compare the performance of our method with two of
the most popular methods in the literature:
\texttt{deamer}~\citep{stirnemann1deconvolution} and
\texttt{decon}~\citep{wang2011deconvolution}. For \texttt{decon}, we compare
with both their methods with and without FFT. We compare the
performance under a range of true and error distributions, including
common examples from \citet{comte2011data} and
\citet{comte2006penalized}. We simulate with a range of different
sample sizes and SNRs, including many cases with smaller sample size
and SNR, which are often excluded from simulations in the literature,
because they highlight a particular weakness of existing methods.

\subsection{Simulation design }

To cover a large range of scenarios of interest, we simulate all
combinations from seven true distributions, three error distributions,
three SNRs, and three sample sizes.
The true distributions used in the simulation are:

\begin{tabular}{ll}
Normal distribution.      &  $X\sim N(0,1)$\\
Chi-square distribution.  &  $X\sim \chi^2(4)/\sqrt{8}$ \\
Beta distribution.        &  $X\sim \sqrt{39.2}\textrm{Beta}(2,5)$ \\
Laplace distribution.     & $f_x(x)=\frac{1}{\sqrt{2}}\exp(-\sqrt{2}|x|)$\\
Mixed normal distribution. & $X\sim\frac{2}{\sqrt{29}}(0.5N(-3,1)+0.5N(2,1))$  \\
Mixed gamma distribution.  & $X\sim(0.4\Gamma(5,1)+0.6\Gamma(13,1))/\sqrt{25.16}$ \\
Cauchy distribution.       & $f_x(x)=(1/\pi)/(1+x^2)$
\end{tabular}

These distributions cover a range of situations including heavy tails,
light tails, symmetric and skew distributions, unimodal and bimodal
distributions, and different levels of smoothness. With the exception
of the Cauchy distribution (which has infinite variance), these
densities have all been normalised to have unit variance. These
distributions have previously been used in the
literature~\citep{comte2006penalized}. However, the standardisation
used in the literature was incorrect for the mixture distributions, so
we have corrected the standardisation constants here.

For the error distribution, we use the following three distributions,
scaled by a factor $C$. When the true distribution has finite variance
(i.e. when it is not Cauchy), $C^2$ is the inverse of the SNR.

\begin{tabular}{ll}
Laplace noise.    & $f_\epsilon(\epsilon)=\frac{1}{\sqrt{2}}\exp(-\sqrt{2}|\epsilon|)$\\
Gaussian noise.   & $f_\epsilon(\epsilon)=\frac{1}{\sqrt{2\pi}}\exp(-0.5\epsilon^2)$\\
Beta noise.       & $f_\epsilon(\epsilon)=30\sqrt{39.2}\epsilon(1-\epsilon)^4$
\end{tabular}

We choose these three distributions because they have different levels
of smoothness. The normal distribution is super smooth and the Laplace
distribution is ordinary smooth. Both the normal distribution and the
Laplace distribution are often used in the literature on measurement
error. The beta distribution often arises as convergence error. The
parameters of the three distributions are chosen to ensure the error
distribution has unit variance. Because the \texttt{decon} package
only permits a limited number of chosen distribution families, which
does not include the beta distribution, we were unable to compare its
performance in the beta noise simulations.

In our simulation we study three sample sizes: 30, 100 and 300. In
each case, we use the same sample size for the noisy data and for the
pure error sample. Note, however, that because the \texttt{decon}
package requires a known error distribution family, we provided the
true error distribution to this package, which gives an unfair
advantage to this method. This unfair advantage is particularly
significant in the small sample size cases.

In each scenario, we simulate 100 replicates. For computational
efficiency, we use the heuristic method from
Section~\ref{TuningSubsection} to choose the smoothness penalty $\lambda_n$.

\subsection{Simulation Results}\label{SimulationResultsSubsection}

We use Mean Integrated Squared Error (MISE) to evaluate the
performance of the estimators in each scenario. This measure is
defined by $\textrm{MISE}=E\int (\hat{f}_x(x)-f_x(x))^2dx$. Here the
MISE is computed as the empirical mean of the approximated ISE $\int
(\hat{f}_x(x)-f_x(x))^2dx$ over 100 simulation replicates. This is a
traditional method for evaluating the performance of deconvolution
methods that is widely used in the literature
\citep{comte2006penalized}. We calculate the integral over an interval
which contains both the support of the underlying true distribution
from the $0.01\%$ quantile to the $99.99\%$ quantile and the estimated
boundaries. \texttt{Deamer} does not give estimated boundaries for the true
density, so we calculate its ISE on the same interval used for
P-MLE. We numerically calculate the integral using the rectangle rule
with the squared error evaluated at evenly-spaced points, where the
spacing is chosen for each method so that there are 1000 (or 1024 for
\texttt{decon} with FFT) points within the interval returned by the method. We
found that changing the number of points used to estimate the ISE did
not noticeably affect the results.

\newcommand{\resultstablecomments}{\parbox{\textwidth}{
    The best overall performance in
  each simulation is highlighted in yellow if it is significantly
  better than other methods and in orange if the difference is not significant.}}

\newcommand{\resultstablecaption}[1]{  \caption[MISE of the estimates when the underlying distribution is
    #1]{MISE of the estimates when the underlying distribution is
    #1 \par\vspace{0.5\baselineskip} \resultstablecomments}
}

\begin{table}
  \resultstablecaption{normal  \label{tab:1}}
  \centering
    \begin{tabular}{llllllll}
    \hline
  \multicolumn{2}{c}{normal-normal}  & \multicolumn{2}{c}{n=30} & \multicolumn{2}{c}{n=100} & \multicolumn{2}{c}{n=300} \\
\cline{3-8}
  \multicolumn{2}{c}{} &mean & se&mean&se&mean&se    \\
\hline
\multirow{4}[0]{*}{SNR=4} & P-MLE   & \cellcolor{yellow}{0.0193} & 0.0015 & \cellcolor{orange}{0.0085} & 0.0007 & 0.0055 & 0.0004 \\
          & deamer & 0.0486 & 0.0021 & 0.0097  & 0.0006 & \cellcolor{yellow}{0.0029} & 0.0003 \\
          & decon & 0.0286 & 0.0015 & 0.0147 & 0.0007 & 0.0082 & 0.0004 \\
          & decon (with FFT) & 0.0283 & 0.0016 & 0.0151 & 0.0008 & 0.0081 & 0.0004 \\
          \cline{2-8}
    \multirow{4}[0]{*}{SNR=1} & P-MLE   & \cellcolor{yellow}{0.0232}  & 0.0018 & \cellcolor{yellow}{0.0121} & 0.0011 &\cellcolor{yellow}{0.0068} & 0.0007 \\
          & deamer & 0.1104  & 0.0025 & 0.0622 & 0.0014 & 0.0353 & 0.0004 \\
          & decon & 0.0650 & 0.0023 & 0.0375 & 0.0012 & 0.0246 & 0.0008 \\
          & decon (with FFT) & 0.0636 & 0.0018 & 0.0382 & 0.0012 & 0.0245 & 0.0009 \\
          \cline{2-8}
    \multirow{4}[0]{*}{SNR=0.25} & P-MLE   & \cellcolor{yellow}{0.0726} & 0.0025 &\cellcolor{yellow}{0.0360} & 0.0019 &  \cellcolor{yellow}{0.0198}     &0.0021  \\
          & deamer & 0.1973 & 0.0001 & 0.1782 & 0.0032 &  0.1247     & 0.0002 \\
          & decon & 0.1066 & 0.0032  & 0.1039 & 0.0015 &   0.0839    &0.0011  \\
          & decon (with FFT) & 0.1073 & 0.0033 & 0.1030 & 0.0015 &   0.0836    & 0.0011 \\
          \hline
    normal-laplace &       &       &       &       &       &       &  \\
    \multirow{4}[0]{*}{SNR=4} & P-MLE   & \cellcolor{yellow}{0.0187} & 0.0015 & \cellcolor{orange}{0.0072} & 0.0006 & 0.0046 & 0.0004 \\
          & deamer & 0.0455 & 0.0019 & 0.0075 & 0.0005 & \cellcolor{yellow}{0.0029} & 0.0003 \\
          & decon & 0.0526 & 0.0052 & 0.0517 & 0.0062 & 0.0391 & 0.0050 \\
          & decon (with FFT) & 0.0571 & 0.0064 & 0.0506  & 0.0065 & 0.0456 & 0.0057 \\
          \cline{2-8}
    \multirow{4}[0]{*}{SNR=1} & P-MLE   & \cellcolor{yellow}{0.0241} & 0.0017 & \cellcolor{yellow}{0.0123} & 0.0009 & \cellcolor{yellow}{0.0064} & 0.0008 \\
          & deamer & 0.0974 & 0.0029 & 0.0462 & 0.0017 & 0.0156 & 0.0005 \\
          & decon & 0.1083 & 0.0121 & 0.1142 & 0.0141 & 0.1010 & 0.0122 \\
          & decon (with FFT) & 0.1125 & 0.0158 & 0.1222 & 0.0149 & 0.0902  & 0.0091 \\
          \cline{2-8}
    \multirow{4}[0]{*}{SNR=0.25} & P-MLE   & \cellcolor{yellow}{0.0682} & 0.0025 & \cellcolor{yellow}{0.0412} & 0.0017 & \cellcolor{yellow}{0.0217} & 0.0015 \\
          & deamer & 0.1814 & 0.0030  & 0.1265 & 0.0015 & 0.0900  & 0.0026 \\
          & decon & 0.2186 & 0.0263 & 0.2251 & 0.0236 & 0.2042 & 0.0185 \\
          & decon (with FFT) & 0.2227 & 0.0295 & 0.2510 & 0.0378 & 0.2409 & 0.0280 \\
          \hline
    normal-beta &       &       &       &       &       &       &  \\
    \multirow{2}[0]{*}{SNR=4} & P-MLE   &\cellcolor{yellow}{0.0307} & 0.0038 & 0.0157 & 0.0012 & 0.0145 & 0.0008 \\
          & deamer & 0.0518 & 0.0020 & \cellcolor{orange}{0.0101} & 0.0057 & \cellcolor{yellow}{0.0032} & 0.0002 \\
          \cline{2-8}
    \multirow{2}[0]{*}{SNR=1} & P-MLE   & \cellcolor{yellow}{0.0572} & 0.0049 & \cellcolor{yellow}{0.0463} & 0.0022 & 0.0431 & 0.0021 \\
          & deamer & 0.1111 & 0.0024 & 0.0641 & 0.0013 & \cellcolor{orange}{0.0411} & 0.0006 \\
          \cline{2-8}
    \multirow{2}[0]{*}{SNR=0.25} & P-MLE   & \cellcolor{yellow}{0.0824} & 0.0055 & \cellcolor{yellow}{0.0515} & 0.0048 & \cellcolor{yellow}{0.0414} & 0.0032 \\
          & deamer & 0.1966 & 0.0008  & 0.1784 & 0.0317 & 0.1242  & 0.0002
\\
          \hline
    \end{tabular}

\end{table}

\begin{table}
  \resultstablecaption{chi-squared  \label{tab:2}}
    \centering
    \begin{tabular}{llllllll}
    \hline
    \multicolumn{2}{c}{chisq-normal} & \multicolumn{2}{c}{n=30} & \multicolumn{2}{c}{n=100} & \multicolumn{2}{c}{n=300} \\
\cline{3-8}
    & &mean & se&mean&se&mean&se\\
\hline
\multirow{4}[0]{*}{SNR=4} &P-MLE& \cellcolor{yellow}{0.0453} & 0.0025 & \cellcolor{yellow}{0.0268} & 0.0013 & \cellcolor{yellow}{0.0175} & 0.0007 \\
& deamer & 0.0704  & 0.0025 & 0.0412 & 0.0012 & 0.0268 & 0.0006 \\
        & decon & 0.0994 & 0.0024 & 0.0673 & 0.0014 & 0.0484 & 0.0008  \\
        &  decon (with FFT)& 0.1015 & 0.0023 & 0.0672 & 0.0014 & 0.0480 & 0.0009  \\
\cline{2-8}
        \cline{2-8}
    \multirow{4}[0]{*}{SNR=1} & P-MLE&\cellcolor{yellow}{0.0737} & 0.0028 & \cellcolor{yellow}{0.0631} & 0.0024  & \cellcolor{yellow}{0.0488} & 0.0015  \\
    &deamer & 0.1424 & 0.0033 & 0.0993 & 0.0020  & 0.0750 & 0.0011  \\
         &decon & 0.1429 & 0.0019 & 0.1050 & 0.0015 & 0.0830 & 0.0011 \\
         &decon (with FFT) & 0.1435 & 0.0022 & 0.1034 & 0.0016 & 0.0826 & 0.0010  \\
         \cline{2-8}
    \multirow{4}[0]{*}{SNR=0.25} &P-MLE & \cellcolor{yellow}{0.1412} & 0.0028 & \cellcolor{yellow}{0.1037} & 0.0028 & \cellcolor{yellow}{0.0706}  & 0.0029  \\
         &deamer & 0.2600  & 0.0024 & 0.2195 & 0.0034 & 0.1962 & 0.0002  \\
          &decon& 0.1890 & 0.0030 & 0.1786 & 0.0013 & 0.1564 & 0.0010   \\
          &decon (with FFT)& 0.1870 & 0.0027 & 0.1779 & 0.0013 & 0.1551 & 0.0011  \\
\hline
    chisq-laplace & & & & & & & \\
    \multirow{4}[0]{*}{SNR=4} &P-MLE & \cellcolor{yellow}{0.0446} & 0.0026 & \cellcolor{yellow}{0.0260} & 0.0014 & \cellcolor{yellow}{0.0144} & 0.0005\\
         & deamer& 0.0684 & 0.0027 & 0.0369 & 0.0009 & 0.0206 & 0.0006 \\
          &decon & 0.0757 & 0.0060 & 0.0564 & 0.0058 & 0.0404  & 0.0036  \\
          &decon (with FFT) & 0.0750 & 0.0060  & 0.0625 & 0.0066 & 0.0434 & 0.0039 \\
          \cline{2-8}
    \multirow{4}[0]{*}{SNR=1} &P-MLE & \cellcolor{yellow}{0.0722} & 0.0025 & \cellcolor{yellow}{0.0516} & 0.0018 & \cellcolor{yellow}{0.0394}  & 0.0011 \\
         &deamer & 0.1244 & 0.0033 & 0.0773 & 0.0019 & 0.0505 & 0.0010  \\
          &decon& 0.1322 & 0.0126 & 0.1318 & 0.0130  & 0.1357 & 0.0189 \\
          &decon (with FFT)& 0.1377 & 0.0130  & 0.1321 & 0.0141 & 0.1151 & 0.0102 \\
          \cline{2-8}
    \multirow{4}[0]{*}{SNR=0.25} &P-MLE& \cellcolor{yellow}{0.1314} & 0.0028  & \cellcolor{yellow}{0.1035} & 0.0024  & \cellcolor{yellow}{0.0781} & 0.0026 \\
          &deamer& 0.2350 & 0.0039 & 0.1811 & 0.0026 & 0.1351 & 0.0017 \\
         &decon & 0.2114 & 0.0170 & 0.2874 & 0.0279 & 0.2968 & 0.0354 \\
          &decon (with FFT)& 0.2291 & 0.0210 & 0.2889 & 0.0280 & 0.3220 & 0.0340  \\
\hline
    chisq-beta &  &  & & & & & \\
    \multirow{2}[0]{*}{SNR=4} &P-MLE& \cellcolor{orange}{0.0659} & 0.0040 & 0.0497 & 0.0044 & 0.0532 & 0.0034 \\
         &deamer & 0.0702  & 0.0022 & \cellcolor{orange}{0.0423} & 0.0012 & \cellcolor{yellow}{0.0279} & 0.0007 \\
         \cline{2-8}
    \multirow{2}[0]{*}{SNR=1} &P-MLE& \cellcolor{yellow}{0.0874} & 0.0043 & \cellcolor{yellow}{0.0727} & 0.0035 & 0.0895 & 0.0038 \\
          &deamer& 0.1435 & 0.0027 & 0.0986 & 0.0019 & \cellcolor{yellow}{0.0724} & 0.0010  \\
          \cline{2-8}
    \multirow{2}[0]{*}{SNR=0.25} &P-MLE& \cellcolor{yellow}{0.1412} & 0.0054 & \cellcolor{yellow}{0.0803} & 0.0047 & \cellcolor{yellow}{0.0744} & 0.0048 \\
          &deamer& 0.2632 & 0.0020  & 0.2137 & 0.0032 & 0.1948 & 0.0006  \\
\hline
    \end{tabular}
\end{table}

\begin{table}
  \resultstablecaption{beta   \label{tab:3}}
  \centering
    \begin{tabular}{llllllll}
    \hline
    \multicolumn{2}{c}{beta-normal} & \multicolumn{2}{c}{n=30} &
    \multicolumn{2}{c}{n=100} & \multicolumn{2}{c}{n=300} \\
    \cline{3-8}
      \multicolumn{2}{c}{} &mean & se&mean&se&mean&se\\
\hline
\multirow{4}[0]{*}{SNR=4} & P-MLE   & \cellcolor{yellow}{0.0212} & 0.0015 & \cellcolor{yellow}{0.0114} & 0.0007 & \cellcolor{yellow}{0.0074} & 0.0004 \\
          & deamer & 0.0328 & 0.0013 & 0.0168 & 0.0006 & 0.0096  & 0.0002 \\
          & decon & 0.0646 & 0.0016 & 0.0395 & 0.0009 & 0.0257 & 0.0005 \\
          & decon (with FFT) & 0.0717  & 0.0019 & 0.0405  & 0.0009 & 0.0262 & 0.0049 \\
          \cline{2-8}
    \multirow{4}[0]{*}{SNR=1} & P-MLE   & \cellcolor{yellow}{0.0341}  & 0.0022 & \cellcolor{yellow}{0.0248} & 0.0020 &   \cellcolor{yellow}{0.0158}    &0.0011  \\
          & deamer & 0.0803 & 0.0023 & 0.0474 & 0.0013 &     0.0311  & 0.001 \\
          & decon & 0.0987  & 0.0019 & 0.0621  & 0.0010 &    0.0439   &0.0007  \\
          & decon (with FFT) & 0.0987 & 0.0018 & 0.0616 & 0.0009 &     0.0443  & 0.0008 \\
          \cline{2-8}
    \multirow{4}[0]{*}{SNR=0.25} & P-MLE   & \cellcolor{yellow}{0.0821} & 0.0025 & \cellcolor{yellow}{0.0456} & 0.0020 &   \cellcolor{yellow}{0.0359}    & 0.0021 \\
          & deamer & 0.1540 & 0.0027 & 0.1479 & 0.0029 &   0.1334    &0.0002  \\
          & decon & 0.1260 & 0.0029 & 0.1188 & 0.0012 &      0.0972 &0.0009  \\
          & decon (with FFT) & 0.1259 & 0.0028 & 0.1186 & 0.0010  &     0.0985  & 0.0009 \\
          \hline
    beta-laplace &  & & &  &  &  &  \\
    \multirow{4}[0]{*}{SNR=4} & P-MLE   & \cellcolor{yellow}{0.0200} & 0.0014 & \cellcolor{yellow}{0.0111} & 0.0007 & \cellcolor{yellow}{0.0061} & 0.0003 \\
          & deamer & 0.0319 & 0.0014 & 0.0154 & 0.0007 & 0.0082 & 0.0003 \\
          & decon & 0.0560 & 0.0052 & 0.0458 & 0.0048 & 0.0374 & 0.0041 \\
          & decon (with FFT) & 0.0605  & 0.0054 & 0.0472 & 0.0046 & 0.0334 & 0.0033 \\
          \cline{2-8}
    \multirow{4}[0]{*}{SNR=1} & P-MLE   & \cellcolor{yellow}{0.0327} & 0.0021 & \cellcolor{yellow}{0.0191} & 0.0011 &   \cellcolor{yellow}{0.0127}    &0.0006  \\
          & deamer & 0.0685 & 0.0022 & 0.0347 & 0.0014 &   0.0246    &0.0007  \\
          & decon & 0.1025 & 0.0010   & 0.1204  & 0.0132 &    0.1184   & 0.0114 \\
          & decon (with FFT) & 0.1012 & 0.0103 & 0.1117 & 0.0121 &     0.1145  & 0.0011 \\
          \cline{2-8}
    \multirow{4}[0]{*}{SNR=0.25} & P-MLE   & \cellcolor{yellow}{0.0759} & 0.0024 & \cellcolor{yellow}{0.0481} & 0.0016 &  \cellcolor{yellow}{0.0336}     & 0.0016 \\
          & deamer & 0.1656 & 0.0037 & 0.1147 & 0.0026 &     0.0758  & 0.0013 \\
          & decon & 0.1973 & 0.0189 & 0.2305 & 0.0342 &   0.2657    & 0.0359 \\
          & decon (with FFT) & 0.2081 & 0.0260 & 0.2104 & 0.0240 &    0.2778   & 0.0286 \\
          \hline
    beta-beta &       &       &       &       &       &       &  \\
    \multirow{2}[0]{*}{SNR=4} & P-MLE   & \cellcolor{yellow}{0.0255} & 0.0021 & \cellcolor{yellow}{0.0127} & 0.0007 & \cellcolor{yellow}{0.0082}      & 0.0006 \\
          & deamer & 0.0336 & 0.0014 & 0.0172 & 0.0006 &    0.0097   & 0.0003 \\
          \cline{2-8}
    \multirow{2}[0]{*}{SNR=1} & P-MLE   & \cellcolor{yellow}{0.0501} & 0.0038 & \cellcolor{yellow}{0.0386}  & 0.0030 &  0.0316     &0.0018  \\
          & deamer & 0.0592 & 0.0023 & 0.0480 & 0.0014 &    \cellcolor{orange}{0.0311}   & 0.0010 \\
          \cline{2-8}
    \multirow{2}[0]{*}{SNR=0.25} & P-MLE   & \cellcolor{yellow}{0.0811} & 0.0042 & \cellcolor{yellow}{0.0436} & 0.0031  &  \cellcolor{yellow}{0.0400}     & 0.0026 \\
          & deamer & 0.1944 & 0.0026 & 0.1437 & 0.0027 &    0.1292   &0.0014  \\
          \hline
    \end{tabular}
\end{table}

\begin{table}
  \resultstablecaption{Laplace   \label{tab:4}}
    \centering
    \begin{tabular}{llllllll}
    \hline
    \multicolumn{2}{c}{laplace-normal} & \multicolumn{2}{c}{n=30} &
    \multicolumn{2}{c}{n=100} & \multicolumn{2}{c}{n=300} \\
    \cline{3-8}
      \multicolumn{2}{c}{} &mean & se&mean&se&mean&se\\
\hline
\multirow{4}[0]{*}{SNR=4} & P-MLE   & \cellcolor{yellow}{0.0441}  & 0.0020 & \cellcolor{yellow}{0.0282} & 0.0011 &  \cellcolor{orange}{0.0282} & 0.0011 \\
          & deamer & 0.1012 & 0.0024 & 0.0496 & 0.0008 &    0.0285   &0.0003  \\
          & decon & 0.0734  & 0.0016 & 0.0510 & 0.0011 &     0.0351  & 0.0008 \\
          & decon (with FFT) & 0.0749 & 0.0020  & 0.0504  & 0.0012 &    0.0364   & 0.0007 \\
          \cline{2-8}
    \multirow{4}[0]{*}{SNR=1} & P-MLE   & \cellcolor{yellow}{0.0723} & 0.0024 & \cellcolor{yellow}{0.0545} & 0.0020 & \cellcolor{yellow}{0.0414} & 0.0015 \\
          & deamer & 0.1767 & 0.0028 & 0.1267 & 0.0019 &  0.0955 & 0.0010 \\
          & decon & 0.1247 & 0.0019 & 0.0920 & 0.0014 &  0.0728&  0.0011\\
          & decon (with FFT) & 0.1260 & 0.0018 & 0.09231 & 0.0015 &  0.0746 & 0.0010 \\
          \cline{2-8}
    \multirow{4}[0]{*}{SNR=0.25} & P-MLE   & \cellcolor{yellow}{0.1343} & 0.0024 & \cellcolor{yellow}{0.0969} & 0.0023 & \cellcolor{yellow}{0.0548}  & 0.0032 \\
          & deamer & 0.2664 & 0.0013 & 0.2479 & 0.0033 &  0.1965 & 0.0007 \\
          & decon & 0.1745 & 0.0035 & 0.1715 & 0.0016 & 0.1523 & 0.0013 \\
          & decon (with FFT) & 0.1758 & 0.0035 & 0.1729 & 0.0015 & 0.1530 & 0.0011 \\
          \hline
    laplace-laplace &       &  & & & & &\\
    \multirow{4}[0]{*}{SNR=4} & P-MLE   & \cellcolor{yellow}{0.0427} &
    0.0021 & \cellcolor{yellow}{0.0250} & 0.0011 & \cellcolor{yellow}{0.0157} & 0.0006  \\
          & deamer & 0.0986 & 0.0027 & 0.0437 & 0.0010 &  0.0221 & 0.0004 \\
          & decon & 0.0704 & 0.0062 & 0.0684 & 0.0084 & 0.0421 & 0.0044 \\
          & decon (with FFT) & 0.0738 & 0.0070 & 0.0674  & 0.0084 & 0.0443 & 0.0046 \\
          \cline{2-8}
    \multirow{4}[0]{*}{SNR=1} & P-MLE   & \cellcolor{yellow}{0.0700} & 0.0026 & \cellcolor{yellow}{0.0477} & 0.0018 & \cellcolor{yellow}{0.0332} & 0.0014 \\
          & deamer & 0.1576 & 0.0032 & 0.1030 & 0.0017 & 0.0658 & 0.0009 \\
          & decon & 0.1283 & 0.0123 & 0.1485 & 0.0226  & 0.1006 & 0.0097 \\
          & decon (with FFT) & 0.1287 & 0.0126 & 0.1256 & 0.0134 & 0.1023 & 0.0096 \\
          \cline{2-8}
    \multirow{4}[0]{*}{SNR=0.25} & P-MLE   & \cellcolor{yellow}{0.1332} & 0.0027 & \cellcolor{yellow}{0.1021} & 0.0024  & \cellcolor{yellow}{0.0673} & 0.0028 \\
          & deamer & 0.2501  & 0.0032 & 0.1971 & 0.0013 & 0.1604  & 0.0027 \\
          & decon & 0.2349 & 0.0279 & 0.3121 & 0.0427 & 0.2972 & 0.0294 \\
          & decon (with FFT) & 0.2231 & 0.0194 & 0.2996   & 0.0446 & 0.3132 & 0.0330 \\
          \hline
    laplace-beta &       &       &       &       &       &       &  \\
    \multirow{2}[0]{*}{SNR=4} & P-MLE   & \cellcolor{yellow}{0.0758} & 0.0042 & 0.0604  & 0.0036 & 0.0441 & 0.0036 \\
          & deamer & 0.1040 & 0.0027 & \cellcolor{yellow}{0.0520} & 0.0009 & \cellcolor{yellow}{0.0293} & 0.0004 \\
          \cline{2-8}
    \multirow{2}[0]{*}{SNR=1} & P-MLE  & \cellcolor{yellow}{0.1068} & 0.0069 & \cellcolor{yellow}{0.1100} & 0.0049  & 0.1361  & 0.0037 \\
          & deamer & 0.1804  & 0.0026 & 0.1282 & 0.0017 & \cellcolor{yellow}{0.0946} & 0.0010 \\
          \cline{2-8}
    \multirow{2}[0]{*}{SNR=0.25} & P-MLE   & \cellcolor{yellow}{0.1478} & 0.0059 & \cellcolor{yellow}{0.1096} & 0.0064 & \cellcolor{yellow}{0.1129} & 0.0063 \\
         & deamer& 0.2667 & 0.0012 & 0.2534 & 0.0030  & 0.1952 & 0.0002 \\
          \hline
    \end{tabular}
\end{table}

\begin{table}
  \resultstablecaption{mixture-normal  \label{tab:5}}
  \centering
    \begin{tabular}{llllllll}
    \hline
    \multicolumn{2}{c}{mixnormal-normal} & \multicolumn{2}{c}{n=30} &
    \multicolumn{2}{c}{n=100} & \multicolumn{2}{c}{n=300} \\
    \cline{3-8}
& &mean & se&mean&se&mean&se\\
\hline
\multirow{4}[0]{*}{SNR=4} & P-MLE   & \cellcolor{yellow}{0.1145} & 0.0027 & \cellcolor{yellow}{0.0780} & 0.0025 & \cellcolor{yellow}{0.0412} & 0.0049 \\
          & deamer & 0.1565 & 0.0019 & 0.1145 & 0.0018 & 0.0800  & 0.0016 \\
          & decon & 0.1466 & 0.0015 & 0.1294 & 0.0008 & 0.1041 & 0.0009 \\
          & decon (with FFT) & 0.1470 & 0.0016 & 0.1292 & 0.0008 & 0.1046 & 0.0008 \\
          \cline{2-8}
    \multirow{4}[0]{*}{SNR=1} & P-MLE  & \cellcolor{yellow}{0.1372} &0.0022 & {0.1501} & 0.0048  & \cellcolor{yellow}{0.1245} & 0.0040 \\
          & deamer &0.2108 & 0.0022 & 0.1642 & 0.0012 &0.1388 & 0.0004 \\
          & decon & 0.1695 & 0.0017 & 0.1465 & 0.0009 &0.1365 &0.0006 \\
          & decon (with FFT) & 0.1681 & 0.0018 & \cellcolor{orange}{0.1464} & 0.0009 &0.1368 &0.0005 \\
          \cline{2-8}
     \multirow{4}[0]{*}{SNR=0.25} & P-MLE  & \cellcolor{yellow}{0.1724} & 0.0022 & \cellcolor{yellow}{0.1456} & 0.0014 & \cellcolor{orange}{0.1800} & 0.0046 \\
          & deamer & 0.2942 & 0.0010  & 0.2731 & 0.0034 & 0.2240 & 0.0007 \\
          & decon & 0.2069 & 0.0030 & 0.2026 & 0.0012  & 0.1849 & 0.0010 \\
          & decon (with FFT) & 0.2096  & 0.0032 & 0.2023 & 0.0014 & 0.1851 & 0.0010 \\
          \hline
    mixnormal-laplace &  &   &  &  & & & \\
    \multirow{4}[0]{*}{SNR=4} & P-MLE   & 0.1085 & 0.0033 & \cellcolor{yellow}{0.0646} & 0.0017 & 0.0462 & 0.0054 \\
          & deamer & 0.1475 & 0.0024 & 0.0848 & 0.0025 & \cellcolor{orange}{0.0404}  & 0.0014 \\
          & decon& 0.1011 & 0.0066 & 0.0811 & 0.0073 & 0.0484 & 0.0043 \\
          & decon (with FFT)& \cellcolor{orange}{0.1010} & 0.0055 & 0.0771 & 0.0058 & 0.0429 & 0.0041 \\
          \cline{2-8}
    \multirow{4}[0]{*}{SNR=1} & P-MLE   & \cellcolor{yellow}{0.1360} & 0.0021 & \cellcolor{yellow}{0.1305} & 0.0032 & \cellcolor{yellow}{0.0983} & 0.0045 \\
          & deamer & 0.1986 & 0.0028 & 0.1524 & 0.0015 & 0.1252 & 0.0006 \\
          & decon  & 0.1735 & 0.0103  & 0.1575 & 0.0119 & 0.1249 & 0.0104 \\
          & decon (with FFT)  & 0.1802  & 0.0124 & 0.1555 & 0.0121 & 0.1206 & 0.0091 \\
          \cline{2-8}
    \multirow{4}[0]{*}{SNR=0.25} & P-MLE    & \cellcolor{yellow}{0.1674} & 0.0020 & \cellcolor{yellow}{0.1486} & 0.0014 & \cellcolor{yellow}{0.1531} & 0.0029 \\
          & deamer  & 0.2819 & 0.0029 & 0.2253 & 0.0012 & 0.1989 & 0.0026 \\
          & decon & 0.3090 & 0.0307 & 0.3337 & 0.0341 & 0.3521 & 0.0281 \\
          & decon (with FFT) & 0.3250 & 0.0312 & 0.3356 & 0.0383 & 0.3689 & 0.0317 \\
          \hline
    mixnormal-beta &       &       &       &       &       &       &  \\
    \multirow{2}[0]{*}{SNR=4} & P-MLE   & 0.1601  & 0.0041 & 0.1527 & 0.0030 & 0.1312 & 0.0026 \\
          & deamer  & \cellcolor{orange}{0.1561} & 0.0022 & \cellcolor{yellow}{0.1159} & 0.0014 & \cellcolor{yellow}{0.0747} & 0.0015 \\
          \cline{2-8}
    \multirow{2}[0]{*}{SNR=1} & P-MLE   & \cellcolor{yellow}{0.1747} & 0.0031 & 0.2095 & 0.0063 & 0.1921 & 0.0033 \\\
          & deamer & 0.2129 & 0.0021 & \cellcolor{yellow}{0.1679} & 0.0008 & \cellcolor{yellow}{0.1385} & 0.0004 \\
          \cline{2-8}
    \multirow{2}[0]{*}{SNR=0.25} & P-MLE   & \cellcolor{yellow}{0.1866} & 0.0033 & \cellcolor{yellow}{0.1776}  & 0.0032 & \cellcolor{yellow}{0.1978} & 0.0034 \\
          & deamer & 0.2951 & 0.0007 & 0.2812 & 0.0029 & 0.2281 & 0.0003 \\
         \hline
    \end{tabular}
\end{table}

\begin{table}
  \resultstablecaption{mixture-gamma  \label{tab:6}}
  \centering
    \begin{tabular}{llllllll}
    \hline
    \multicolumn{2}{c}{mixgamma-normal} & \multicolumn{2}{c}{n=30} & \multicolumn{2}{c}{n=100} & \multicolumn{2}{c}{n=300} \\
\cline{3-8}
    & &mean & se&mean&se&mean&se\\
\hline
\multirow{4}[0]{*}{SNR=4} & P-MLE   & \cellcolor{yellow}{0.0328} & 0.0016 & \cellcolor{yellow}{0.0221} & 0.0009 & \cellcolor{yellow}{0.0161} & 0.0008 \\
          & deamer & 0.0412 & 0.0015 & 0.0258 & 0.0005 & 0.0209 & 0.0002 \\
          & decon & 0.0756 & 0.0017 & 0.0458 & 0.0006 & 0.0335 & 0.0004 \\
          & decon (with FFT) & 0.0795  & 0.0023 & 0.0460 & 0.0007 & 0.0338 & 0.0004 \\
          \cline{2-8}
    \multirow{4}[0]{*}{SNR=1} & P-MLE   & \cellcolor{yellow}{0.0381} & 0.0021 & \cellcolor{yellow}{0.0362} & 0.0028  & \cellcolor{yellow}{0.0267} & 0.0012 \\
          & deamer & 0.0789 & 0.0024 & 0.0494 & 0.0012 & 0.0358 & 0.001 \\
          & decon  & 0.1020 & 0.0018 & 0.0655 & 0.0010  & 0.0500 & 0.0006 \\
          & decon (with FFT)  & 0.1012   & 0.0018 & 0.0656 & 0.0011 & 0.0497  & 0.0005 \\
          \cline{2-8}
    \multirow{4}[0]{*}{SNR=0.25} & P-MLE   & \cellcolor{yellow}{0.0780} & 0.0026 & \cellcolor{yellow}{0.0469} & 0.0014 & \cellcolor{yellow}{0.0462} & 0.0018 \\
          & deamer& 0.1908 & 0.0029 & 0.1428 & 0.0026 & 0.1314 & 0.0005 \\
          & decon & 0.1232 & 0.0028 & 0.1193 & 0.0011  & 0.0996 & 0.0085 \\
          & decon (with FFT)& 0.1236 & 0.0027 & 0.1191 & 0.0010  & 0.0990 & 0.0094 \\
          \hline
 mixgamma-laplace &       &       &       &       &       &       &  \\
    \multirow{4}[0]{*}{SNR=4} & P-MLE   & \cellcolor{yellow}{0.0321} & 0.0015 & \cellcolor{yellow}{0.0210} & 0.0008 & \cellcolor{yellow}{0.0133} & 0.0005 \\
          & deamer & 0.0394 & 0.0013 & 0.0252 & 0.0005 & 0.0209 & 0.0002 \\
          & decon & 0.0650 & 0.0052  & 0.0519 & 0.0059 & 0.0469 & 0.0041 \\
          & decon (with FFT) & 0.0685  & 0.0070 & 0.0542 & 0.0068 & 0.0433 & 0.0039 \\
          \cline{2-8}
    \multirow{4}[0]{*}{SNR=1} & P-MLE    & \cellcolor{yellow}{0.0356} & 0.0017 & \cellcolor{yellow}{0.0301} & 0.0011 & \cellcolor{yellow}{0.0243} & 0.0013 \\
          & deamer& 0.0698  & 0.0022 & 0.0418 & 0.001  & 0.0322 & 0.0005 \\
          & decon & 0.1101 & 0.0119 & 0.1165 & 0.0146 & 0.0988 & 0.0104 \\
          & decon (with FFT)& 0.1104 & 0.0118 & 0.1141 & 0.0114 & 0.1011 & 0.0110 \\
          \cline{2-8}
    \multirow{4}[0]{*}{SNR=0.25} & P-MLE   & \cellcolor{yellow}{0.0766} & 0.0023 & \cellcolor{yellow}{0.0523} & 0.0016 & \cellcolor{yellow}{0.0417} & 0.0019 \\
          & deamer & 0.1599  & 0.0039 & 0.1140 & 0.0025 & 0.0873 & 0.0104 \\
          & decon & 0.2709 & 0.0446 & 0.2323 & 0.0285 & 0.2764 & 0.0287 \\
          & decon (with FFT) & 0.2998 & 0.0450 & 0.2382 & 0.0233 & 0.2703 & 0.0254 \\
          \hline
    mixgamma-beta &       &       &       &       &       &       &  \\
    \multirow{2}[0]{*}{SNR=4} & P-MLE   & 0.0423 & 0.0020 & 0.0315 & 0.0011 &  0.0246     & 0.0007 \\
          & deamer & \cellcolor{orange}{0.0394} & 0.0012 & \cellcolor{yellow}{0.0259} & 0.0005 &    \cellcolor{yellow}{0.0209}   & 0.0002 \\
          \cline{2-8}
    \multirow{2}[0]{*}{SNR=1} & P-MLE  & \cellcolor{yellow}{0.0644} & 0.0030 & 0.0614 & 0.0029 &   0.0503    & 0.0016 \\
          & deamer & 0.0802  & 0.0021 & \cellcolor{yellow}{0.0505}  & 0.0014 &   \cellcolor{yellow}{0.0420}    & 0.0010 \\
          \cline{2-8}
    \multirow{2}[0]{*}{SNR=0.25} & P-MLE   {2} & \cellcolor{yellow}{0.0884} & 0.0041  & \cellcolor{yellow}{0.0623} & 0.0035  & \cellcolor{yellow}{0.0565} & 0.0022 \\
          & deamer& 0.1927 & 0.0026 & 0.1403  & 0.0024 & 0.1344 & 0.0013 \\
          \hline
    \end{tabular}
\end{table}

\begin{table}
  \resultstablecaption{Cauchy\label{tab:7}}
{
  \centering
    \begin{tabular}{llllllll}
    \hline
     \multicolumn{2}{c}{cauchy-normal} & \multicolumn{2}{c}{n=30} &
     \multicolumn{2}{c}{n=100} & \multicolumn{2}{c}{n=300} \\
     \cline{3-8}
      \multicolumn{2}{c}{} &mean & se&mean&se&mean&se\\
\hline
\multirow{4}[0]{*}{C=0.5} & P-MLE  & 0.0294 & 0.0024 & \cellcolor{yellow}{0.0142} & 0.0007 & \cellcolor{yellow}{0.0100} & 0.0004 \\
          & deamer& 0.0513 & 0.0012  & 0.0354 & 0.0014 & 0.0106 & 0.0003 \\
          & decon & 0.0259 & 0.0018 & 0.1356 & 0.0581 & 0.417 & 0.1905 \\
          & decon (with FFT) & \cellcolor{orange}{0.0244} & 0.0010 & 0.0187 & 0.0016 & 0.0196 & 0.0027 \\
          \cline{2-8}
    \multirow{4}[0]{*}{C=1} & P-MLE   & 0.0389 & 0.0080 & \cellcolor{yellow}{0.0153} & 0.0007 & \cellcolor{yellow}{0.0092} & 0.0003 \\
          & deamer & 0.0649 & 0.0013 & 0.0369 & 0.0008 & 0.0239 & 0.0004 \\
          & decon & 0.0392 & 0.0014  & 0.0891 & 0.0031 & 0.4274 & 0.1606 \\
          & decon (with FFT) & \cellcolor{orange}{0.0368} & 0.0010  & 0.0803 & 0.0064 & 0.0315 & 0.0056 \\
          \cline{2-8}
    \multirow{4}[0]{*}{C=2} & P-MLE   & \cellcolor{orange}{0.0473} & 0.0032 & \cellcolor{yellow}{0.0285} & 0.0010  & \cellcolor{yellow}{0.0228} & 0.0007 \\
          & deamer & 0.0944 & 0.0001 & 0.0903  & 0.0004 & 0.0658 & 0.0018 \\
          & decon & 0.0512 & 0.0017 & 0.2189 & 0.0221 & 0.1551 & 0.0603 \\
          & decon (with FFT) & 0.0505 & 0.0016 & 0.2281 & 0.0256 & 0.0494 & 0.0034 \\
          \hline
    cauchy-laplace &       &       &       &       &       &       &  \\
    \multirow{4}[0]{*}{SNR=4} & P-MLE   & \cellcolor{yellow}{0.0308} & 0.0025 & \cellcolor{yellow}{0.0135} & 0.0006 & \cellcolor{orange}{0.0104} & 0.0004 \\
          & deamer & 0.0503  & 0.0012  & 0.0347  & 0.0016 & 0.0110 & 0.0005 \\
          & decon & 0.0535 & 0.0050  & 0.0527 & 0.0053 & 0.0974 & 0.0325 \\
          & decon (with FFT) & 0.0692 & 0.0077 & 0.0521 & 0.0047 & 0.0472 & 0.0061 \\
          \cline{2-8}
    \multirow{4}[0]{*}{SNR=1} & P-MLE  & \cellcolor{yellow}{0.0330} & 0.0048 & \cellcolor{yellow}{0.0146} & 0.0006 & \cellcolor{yellow}{0.0093}  & 0.0004 \\
          & deamer & 0.0631 & 0.0014 & 0.0337 & 0.0006 & 0.0202  & 0.0003 \\
          & decon & 0.0850 & 0.0079 & 0.1286 & 0.0173 & 0.1124 & 0.0146 \\
          & decon (with FFT)  & 0.0833 & 0.0073 & 0.1294 & 0.014  & 0.1016 & 0.0096 \\
          \cline{2-8}
    \multirow{4}[0]{*}{SNR=0.25} & P-MLE   & \cellcolor{yellow}{0.0354} & 0.0016 & \cellcolor{yellow}{0.0236} & 0.0009 & \cellcolor{yellow}{0.0173} & 0.0009 \\
          & deamer & 0.0934 & 0.0005 & 0.0695 & 0.0019 & 0.0523 & 0.0005 \\
          & decon  & 0.2115 & 0.0262 & 0.3083 & 0.1464 & 0.3148  & 0.0415 \\
          & decon (with FFT) & 0.1822 & 0.0180 & 0.2157 & 0.0266 & 0.3105 & 0.0581 \\
          \hline
    cauchy-beta &       &       &       &       &       &       &  \\
    \multirow{2}[0]{*}{SNR=4} & P-MLE   & \cellcolor{yellow}{0.0428} & 0.0041 & \cellcolor{yellow}{0.0164} & 0.0007 & \cellcolor{orange}{0.0108} & 0.0004 \\
          & deamer & 0.0783 & 0.0020  & 0.0357 & 0.0004 & 0.0108 & 0.0003 \\
          \cline{2-8}
    \multirow{2}[0]{*}{SNR=1} & P-MLE   & \cellcolor{yellow}{0.0557} & 0.0049  & \cellcolor{yellow}{0.0297} & 0.0013 & \cellcolor{orange}{0.0217} & 0.0010 \\
          & deamer & 0.0899 & 0.0017 & 0.0349 & 0.0007 & 0.0238 & 0.0004 \\
          \cline{2-8}
    \multirow{2}[0]{*}{SNR=0.25} & P-MLE   &\cellcolor{yellow}{ 0.0708} & 0.0062 & \cellcolor{yellow}{0.0532} & 0.0022 & \cellcolor{yellow}{0.0516} & 0.0011 \\
          & deamer & 0.0914 & 0.0001 & 0.0901  & 0.0005 & 0.0620 & 0.0016 \\
          \hline
    \end{tabular}
}
\end{table}

MISE for each method in each scenario is given in
Tables~\ref{tab:1}--\ref{tab:7}. We see that P-MLE significantly
outperforms both other methods in 153 out of 189
scenarios. \texttt{Deamer} significantly outperforms P-MLE in 16 out of
189 scenarios, and there are 14 scenarios where there is no
significant difference between \texttt{deamer} and P-MLE. \texttt{Decon} does
not significantly outperform P-MLE in any scenario, but there are 6
scenarios where P-MLE significantly outperforms \texttt{deamer} but not
\texttt{decon}.  Outperforming the state-of-the-art existing methods
in such a wide range of scenarios is extremely impressive. It is also
worth noting that because of the scale of the simulations, we used the
heuristic approach to selecting the penalty parameter $\lambda_n$. For
analysing a single data set, we would select $\lambda_n$ more carefully
using cross-validation, which would be expected to lead to better
results. Furthermore, despite the fact that \texttt{decon} is using the true
error distribution, while P-MLE and \texttt{deamer} are using an empirical
estimate of the error distribution, \texttt{decon} never significantly
outperforms P-MLE.

To see how much potential improvement could be made by choosing
$\lambda_n$ more carefully, we reran a number of scenarios where P-MLE
was not the best method with a range of different values of
$\lambda_n$. We found that in 6 of the 16 scenarios where
\texttt{deamer} performed significantly better than P-MLE in the
original results, another choice for $\lambda_n$ using a different
constant in the heuristic method would give an MISE that is not
significantly different from \texttt{deamer}.  In 2 more scenarios,
different choices of $\lambda_n$ improve results not significantly
different from \texttt{deamer} to results that are significantly
better than \texttt{deamer}. In one scenario, choosing a different
constant for the heuristic choice of $\lambda_n$ changes a result that
is not significantly different from \texttt{decon} to a result that is
significantly better. These results are based on choosing the value of
$\lambda_n$ which actually performs best for the scenario, rather than
by cross-validation, so are not totally reliable. On the other hand,
the results are using the heuristic approach, rather than tuning
$\lambda_n$ for each dataset, so it is possible that better tuning of
$\lambda_n$ might produce even better results. Full details of these
improvements are in the supplemental appendices.

\begin{table}
  \caption[{Summary of outcomes for each scenario}]{Summary of
    outcomes for each scenario

A ---    P-MLE significantly outperforms both other methods.
 
B ---  P-MLE significantly outperforms \texttt{deamer} but not \texttt{decon}.

C ---  No significant difference between P-MLE and \texttt{deamer}.

D ---  \texttt{Deamer} significantly outperforms P-MLE.
 \label{SimulationSummaryTable}
}

 \begin{tabular}{llrrrr}
Outcome  && A    & B    & C & D \\
\hline
\multirow{7}{*}{True distribution}
& Normal      & 20 & 0 & 4 & 3 \\
& Chi-squared & 23 & 0 & 2 & 2 \\
& Beta        & 26 & 0 & 1 & 0 \\
& Laplace     & 23 & 0 & 1 & 3 \\
& Mixnormal   & 18 & 3 & 2 & 4 \\
& Mixgamma    & 22 & 0 & 1 & 4 \\
& Cauchy      & 21 & 3 & 3 & 0 \\
\hline
\multirow{3}{*}{Error distribution}
& Normal & 55 & 5 & 2 & 1\\
& Laplace & 58 & 1  & 3 & 1 \\
& Beta & 40 & NA & 9 & 14
\\
\hline
\multirow{3}{*}{SNR}
& $C=0.5$ & 40 & 2 & 11 & 10 \\
& $C=1$   & 52 & 2 &  3 &  6 \\
& $C=2$   & 61 & 2 &  0 &  0 \\
\hline
\multirow{3}{*}{Sample size}
&  30  & 56 & 4 & 3 &  0 \\
& 100  & 53 & 1 & 4 &  5 \\
& 300  & 44 & 1 & 7 & 11 \\
\hline
 \end{tabular}

 \end{table} 

Table~\ref{SimulationSummaryTable} gives a breakdown of these
simulation outcomes. We see that P-MLE outperforms the other methods
for all true distributions, but the level of outperformance is less
when the true distribution is a mixture of normal distributions. This
is not surprising, since the mixture of normal distributions is
bimodal, so the smoothness penalty will be larger than for the other
distributions. However, even in this more challenging case, P-MLE still
outperforms the other methods in most scenarios. The distribution of
the noise has a bigger influence, with \texttt{deamer} performing relatively
better when the noise follows a beta distribution. SNR
is another important factor, with P-MLE performing much better than the
other methods in the low SNR case.  Finally, P-MLE outperforms \texttt{deamer}
and \texttt{decon} in more scenarios with low sample size. 

We also compare the distribution of the integrated squared error (ISE)
for the different methods in each scenario. Figure~\ref{f2}
compares boxplots of the ISE for the three methods for each
combination of true and error distributions for the case where sample
size=30 and $C=1$. While there are some outliers, and sometimes large
variance, the results in most cases are fairly consistent, rather than
being caused by a few outliers where one method gives a very poor
estimate. An exception to this is the case where the true distribution
is Cauchy and the error is normal. We see that P-MLE generally produced
slightly better estimates than \texttt{decon}, but there was one simulation
where the estimate from P-MLE was very poor, with ISE about 0.8,
causing the overall MISE to be larger for P-MLE than \texttt{decon}. For
$C=0.5$, the situation is less clear-cut, but there are several
outliers for P-MLE, so for most simulations the results are much closer
between P-MLE and \texttt{decon}. Similar boxplots for other scenarios are in
the Supplemental Appendices, Figures~\ref{f1}--\ref{f9}.

\newcommand{\simulationresultsfigurecaption}[2]{
  \caption{ Sample distribution of ISE for sample size #1 and SNR #2.}
  \parbox{\textwidth}{
    Each column corresponds to one of the seven true distributions in the
    simulation. Rows correspond to the error distribution.
    The \texttt{decon} package only allows a limited selection of
    error families, so could not be compared for simulations with a
    beta error. Some outliers where \texttt{decon} produced a large
    ISE are truncated from these plots. No P-MLE results have been
    truncated.
\\
}}

\begin{figure}
\centering
\includegraphics[width=14cm]{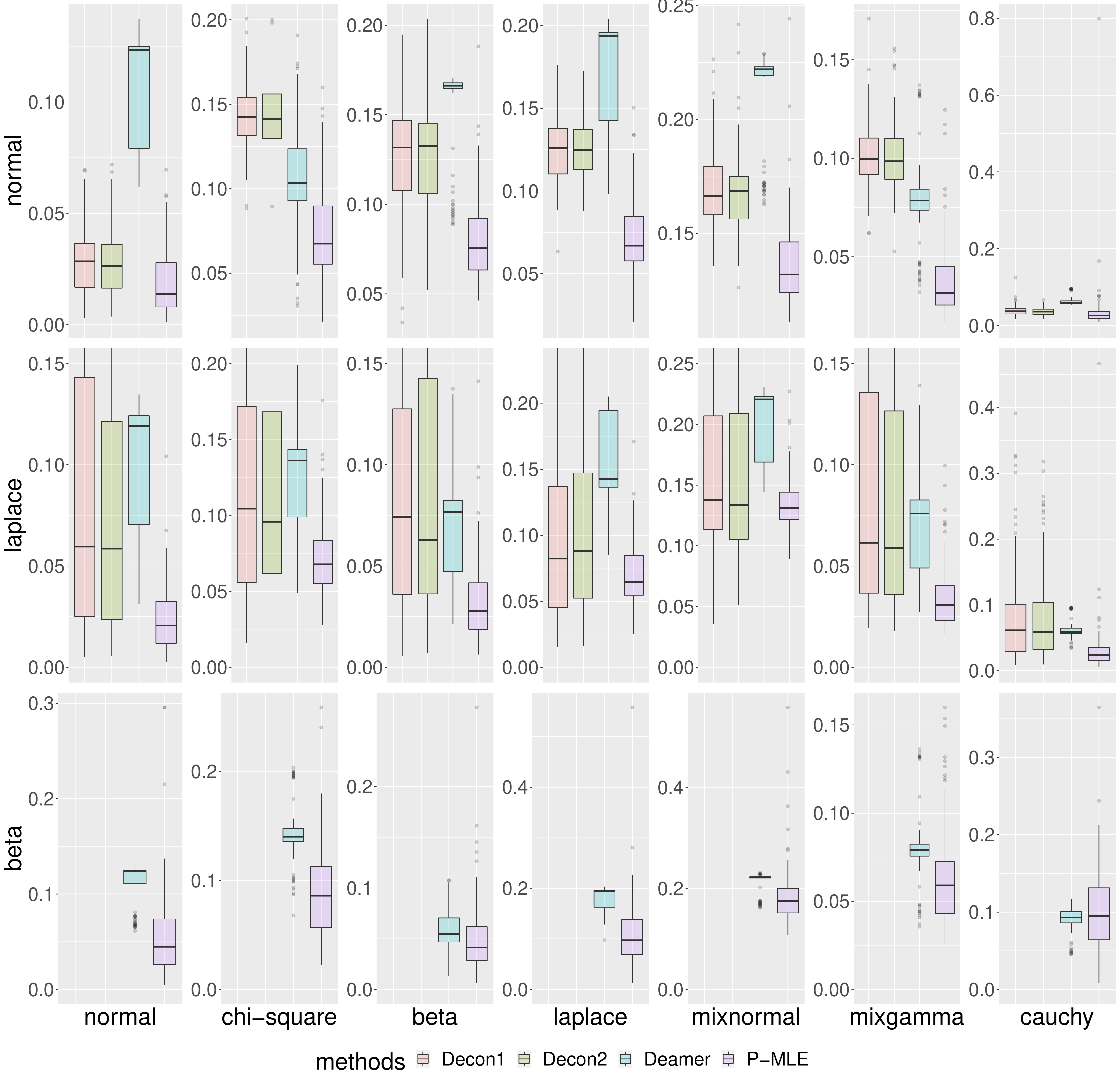}
\simulationresultsfigurecaption{30}{1}
\label{f2}
\end{figure}


\section{Real data analysis}\label{RealDataSection}

\begin{figure}

\begin{subfigure}[\makebox{Histogram}]

  \centering
    \includegraphics[width=7cm]{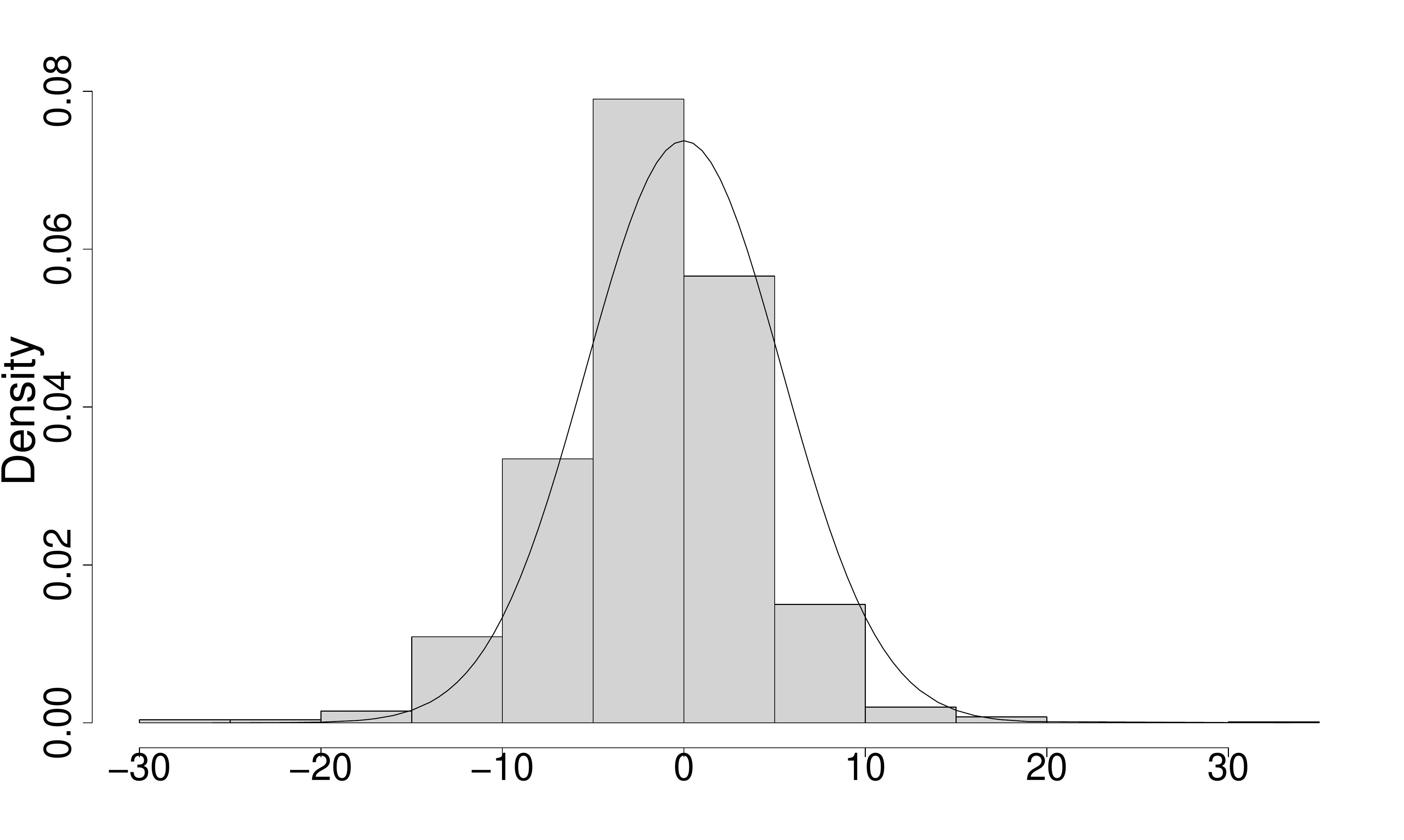}
\end{subfigure}
\begin{subfigure}[\makebox{Q-Q plot}]
  \centering
\makebox{\includegraphics[width=7cm]{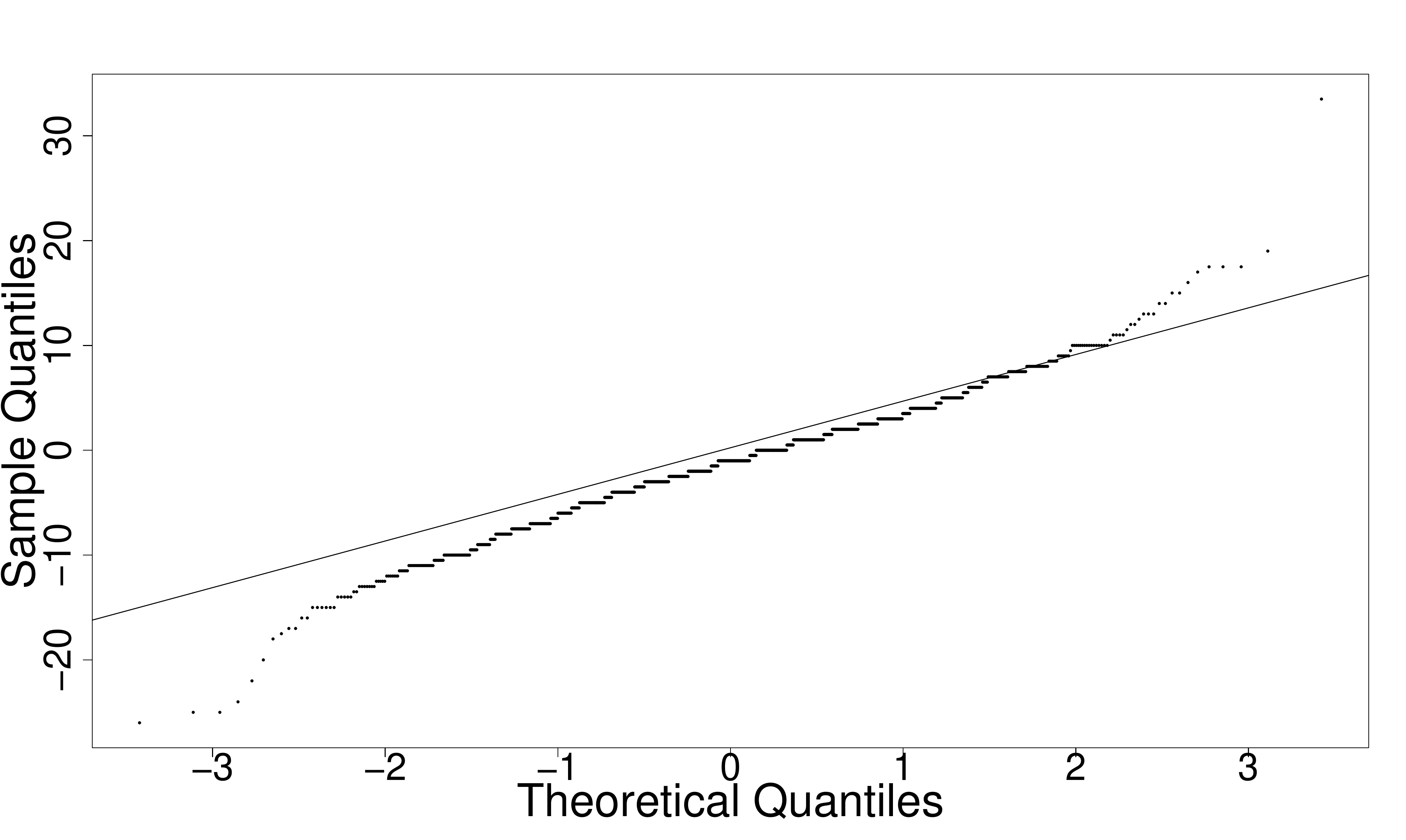}}
\end{subfigure}
\caption{Comparison of real data error distribution with a normal distribution}
\label{ErrorSampleSBP}
\end{figure}

Next we apply our P-MLE method to a real data set. The Framingham data
\citep{carroll2006measurement} records the systolic blood pressure
measured for 1615 male subjects. Each subject's blood pressure was
measured twice at a first visit and twice at a second visit eight
years later. We are going to use the measurements at the second visit
only.  Let $\textrm SBP_{21}$ and $\textrm SBP_{22}$ denote the two
observations at the second visit. $\textrm SBP_{2}$ is the average of
$\textrm SBP_{21}$ and $\textrm SBP_{22}$. We are going to estimate
the density of the underlying true blood pressure X from $\textrm
SBP_{2}$. We have $\textrm SBP_{2}=X+e$,
$e=\frac{e_{22}+e_{21}}{2}$. $e_{21}$ and $e_{22}$ are measurement
errors of $\textrm SBP_{21}$ and $\textrm SBP_{22}$. We also have
$\frac{\textrm SBP_{22}-\textrm SBP_{21}}{2}=\frac{e_{22}-e_{21}}{2}$.
Assume that the error distribution is symmetric, which is a common
assumption for measurement error. Then $\frac{e_{22}+e_{21}}{2}$ and
$\frac{e_{22}-e_{21}}{2}$ follow the same distribution. Therefore, we
can use $\frac{\textrm SBP_{22}-\textrm SBP_{21}}{2}$ as the pure
error sample for P-MLE and \texttt{deamer}.  For \texttt{decon}, we
assume the error distribution is normal with mean zero and estimated
variance. Figure~\ref{ErrorSampleSBP} shows the observed error
distribution compared with an estimated normal density. We see that
the normal assumption is not unreasonable for this distribution,
though the error distribution appears to have heavier tails than
the normal distribution.

From the empirical distributions, the variance of the error sample is
$29.274$, and the variance of the observed sample is $395.6506$. This
suggests an SNR of about $12.5154$.

\begin{figure}
\centering
\begin{subfigure}[\makebox{\parbox{\textwidth}{Deconvolved estimate for density of
      distribution of SBP. A kernel density estimate of the observed
      values is shown for comparison, but this is with measurement
      error, so estimates close to the data estimate have not
      adequately removed the measurement error.}}\label{sbpdecon}]
  
\includegraphics[width=12cm]{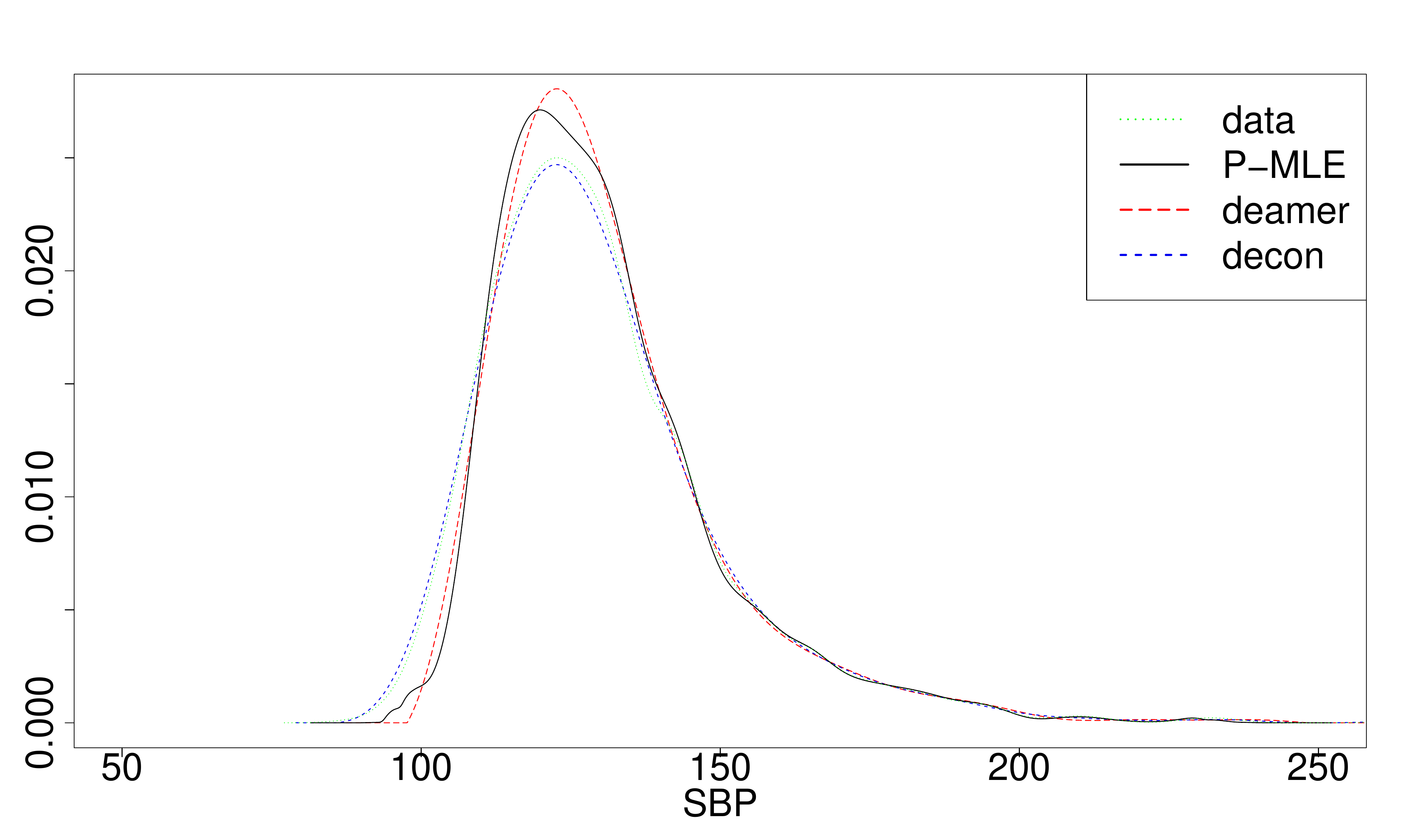}
\end{subfigure}

\begin{subfigure}[\makebox{\parbox{\textwidth}{Estimated density of SBP distribution
      convolved with error distribution, compared with a kernel
      density estimate from the observed data.}}\label{dbpconv}]

\includegraphics[width=12cm]{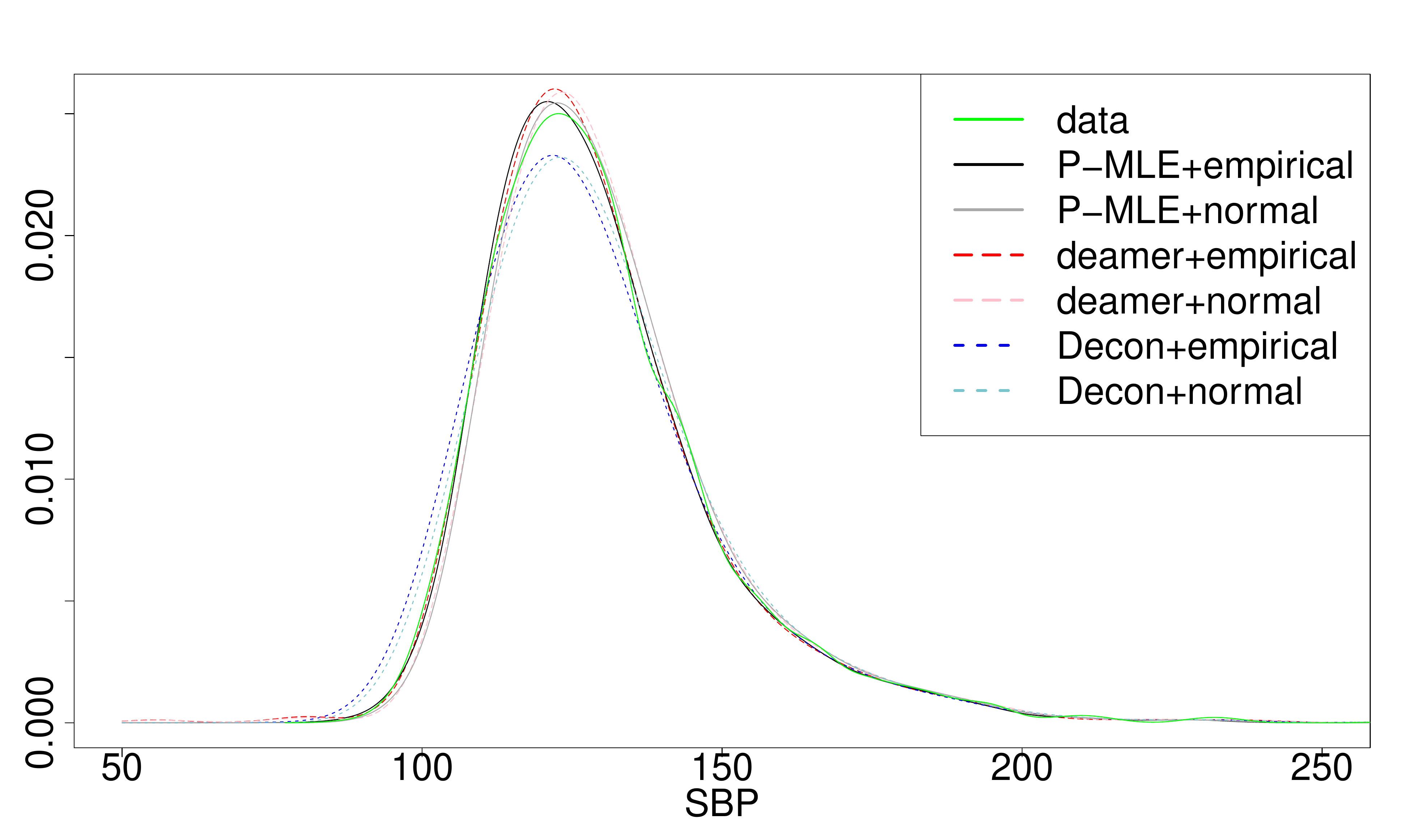}
\end{subfigure}
\caption{Real data results}
\end{figure}


Figure~\ref{sbpdecon} shows the estimated true distributions by P-MLE
and by \texttt{deamer} and \texttt{decon}. We see that P-MLE and \texttt{deamer}
both select a much sharper peak and lower variance than the observed
data, which is what we should expect to see, since adding noise should
increase the variance and produce a less sharp peak. The density
estimated by \texttt{decon} is extremely close to the observed data, suggesting
that \texttt{decon} has not removed most of the measurement error.

To give a better sense of how well the methods estimate the latent
true distribution, we convolve the estimated distributions with the
error distribution and compare the results with a kernel density
estimate from the observed data.  Figure~\ref{dbpconv} compares the
convolved estimated distribution for each method, convolving with both
a normal measurement-error distribution, and the empirical error
distribution.  As can be seen, the distributions estimated by P-MLE
and \texttt{deamer}, when convolved with the empirical error
distribution, produce something close to the original
data. \texttt{Decon} estimated a density much closer to the observed
data, and as a result, the convolved estimator has higher variance
than the observed data. We also see that when a normal error is used,
the right tail of the distribution is estimated well by all methods,
but the left tail is more challenging.

To give a quantitative measurement of the performance, we calculate
the distance between the observed data and the estimated convolved
distribution. We use both the Anderson-Darling (AD) distance and the
Kolmogorov-Smirnov (KS) distance for measuring the difference. We also
calculate the integrated squared difference between the estimated
densities and the kernel density estimate from the real data in
Figure~\ref{dbpconv}. Because \texttt{deamer} and \texttt{decon} do
not guarantee that the deconvolved ``densities'' returned have
integral 1, we need to rescale them so that the AD and KS distances
can be computed. We do not use Kullback-Leibler divergence because
P-MLE directly optimises likelihood, so choosing a quantity so closely
related to the objective function of P-MLE might be considered an
unfair advantage to P-MLE, particularly considering that these results
are on training data, so there is potential overfitting. The results
are in Table~\ref{tab:real}. We see that, as expected, \texttt{decon}
produces bad results, and that convolving with the empirical error
distribution gives a closer result to the observed data than
convolving with the normal distribution. For the empirical error
distribution, \texttt{deamer} and P-MLE both perform similarly with
P-MLE being better by some measures and \texttt{deamer} better under
other measures. This is consistent with the simulation results, where for
larger sample sizes, the difference in performance between
\texttt{deamer} and P-MLE was small. We also see that the density
functions estimated by P-MLE and \texttt{deamer} are quite different,
but that the convolutions with the error density are much closer. This
is the identifiability issue in the convolution problem, with two
distributions having a very similar convolution with the measurement
error distribution.


\begin{table}
  \caption{Difference between empirical distribution
    and convolved estimated distribution\label{tab:real}}

    \begin{tabular}{lrrrlrrr}
    \hline
& \multicolumn{3}{c}{Normal Error} & & \multicolumn{3}{c}{Empirical Error}\\
    \cline{2-4}
    \cline{6-8}
    &A.D.&K.S. & ISE & &A.D.&K.S. & ISE \\
    \hline
        P-MLE &\cellcolor{BurntOrange}{2.632}&0.0267 &        \cellcolor{BurntOrange}{0.0135} &   &\cellcolor{BurntOrange}{0.778} & 0.0085 & \cellcolor{BurntOrange}{0.0051}\\
        \hline
        Deamer&3.140. & 0.0302 & 0.0168 & &0.974 &\cellcolor{BurntOrange}{0.0071} & 0.0056\\
        \hline
       Decon&4.602&\cellcolor{BurntOrange}{0.0228} & 0.0272 & &10.598&0.0412 & 0.0330\\
       \hline
       Decon (with FFT)&4.722&0.0237 & 0.0277 & &10.648&0.0414 & 0.0331 \\
          \hline
  \end{tabular}

\end{table}

\section{Conclusion}\label{ConclusionSection}

We have developed a deconvolution method for additive error based on
penalised maximum log-likelihood estimation with a smoothness penalty
on the estimated density. The smoothness penalty we use has previously been
used to good effect in smoothing spline fittings. Our method is
applicable to either continuous or discrete error distributions. In
cases where the error distribution is unknown, this allows us to
substitute the empirical distribution from a pure-error sample.

We have proved that our P-MLE method is consistent. We have also
provided methods to address the practical optimisation difficulties
which arise. In extensive simulation studies, and a real-data example,
we have shown that our method has much better performance than
existing methods, particularly when sample size and SNR are small. If
faster computation is necessary, we provided a quick heuristic to
choose the tuning parameter without cross-validation, and showed that
with this heuristic, P-MLE produces good results, but with additional
time to tune the penalty parameter $\lambda_n$ by cross-validation, it
will perform even better.

There are a number of directions in which the method can potentially
be improved in future. Firstly, for practical purposes, we replaced
the infinite number of non-negativity constraints by a finite subset
of non-negativity constraints at a set of evenly spaced points. The
solution to this constrained optimisation problem might not satisfy
all the non-negativity constraints. It seems likely that with
carefully chosen constraints, we should be able to ensure that the
estimated density satisfies all the non-negativity constraints.  If
this is the case, then it should be possible to develop an adaptive
algorithm for choosing the correct points at which to impose
non-negativity constraints.

Another issue that could be studied in the future is truncation. From
the theory developed, we see that convergence depends upon the
estimated support $|u-l|$ not increasing too fast as
$n\rightarrow\infty$. For common light-tailed distributions, this will
almost surely happen. However, for heavy-tailed distributions, the
estimated support could grow too fast to achieve consistency. This
problem could be resolved via an appropriate truncation method where
certain data points are removed from the sample so that the rate of
growth of the estimated support is controlled. This would be expected
to improve large-sample performance in cases where the underlying true
distribution is heavy-tailed. Given that P-MLE performed well in the Cauchy
simulation results in Table~\ref{tab:7}, it seems that this is more of
a theoretical issue than a practical concern.

\bibliographystyle{rss}
\bibliography{bmc_article}

\end{document}


\title{Supplementary Appendices to Deconvolution density estimation with penalized MLE}

\author{Yun Cai, Hong Gu \& Toby Kenney}

\maketitle

\section{Theory}

In this section we prove the consistency theorem from the main paper

\begin{theorem}\label{MainTheorem}
  Let the true density $f_x$ be twice continuously differentiable, and
  let the convolved density be $f_y=f_x*f_e$. Let $\hat{f}_x$ be the
  P-MLE estimate for $f_x$ and $\hat{f}_y=\hat{f}_x*f_e$.  If the
  smoothness penalty parameter for each estimate is given by
  $\lambda_n=C_1 n^{\frac{7}{8}}\log(n)^{\frac{1}{8}}\sqrt{|u-l|}$,
  for a certain constant $C_1$, then almost surely, for all
  sufficiently large $n$, $\lVert
  \hat{f}_y-f_y\rVert_\infty<C_2n^{-\frac{1}{32}}|u-l|^{\frac{1}{8}}\log(n)^{\frac{1}{32}}$
  for some constant $C_2$. The constants $C_1$ and $C_2$ are given by
  \begin{align*}
    C_1&=\frac{2^{\frac{31}{20}}3^{-\frac{1}{10}}5^{\frac{3}{20}}}{\psi(f_x)^{\frac{4}{5}}\psi(f_y)^{\frac{1}{10}}}\\
    C_2&=\left(
  1+\sqrt{1+\frac{3^{\frac{2}{5}}}{16}}\right)^{\frac{1}{4}}2^{\frac{179}{80}}3^{\frac{9}{40}}
  {5^{\frac{27}{80}}}\psi(f_x)^{\frac{1}{4}}\psi(f_y)^{-\frac{1}{40}}
    \end{align*}
\end{theorem}

In this section, we will change the notation slightly to avoid
excessive subscripts.

For a twice continuously differentiable function
$f$ with support $S$, let $\psi(f)=\int_S (f''(x))^2\,dx$ be the
smoothness penalty. Our estimator is the function $\hat{f}_n$ that
maximises the penalised log-likelihood.
$$\sum_{i=1}^n \log\left(\left(\hat{f}_n*f_e\right)(x_i)\right)-\lambda_n\psi\left(\hat{f}_n\right)$$ We
want to prove consistency of $\hat{f}_n$. That is, more formally,
suppose we have a sequence of i.i.d. observations $x_1,x_2,\ldots$
from a distribution with density function $f*f_e$, where $f$ is twice
differentiable with finite smoothness.  We use the notation $f_e$ for
the density of the error distribution, but this is a slight abuse of
notation, because the proof works equally well for non-continuous
error distribution. Let $(\lambda_n)_{n=1,\ldots}$ be a chosen
sequence of penalty terms and $S_1\subseteq S_2\subseteq\cdots$ be a
chosen increasing sequence of measurable subsets of $\mathbb R$ with
union $\mathbb R$. Let $\hat{f}_n$ be the twice continuous function
with support $S_n$ that maximises the penalised log-likelihood
function
$$\sum_{i=1}^n \log((\hat{f}_n*f_e)(x_i))-\lambda_n\psi(\hat{f}_n)$$
for the first $n$ observations. We want to prove that $\hat{f}_n$
converges to $f$. Because of the difficulties that can arise when
$f_e$ is supersmooth, we will show that $\hat{f}_n*f_e$ converges to
$f*f_e$.  That is, we will show that for some $\gamma,\xi>0$, and some
$\zeta$, with probability 1, there is some $N$ such that for all $n>N$
we have $\lVert (\hat{f}_n-f)*f_e\rVert_\infty<\gamma
n^{-\xi}|S_n|^\zeta$.

In addition to the smoothness assumptions needed to state the problem,
we will assume that $g=f*f_e$ has only a finite number $M$ of local
maxima. This eliminates a number of pathological distributions without
eliminating any distributions of real interest. It simplifies the
proof, but we believe that the method is still consistent without this
assumption. This assumption is needed for
Proposition~\ref{logLowerBoundIntegral}, which allows a relatively
simple proof of Proposition~\ref{gStarLogLikelihood}. However, we
believe that a version of Proposition~\ref{gStarLogLikelihood}
(possibly with the lower bound changed slightly) is true without the
need for this assumption.

In our proof, we will change between working on the noiseless scale of
the underlying distribution we are trying to estimate, and working on
the convolved scale, with the noise distribution added. We will
consistently use $f$ to represent densities on the noiseless scale,
and $g$ to represent densities on the convolved scale so for example
$g=f*f_e$ refers to the convolved density of the true
distribution. Similar notation will be used throughout. For example
$\hat{g}_n=\hat{f}_n*f_e$ is the convolution of the penalised MLE
estimate with the error distribution. For this proof, we will assume
that $f_e$ is known. 

We will use the notation $g$ for densities that are related
to the convolved densities, even if they do not themselves arise as
convolutions with $f_e$. For example, in the proof, we will introduce
a density $\tilde{g}_n$, where we use the notation $g$ because it is
an estimator for the convolved density $g=f*f_e$, but the density
$\tilde{g}_n$ does not necessarily arise as a convolution of any
density function with $f_e$.

To prove consistency, we will first note that the
Dvoretzky-Kiefer-Wolfowitz~\citep{DvoretzkyKieferWolfowitz} inequality gives that
for any i.i.d. sample from a distribution with c.d.f. $G$, the
c.d.f. $G_n$ of the empirical distribution satisfies
$$P\left(\sqrt{n}\lVert
G_n-G\rVert_\infty>\sqrt{\log(n)}\right)<Cn^{-2}$$ for some positive
constant $C$ (later, \citet{Massart1990} showed that $C=1$), so the
expected number of $n$ for which
$$\lVert G_n-G\rVert_\infty>\sqrt{\frac{\log(n)}{n}}$$ is at most
$C\sum_{i=1}^\infty n^{-2}<\infty$. Thus with probability 1, there is
some $N$ such that for all $n>N$ we have $\lVert
G_n-G\rVert_\infty\leqslant\sqrt{\frac{\log(n)}{n}}$. In our case, we
have that $G$ is the c.d.f. for $g=f*f_e$, and $G_n$ is the empirical
c.d.f. of the observed sample.

Our approach for proving consistency will be as follows:

\begin{enumerate}[1.]

\item Construct a sequence of estimators $\tilde{g}_n(x)$ for $g$ (we
  use kernel density estimators) such that the following conditions
  hold whenever $\lVert
  G_n-G\rVert_\infty\leqslant\sqrt{\frac{\log(n)}{n}}$.

  \begin{enumerate}[(a)]

    \item $\lVert
      \tilde{g}_n-g\rVert_\infty<C_3\frac{\psi(g)^{\frac{1}{5}}\log(n)^{\frac{1}{4}}}{n^{\frac{1}{4}}}$
      for some constant $C_3$.

    \item For any density function $g^*_L$ that is Lipschitz with
      constant $L$ and has support $S_n$, and has
      $g^*_L(x_i)\geqslant\epsilon$ for $i=1,\ldots,n$, 
      $$\left|\frac{1}{n}\sum_{i=1}^n \log(g_L^*(x_i))-\int_{S_n} \tilde{g}_n(x)\log(g_L^*(x))\,dx\right|<\frac{\log(n)^{\frac{1}{4}}L}{{C_3}\psi(g)^{\frac{1}{5}}n^{\frac{1}{4}}\epsilon}$$
      Intuitively, this is saying that the empirical mean of
      $\log(g^*_L(x))$ is close to the mean over the distribution
      $\tilde{g}$.

  \end{enumerate}
      
\item Construct a sequenece of distributions $f_n^*$ that converge to
  $f$, (based partly on the data, and partly on the true function $f$)
  with the following properties:

  \begin{enumerate}[(a)]

  \item $\psi(f_n^*)\leqslant 2\psi(f)$

  \item Whenever, $\lVert G_n-G\rVert<\sqrt{\frac{\log(n)}{n}}$, and
    $n$ is sufficiently large,
    $\frac{1}{n}\sum_{i=1}^n \log(g_n^*(x_i))> -H(g)-4Mn^{-\frac{1}{2}}\log(n)^{\frac{3}{2}}$, where
    $H(g)$ is the entropy $-\int_{-\infty}^\infty g(x)\log(g(x))$.

  \end{enumerate}

\item Since the penalised likelihood of $\hat{f}_n$ is bounded below by
  the penalised likelihood of $f^*$, using property 1(b), we can bound
  the likelihood of $\hat{g}_n$, which allows us to prove
  $\psi(\hat{f}_n)<3\psi(f)$ for all sufficiently large $n$.

\item Having found a uniform Lipschitz constant for all $\hat{g}_n$,
  we will bound $\sum_{i=1}^n\log\left(\hat{g}_n(x_i)\right)$ as a
  function of $\left\lVert 
  \hat{g}_n-g\right\rVert_\infty$, and deduce that  
$\left\lVert
  \hat{g}_n-g\right\rVert_\infty\rightarrow 0$ almost surely.
  
\end{enumerate}

\subsection{Preliminary Results}

We begin with some general inequalities and results about the
smoothness penalty that will be necessary for proving our main
results. 

\subsubsection{Smoothness and $L^\infty$ norms}
\begin{lemma}\label{psiBoundConvolution}
If $g$ is a density function, then $\psi(f*g)\leqslant\psi(f)$.
\end{lemma}

\begin{proof}
  \begin{align*}
    \psi(f*g)&=\int_{-\infty}^\infty ((f*g)''(x))^2\,dx\\
    &=\int_{-\infty}^\infty \left(\frac{d^2}{dx^2}\int_{-\infty}^{\infty}g(t)f(x-t)\,dt\right)^2\,dx\\
    &=\int_{-\infty}^\infty \left(\int_{-\infty}^{\infty}g(t)\frac{d^2f(x-t)}{dx^2}\,dt\right)^2\,dx\\
    &=\int_{-\infty}^\infty \left(\int_{-\infty}^{\infty}g(t)f''(x-t)\,dt\right)\left(\int_{-\infty}^{\infty}g(s)f''(x-s)\,ds\right)\,dx\\
    &=\int_{-\infty}^\infty\int_{-\infty}^\infty g(t)g(s) \int_{-\infty}^{\infty}f''(x-t)f''(x-s)\,dx\,ds\,dt\\
    &\leqslant \int_{-\infty}^\infty\int_{-\infty}^\infty g(t)g(s) \int_{-\infty}^{\infty}f''(x-t)^2\,dx\,ds\,dt\\
    &=\psi(f)
  \end{align*}

\end{proof}

\begin{remark}
  It is straightforward to modify this proof in the case where the
  error distribution is not continuous.
  \end{remark}

\begin{lemma}\label{psiBound}
For any twice continuously differentiable density function $g$,
$\lVert g\rVert_\infty\leqslant
\left(\frac{5^4\psi(g)}{3\times 2^{12}}\right)^{\frac{1}{5}}$
\end{lemma}

\begin{proof}
Suppose $g$ attains its maximum at $x=0$. (Since $\lVert
g\rVert_\infty$ and $\psi(g)$ are translation invariant, there is no
loss of generality). We first prove the result under the assumption
that $g$ is symmetric about 0.

For any constant $c>0$, the function $h(x)=cg(cx)$ satisfies
$\int_{-\infty}^\infty h(x)\,dx=\int_{-\infty}^\infty g(x)\,dx$, so
$h$ is a twice continuously differentiable density function. We have
that $\lVert h\rVert_\infty=h(0)=cg(0)=c\lVert g\rVert_\infty$, and
$h''(x)=c^{3}g''(cx)$, so $\psi(h)=c^5\psi(g)$. This means that
$\lVert g\rVert_\infty \psi(g)^{-\frac{1}{5}}=\lVert h\rVert_\infty
\psi(h)^{-\frac{1}{5}}$. Thus, in trying to maximise $\lVert
g\rVert_\infty \psi(g)^{-\frac{1}{5}}$, we may without loss of
generality assume that $\int_{0}^1g(x)\,dx=0.4g(0)$.

Let $m=\sup g(x)=g(0)$. Let $a=g(1)$, $b=g'(1)$. We have $\int_0^1
g(x)\,dx=0.4m$, $g(0)=m$ and $g'(0)=0$. Now on the Hilbert space
$L^2([0,1])$, we have

\begin{align*}
\langle g'',x^2\rangle&=\int_0^1
x^2g''(x)\,dx=[x^2g'(x)]_0^1-[2xg(x)]_0^1+2\int_0^1 g(x)\,dx\\
&=b-2a+0.8m\\
\langle g'',x\rangle&=[xg'(x)]_0^1-\int_0^1g'(x)\,dx\\
&=b+m-a\\
\langle g'',1\rangle&=b\\
\langle x^\alpha,x^\beta\rangle&=\frac{1}{\alpha+\beta+1}
\end{align*} 
Clearly subject to these conditions, $\lVert g''\rVert_2$ is minimised
by setting $g''$ to be a linear
 combination $\beta_0+\beta_1x+\beta_2 x^2$. Let
 $v=(\beta_0,\beta_1,\beta_2)^T$ and let
 $$M=\left(\begin{array}{lll}
   1& \frac{1}{2}&   \frac{1}{3}\\
   \frac{1}{2}&   \frac{1}{3} & \frac{1}{4} \\
   \frac{1}{3} & \frac{1}{4} &   \frac{1}{5}
 \end{array}\right)$$
Be given by $M_{ij}=\langle x^{i-1},x^{j-1}\rangle$. Then our
equations become $Mv=w$, where $w=(b,b+m-a,b-2a+0.8m)^T$.
We have 
 $$M^{-1}=\left(\begin{array}{rrr}
   9& -36 & 30\\
   -36 &   192 & -180 \\
   30 & -180 & 180
 \end{array}\right)$$
Thus $\lVert g''\rVert_2^2\geqslant v^TMv=w^TM^{-1}MM^{-1}w=w^TM^{-1}w$. Let
$t=(m,a,b)^T$. Then $w=Bt$ where
$$B=\left(\begin{array}{rrr}
  0  &  0 & 1\\
  1  & -1 & 1\\
  0.8 & -2 & 1
\end{array}
  \right)$$

  This gives
  $$\lVert
  g''\rVert_2^2\geqslant t^TB^TM^{-1}Bt=(m,a,b)\left(\begin{array}{rrr}
19.2 &  24 & 0 \\ 
24   & 192 & -36 \\
0    & -36 &   9 
\end{array}
\right)\left(\begin{array}{r}m\\a\\b\end{array}\right)   $$
For fixed $a$, $\lVert g''\rVert_2$ is minimised when $b=4a$. In this case
$t^TB^TM^{-1}Bt$ is an increasing function of $a$ for all non-negative
$a$. Since $a\geqslant 0$, $t^TB^TM^{-1}Bt$ is minimised when $a=b=0$. 
Thus $\lVert g''\rVert_2^2\geqslant 19.2m^2$. 

Now since $g$ is an even function, we have $\int_{-1}^1
g(x)\,dx=0.8m\leqslant 1$. It follows that $m\leqslant 1.25$ and
$\psi(g)\geqslant 2t^TB^TM^{-1}Bt\geqslant 38.4m^2\geqslant
\frac{38.4m^5}{1.25^3}=\frac{3\times 2^{12}}{5^4}m^5$.
Therefore, $\lVert g\rVert_\infty\psi(g)^{-\frac{1}{5}}\leqslant \left(\frac{5^4}{3\times 2^{12}}\right)^{\frac{1}{5}}$.

We have proved the result when $g$ is symmetric about $0$. Suppose $g$
is not symmetric about 0. Let $s_+(x)=\frac{g(|x|)}{c_+}$ and
$s_-(x)=\frac{g(-|x|)}{c_-}$ be the symmetric density functions
obtained by reflecting the positive and negative parts of $g$
respectively in the $y$ axis, and rescaling. The positive constants
$c_+$ and
$c_-$ are chosen so that $\int_{-\infty}^\infty s_+(x)\,dx=1$ and 
$\int_{-\infty}^\infty s_-(x)\,dx=1$. That is, $c_-=2\int_{-\infty}^0
g(x)\,dx$ and $c_+=2\int_0^{\infty} g(x)\,dx$. It is easy to see that
$s_+$ and $s_-$ are twice continuously differentiable density
functions (because $g'(0)=0$), and are symmetric, so applying the
result for symmetric functions, we get $\psi(s_+)\geqslant
\frac{3\times 2^{12}}{5^4}\frac{\lVert g\rVert_\infty^5}{{c_+}^5}$ and $\psi(s_-)\geqslant
\frac{3\times 2^{12}}{5^4}\frac{\lVert g\rVert_\infty^5}{{c_-}^5}$. Also, we have
$\psi(g)=\frac{1}{2}\left({c_-}^2\psi(s_-)+{c_+}^2\psi(s_+)\right)\geqslant
\frac{3\times 2^{11}}{5^4}\lVert g\rVert_\infty^5\left(\frac{1}{{c_-}^3}+\frac{1}{{c_+}^3}\right)$. Since
$\int_{-\infty}^\infty g(x)\,dx=1$, we have $c_-+c_+=2$, so
${c_-}^{-3}+{c_+}^{-3}$ is minimised when $c_-=c_+=1$. Thus, the result
holds for any $g$.
\end{proof}

\begin{corollary}\label{psiBoundStar}
For any density function $f$ and any distribution, $f_e$, $\lVert
f*f_e\rVert_\infty\leqslant \left(\frac{5^4\psi(f)}{3\times 2^{12}}\right)^{\frac{1}{5}}$
\end{corollary}

\begin{proof}
This is immediate from Lemma~\ref{psiBoundConvolution} and Lemma~\ref{psiBound}
\end{proof}

\begin{lemma}\label{psiBoundLipschitz}
  If a density function $g:{\mathbb R}\rightarrow{\mathbb
    R}_{\geqslant 0}$ is twice continuously differentiable and has
  finite support, then $\lVert g'\rVert_\infty \leqslant
  \left(\frac{125\psi(g)^2}{144}\right)^{\frac{1}{5}}$
\end{lemma}

\begin{proof}
Translating if necessary, we may assume that $|g'(x)|$ is maximised by
$x=0$. By reflecting in the line $x=0$ if necessary, we may also
assume $g'(0)>0$. We may also rescale (that is, replace $g$ by
$h(x)=cg(cx)$ as in the proof of Proposition~\ref{psiBound}: it is
easy to check that $\lVert
g'\rVert_\infty\psi(g)^{-\frac{2}{5}}=\lVert
h'\rVert_\infty\psi(h)^{-\frac{2}{5}}$) so that $g(1)=g'(1)=0$.  We
will prove the result by finding a suitable function $h$ that
minimises $\psi(h)$ subject to the constraints
\begin{enumerate}[(i)]
\item $\lVert h'\rVert_\infty=h'(0)=s>0$

\item $\int_{-\infty}^1 h(x)\,dx\leqslant 1$,

\item $h$ is twice continuously differentiable
everywhere except for one value $x\leqslant 0$ at which $h(x)=0$.
\end{enumerate}
Since these conditions on $h$ are slightly weaker than the conditions
for $g$ in the proposition, any $g$ satisfying the conditions of the
proposition also satisfies the conditions for $h$. It follows that
$\psi(g)^{-\frac{2}{5}}\lVert g'\rVert_\infty\leqslant
\psi(h)^{-\frac{2}{5}}\lVert h'\rVert_\infty$, so if this $h$
satisfies the inequality in the proposition, the result will be proved. 

Let
$$k(x)=\left\{\begin{array}{ll} 0 & \textrm{if
}x<-\frac{h(0)}{h'(0)}\\ h'(0)x+h(0) & \textrm{if
}-\frac{h(0)}{h'(0)}\leqslant x<0\\ h(x) & \textrm{if }x\geqslant
0\end{array}\right.$$ That is, $k$ is a linear extension of the
restriction of $h$ to positive real numbers.  By inspection, $\lVert
k'\rVert_\infty\leqslant h'(0)=s$, so $k$ satisfies Condition~(i);
also by the mean value theorem, $k(x)\leqslant h(x)$ for all $x$, so
$\int_0^1 k(x)\,dx\leqslant \int_0^1h(x)\,dx\leqslant 1$, so $k$
satisfies Condition~(ii). Finally, Condition~(iii) is obvious on all
intervals, $\left(-\infty,-\frac{h(0)}{h'(0)}\right]$,
$\left(-\frac{h(0)}{h'(0)},0\right]$ and $(0,\infty)$, so the
condition only needs to be tested at $0$. At $0$, we obviously have
$k'(0)=h'(0)=s$, and $k''(0)=h''(0)=0$ because $h'(x)$ attains its
maximum value at $x=0$.

Since $k$ satisfies the same conditions as $h$, we have
$\psi(k)\geqslant \psi(h)$.  On the other hand $|k''(x)|\leqslant
|h''(x)|$ for all $x$, so $\psi(k)\leqslant \psi(h)$. Therefore,
$k\left(x+\frac{h(0)}{h'(0)}\right)$ is an alternative minimiser of
$\psi(k)$ subject to Conditions~(i)--(iii), and it satisfies the additional
condition that $k(0)=0$. Thus, we may assume w.l.o.g. that $h(0)=0$.

Now since $h(0)=0$, $h'(0)=s$, $h(1)=0$, $h'(1)=0$ and $\int_0^1
h(x)\,dx=r\leqslant 1$, we have

\begin{align*}
  \langle h'',1\rangle&=\int_0^1h''(x)\,dx=h'(1)-h'(0)=-s\\
  \langle h'',x\rangle&=[h'(x)x]_0^1-\int_0^1h'(x)\,dx=h'(1)-(h(1)-h(0))=0\\
  \langle
  h'',x^2\rangle&=[h'(x)x^2]_0^1-2[h(x)x]_0^1+2\int_0^1h(x)\,dx=2r\leqslant
  2\\
\end{align*}

It is easy to see that subject to these conditions, $\psi(h)$ is
minimised by setting $h''$ as a linear combination of $1$, $x$ and
$x^2$. Let $h''(x)=a+bx+cx^2$. Then we have
$$\left(
\begin{array}{rrr}
1 & \frac{1}{2} & \frac{1}{3}\\ 
\frac{1}{2} & \frac{1}{3} & \frac{1}{4} \\
\frac{1}{3} & \frac{1}{4} & \frac{1}{5} \end{array}\right)
\left(\begin{array}{r}a\\b\\c\end{array}\right)=\left(\begin{array}{r}-s\\0\\2r\end{array}\right)$$
which gives 
$$
\left(\begin{array}{l}a\\b\\c\end{array}\right)=
  \left(\begin{array}{rrr}
9 & -36 & 30 \\ 
-36 & 192 & -180 \\
30 & -180 & 180 \end{array}\right)
\left(\begin{array}{r}-s\\0\\2r\end{array}\right)$$
  and $\psi(h)=9s^2-120sr+720r^2$.

Recall that by rescaling, we are aiming to minimise $$\psi(h)\lVert
h'\rVert_\infty^{-\frac{5}{2}}=9s^{-\frac{1}{2}}-120rs^{-\frac{3}{2}}+720r^2s^{-\frac{5}{2}}$$ 
It is easy to see that this is minimised when $s$ is a solution to
\begin{align*}
 -\frac{1}{2}9s^{-\frac{3}{2}}+120\frac{3}{2}rs^{-\frac{5}{2}}-720\frac{5}{2}r^2s^{-\frac{7}{2}}&=0\\
  -s^{2}+40rs-400r^2&=0\\
  s&=20r
\end{align*}
For $s=20r$, we have $$\frac{\psi(h)^2}{s^5}=\frac{(3600-2400+720)^2r^2}{20^5r^5}=\frac{1920^2}{20^5r^3}=\frac{144}{125}r^{-3}$$
Since $r\leqslant 1$, we deduce $\frac{\psi(h)^2}{\lVert h'\rVert_\infty^5}\leqslant\frac{144}{125}$.
  
\end{proof}

\subsubsection{General Inequalities and probability theory}

We start by showing that for any random variable, there is at
least one point at which it has positive density.

\begin{lemma}\label{DensityPoint}
  For any random variable $E$, there is some $b$, $\delta>0$ and
  $\epsilon>0$ such that for any $t<\delta$, $P(|E-b|<t)>\epsilon t$.
\end{lemma}

\begin{proof}
Pick a finite interval $[l_0,u_0]$ such that $E$ has finite probability
$p$ on the interval $[l_0,u_0]$, and let $d=\frac{p}{u_0-l_0}$. Let $m_0$ be the
midpoint of $[l_0,u_0]$, For the intervals $[l_0,m_0]$ and $[m_0,u_0]$, we have
$P(E\in[l_0,m_0])+P(E\in[m_0,u_0])=p$, so one of $P(E\in[l_0,m_0])$ and
$P(E\in[m_0,u_0])$ must be at least $\frac{p}{2}$. Without loss of
generality, suppose $P(E\in[l_0,m_0])\geqslant\frac{p}{2}$. We now let
$l_1=l_0$ and $u_1=m_0$ to create a new interval whose average density
is at least $\frac{p}{u_0-l_0}$. Repeating this interval bisection, we
get a decreasing sequence of intervals $[l_k,u_k]$ whose intersection
is a single point $b$. We will show that this $b$ with
$\delta=(u_0-l_0)$ satisfies the result. To do this, for any $t<\delta$,
we will show that there is some $k$ such that
$[l_k,u_k]\subseteq [b-t,b+t]$ and
$u_{k}-l_k>\frac{t}{2}$. It will follow that $P(|E-b|<t)\geqslant
P(E\in[l_k,u_k])\geqslant d(u_k-l_k)\geqslant d\frac{t}{2}$. For
$t\geqslant \max(u_0-b,b-l_0)$, $k=0$ obviously satisfies the
conditions. Otherwise, since $l_k\rightarrow b$ and $u_k\rightarrow
b$, there must be some smallest $k$ such that $[l_k,u_k]\subseteq
[b-t,b+t]$. By minimality of this $k$, we have that either $l_{k-1}<b-t$
or $u_{k-1}>b+t$. In either case, $u_k-l_k=\frac{u_{k-1}-l_{k-1}}{2}>\frac{t}{2}$.
\end{proof} 

\begin{proposition}\label{logLowerBoundIntegral}
Let $g$ be a twice continuously differentiable function with support
$[l,u]$ and with at most
$M$ local maxima. Let $r>0$ and set $b(x)=g(x)\lor r$. Then
$$\int_{l}^u \left|\frac{b'(x)}{b(x)}\right|\,dx\leqslant
{\frac{2M}{5}}\log\left(\frac{5^4\psi(g)}{3\times 2^{12}r^5}\right)$$
\end{proposition}

\begin{proof}
  Let the local maxima of $g(x)$ be at $a_1<a_3<\ldots<a_{2M-1}$, and
  let the local minima be at $a_2<a_4<\ldots<a_{2M-2}$, and let
  $a_0=l,a_{2M}=u$. We have 
  \begin{align*}
  \int_{l}^u \left|\frac{b'(x)}{b(x)}\right|\,dx
  &=\sum_{i=0}^{M-1}\left( \int_{a_{2i}}^{a_{2i+1}}\frac{b'(x)}{b(x)}\,dx-
  \int_{a_{2i+1}}^{a_{2i+2}}\frac{b'(x)}{b(x)}\,dx\right)\\
  &=\sum_{i=0}^{M-1}
  \left(2\log(b(a_{2i+1}))-\log(b(a_{2i}))-\log(b(a_{2i+2}))\right)\\
  &\leqslant 2\sum_{i=0}^{M-1}\log\left(\frac{b(a_{2i+1})}{r}\right)\\
  \end{align*}

  By Lemma~\ref{psiBound}, we have $\lVert g\rVert_\infty\leqslant
\left(\frac{5^4\psi(g)}{3\times 2^{12}}\right)^{\frac{1}{5}}$, so
$\log(b(a_{2i+1}))\leqslant \log\left(\left(\frac{5^4\psi(g)}{3\times 2^{12}}\right)^{\frac{1}{5}}\right)$
This gives 
$$  \int_{l}^u \left|\frac{b'(x)}{b(x)}\right|\,dx
\leqslant {\frac{2M}{5}}\log\left(\frac{5^4\psi(g)}{3\times 2^{12}r^5}\right)$$
\end{proof}

\subsubsection{Kulback-Leibler Divergence}

\begin{lemma}\label{KLbound}
  If $g$ and $h$ are density functions 
  and $h(x)>g(x)+\epsilon$ for all $x\in[x_0-\delta,x_0+\delta]$, then
  the Kulback-Leibler divergence is bounded below by $\int
  g(x)\log\left(\frac{g(x)}{h(x)}\right)\,dx>2\delta^2\epsilon^2$
\end{lemma}

\begin{proof}
Let $h(x)=k(x)+l(x)$, where
$$l(x)=\left\{\begin{array}{ll}\epsilon&\textrm{if }|x-x_0|<\delta\\0&\textrm{otherwise}\end{array}\right.$$
Since $h(x)>g(x)+\epsilon$ for all $x\in [x_0-\delta,x_0+\delta]$, it
follows that $k(x)>g(x)$ for all $x\in [x_0-\delta,x_0+\delta]$.
Now $\log(h(x))=\log(k(x))+\log\left(1+\frac{l(x)}{k(x)}\right)$, so
\begin{align*}
  \int g(x)\log(h(x))\,dx&=  \int g(x)\log(k(x))\,dx+
  \int_{x_0-\delta}^{x_0+\delta}
  g(x)\log\left(1+\frac{\epsilon}{k(x)}\right)\,dx \\
  &\leqslant  \int g(x)\log(k(x))\,dx+
  \int_{x_0-\delta}^{x_0+\delta}
  g(x)\log\left(1+\frac{\epsilon}{g(x)}\right)\,dx
\end{align*}
To bound the first integral, note that
$\frac{k(x)}{1-2\delta\epsilon}$ is a density function, so
\begin{align*}
\int g(x)\log(k(x))\,dx
&=\int
g(x)\left(\log\left(\frac{k(x)}{1-2\delta\epsilon}\right)+\log(1-2\delta\epsilon)\right)\,dx\\
&\leqslant \int
g(x)\left(\log\left(g(x)\right)\,dx+\log(1-2\delta\epsilon\right)
\end{align*}
Thus
\begin{align*}
\int g(x)\log(h(x))\,dx &\leqslant \int
g(x)\log\left(g(x)\right)\,dx+\log\left(1-2\delta\epsilon\right)
+  \int_{x_0-\delta}^{x_0+\delta}
g(x)\log\left(1+\frac{\epsilon}{g(x)}\right)\,dx
\\
&\leqslant \int
g(x)\log\left(g(x)\right)\,dx+\log\left(1-2\delta\epsilon\right)
+  2\delta\epsilon
\\
&\leqslant \int
g(x)\log\left(g(x)\right)\,dx
-  2\delta^2\epsilon^2
\end{align*}
\end{proof}

\begin{lemma}\label{KLbound2}
  If $g$ and $h$ are density functions 
  and $h(x)<g(x)-\epsilon$ for all $x\in[x_0-\delta,x_0+\delta]$, then
  the Kulback-Leibler divergence is bounded below by $\int
  g(x)\log\left(\frac{g(x)}{h(x)}\right)\,dx>2\delta^2\epsilon^2-\frac{8}{3}\delta^3\epsilon^3$
\end{lemma}

\begin{proof}
Let $h(x)=k(x)-l(x)$, where
$$l(x)=\left\{\begin{array}{ll}\epsilon&\textrm{if }|x-x_0|<\delta\\0&\textrm{otherwise}\end{array}\right.$$
Since $h(x)<g(x)-\epsilon$ for all $x\in [x_0-\delta,x_0+\delta]$, it
follows that $k(x)<g(x)$ for all $x\in [x_0-\delta,x_0+\delta]$.
Now $\log(h(x))=\log(k(x))+\log\left(1-\frac{l(x)}{k(x)}\right)$, so
\begin{align*}
  \int g(x)\log(h(x))\,dx&=  \int g(x)\log(k(x))\,dx+
  \int_{x_0-\delta}^{x_0+\delta}
  g(x)\log\left(1-\frac{\epsilon}{k(x)}\right)\,dx \\
  &\leqslant  \int g(x)\log(k(x))\,dx+
  \int_{x_0-\delta}^{x_0+\delta}
  g(x)\log\left(1-\frac{\epsilon}{g(x)}\right)\,dx
\end{align*}
To bound the
first integral, note that $\frac{k(x)}{1+2\delta\epsilon}$ is a density
function, so
\begin{align*}
\int g(x)\log(k(x))\,dx&=\int
g(x)\left(\log\left(\frac{k(x)}{1+2\delta\epsilon}\right)+\log(1+2\delta\epsilon)\right)\,dx\\
&\leqslant \int
g(x)\left(\log\left(g(x)\right)\,dx+\log(1+2\delta\epsilon\right)
\end{align*}
Thus
\begin{align*}
\int g(x)\log(h(x))\,dx &\leqslant \int
g(x)\log\left(g(x)\right)\,dx+\log\left(1+2\delta\epsilon\right)
+  \int_{x_0-\delta}^{x_0+\delta}
g(x)\log\left(1-\frac{\epsilon}{g(x)}\right)\,dx
\\
&\leqslant \int
g(x)\log\left(g(x)\right)\,dx+\log\left(1+2\delta\epsilon\right)
-2\delta\epsilon
\\
&\leqslant \int
g(x)\log\left(g(x)\right)\,dx
-  2\delta^2\epsilon^2+\frac{8}{3}\delta^3\epsilon^3
\end{align*}
\end{proof}

\begin{corollary}\label{LinftyBound}
  If density functions $g^!$ and $g$ are both Lipschitz with constant
  $L$, and $\lVert g^!-g\rVert_\infty >\rho$, where $\rho<\sqrt{2L}$, then the
  Kulback-Leibler divergence is bounded below by $\int
  g(x)\log\left(\frac{g(x)}{g^!(x)}\right)\,dx>\frac{\rho^4}{48L^2}$
\end{corollary}

\begin{proof}
Since $\lVert g^!-g\rVert_\infty >\rho$, there is some $x_0$ such that
$|g^!(x_0)-g(x_0)|>\rho$. By the Lipschitz condition, if we let
$\delta=\frac{\rho}{4L}$, then for $x\in[x_0-\delta,x_0+\delta]$, we
have
\begin{align*}
\rho<|(g^!(x_0)-g(x_0))|&=|(g^!(x_0)-g^!(x))+g^!(x)-g(x)+(g(x)-g(x_0))|\\
&\leqslant L|x-x_0|+|g^!(x)-g(x)|+L|x-x_0|\end{align*}
so $|g^!(x)-g(x)|>\rho-2L\delta=\frac{\rho}{2}$.
By Lemma~\ref{KLbound} and Lemma~\ref{KLbound2}, we therefore have 
$$\int
  g(x)\log\left(\frac{g(x)}{g^!(x)}\right)\,dx>2\left(\frac{\rho^2}{8L}\right)^2-\frac{8}{3}\left(\frac{\rho^2}{8L}\right)^3$$
It is easy to see that for $\rho<\sqrt{2L}$, this is bounded below by $\frac{\rho^4}{48L^2}$.
\end{proof}

\subsection{Constructing $\tilde{g}_n$}

The function $\tilde{g}_n$ is constructed as a kernel density estimate
of $g$ with uniform kernel and bandwidth $\delta_n=\left(\frac{\log(n)}{n}\right)^{\frac{1}{4}}
\left(\frac{3}{\psi(g)}\right)^{\frac{1}{5}}2^{0.9}5^{-0.3}$. That is
$$\tilde{g}_n(x)=\frac{1}{2n\delta_n}\left|\left\{i\in\{1,\ldots,n\}\middle|\vphantom{|^|}|x_i-x|<\delta_n\right\}\right|=\frac{1}{2\delta_n}\left(\vphantom{|^|}G_n(x+\delta_n)-G_n(x-\delta_n)\right)$$
We want to show that $\tilde{g}_n$ has properties (a) and (b).

\begin{proposition}\label{Condition1a}
  If $\lVert
G_n-G\rVert_\infty\leqslant\sqrt{\frac{\log(n)}{n}}$, then $\lVert
\tilde{g}_n-g\rVert_\infty<C_3\psi(g)^{\frac{1}{5}}
\left(\frac{\log(n)}{n}\right)^{\frac{1}{4}}$ where $C_3=2^{0.1}3^{-0.2}5^{0.3}$.
\end{proposition}  

\begin{proof}
By Lemma~\ref{psiBoundLipschitz}, $g$ is Lipschitz with constant
$L=\frac{5^{\frac{3}{5}}\psi(g)^{\frac{2}{5}}}{3^{\frac{2}{5}}2^{\frac{4}{5}}}$.  Since
$\tilde{g}_n(x)=\frac{1}{2\delta_n}\left(\vphantom{|^|}G_n(x+\delta_n)-G_n(x-\delta_n)\right)$,
For any $x$,
\begin{align*}
  |\tilde{g}_n(x)-g(x)|&=\left|\frac{1}{2\delta_n}\left(\vphantom{|^|}G_n(x+\delta_n)-G_n(x-\delta_n)\right)-g(x)\right|\\
  &\leqslant\left|\frac{1}{2\delta_n}\left(\vphantom{|^|}(G_n(x+\delta_n)-G(x+\delta_n))-(G_n(x-\delta_n)-G(x-\delta_n))\right)\right|\\
  &\qquad\qquad\qquad+\left|\frac{1}{2\delta_n}\left(\vphantom{|^|}(G(x+\delta_n))-G(x-\delta_n)\right)-g(x)\right|\\
  &\leqslant \frac{\lVert G_n-G\rVert_\infty}{\delta_n}+\left|\frac{1}{2\delta_n}\int_{-\delta_n}^{\delta_n}g(x+t)\,dt-g(x)\right|\\
  &\leqslant \frac{1}{\delta_n}\sqrt{\frac{\log(n)}{n}}+\left|\frac{1}{2\delta_n}\int_{-\delta_n}^{\delta_n}(g(x+t)-g(x))\,dt\right|\\
  &\leqslant \frac{1}{\delta_n}\sqrt{\frac{\log(n)}{n}}+\frac{1}{2\delta_n}\int_{-\delta_n}^{\delta_n}L|t|\,dt\\
  &= \frac{1}{\delta_n}\sqrt{\frac{\log(n)}{n}}+\frac{L\delta_n}{2}
\end{align*}
In particular, since
$\delta_n=
\left(\frac{\log(n)}{n}\right)^{\frac{1}{4}}
\left(\frac{3}{\psi(g)}\right)^{\frac{1}{5}}2^{0.9}5^{-0.3}$,
we get
$$|\tilde{g}_n(x)-g(x)|\leqslant 2^{0.1}5^{0.3}3^{-0.2}\psi(g)^{0.2}\left(\frac{\log(n)}{n}\right)^{\frac{1}{4}}$$
\end{proof}

\begin{proposition}\label{Condition1b} If $\lVert
G_n-G\rVert_\infty\leqslant\sqrt{\frac{\log(n)}{n}}$, then for any
function $g^*_L$ that is Lipschitz with constant $L$, has
support $S_n$, and has $g_L^*(x_i)\geqslant \epsilon>0$, 
      $$\frac{1}{n}\sum_{i=1}^n \log(g_L^*(x_i))-\int_{S_n}
\tilde{g}_n(x)\log(g_L^*(x))\,dx<{C_3}^{-1}\psi(g)^{-0.2}\left(\frac{\log(n)}{n}\right)^{\frac{1}{4}}
\epsilon^{-1}L$$
\end{proposition}  

\begin{proof}
  We have
  \begin{align*}
\frac{1}{n}\sum_{i=1}^n    \log(g_L^*(x_i))&-\int_{S_n}    \tilde{g}_n(x)\log(g_L^*(x))\,dx\\
    &=\frac{1}{n}\sum_{i=1}^n \log(g_L^*(x_i))-\frac{1}{2\delta_n}\int_{S_n} \left(\frac{1}{n}\sum_{i=1}^n1_{\{|x_i-x|<\delta_n\}}\right)\log(g_L^*(x))\,dx\\
    &=\frac{1}{n}\sum_{i=1}^n \frac{1}{2\delta_n}\int_{x_i-\delta_n}^{x_i+\delta_n} \left(\log(g_L^*(x_i))-\log(g_L^*(x))\right)\,dx\\
&\leqslant \frac{1}{n}\sum_{i=1}^n \frac{1}{2\delta_n}\int_{x_i-\delta_n}^{x_i+\delta_n} \log\left(1+\frac{L|x-x_i|}{g^*_L(x)}\right)\,dx\\
&\leqslant \frac{1}{n}\sum_{i=1}^n \frac{1}{2\delta_n}\int_{x_i-\delta_n}^{x_i+\delta_n} \frac{L|x-x_i|}{g^*_L(x)}\,dx\\
    &\leqslant\frac{\delta_nL}{2\epsilon}
  \end{align*}

\end{proof}

\subsection{Constructing $f^*_n$}

Controlling the P-MLE is based on finding a lower bound for the
penalised maximum likelihood. We do this by exhibiting a good
candidate density function that gives high penalised likelihood. A
first attempt is the true density function $f$. However, $f$ can
have low likelihood because of a small number of outliers with very
low likelihood. To correct for this, we will take a mixture of the
true likelihood $f$ and a data-driven likelihood $f^\dagger_n$ that guarantees
$g^\dagger_n(x_i)>\epsilon_n$ for $i=1,\ldots,n$.

The easiest way to get an estimator with likelihood bounded below on
observed data points is with a kernel density estimate. Since we are
trying to find a deconvolved density estimator, this approach does not
work --- there is no guarantee that the kernel density estimator
arises as a convolution with the error distribution. However, by
Lemma~\ref{DensityPoint}, there is a basepoint $b$ for the error
distribution, and an $\epsilon>0$, such that for all $\delta<c$,
$P(|E-b|<\delta)>\delta\epsilon$. We let $f^\dagger$ be a
kernel estimate from $x_i-b$. Using the condition from
Lemma~\ref{DensityPoint}, we get that if there is an interval about
$x_0$ such that $f^\dagger(x)>a$ for all
$x\in[x_0-\delta,x_0+\delta]$, then $g^\dagger(x_0+b)>\epsilon\delta a$.

We want to control the smoothness penalty $\psi(f^\dagger)$. To do
this, we set the kernel $$k_\delta(x)=\delta^{-4}\left\{\begin{array}{ll}
3\delta(x+\delta)^2-2(x+\delta)^3& \textrm{if }-\delta<x\leqslant 0\\
3\delta(x-\delta)^2+2(x-\delta)^3& \textrm{if }0<x<\delta\\
0&\textrm{otherwise}
\end{array}\right.$$

It is easy to check that this kernel has the following properties:
\begin{lemma}
  \begin{enumerate}[(i)]

  \item $\int_{-\infty}^\infty k_\delta(x)\,dx=1$.

  \item $\psi(k_\delta)=24\delta^{-5}$

  \item $k_\delta(0)=\delta^{-1}$
  \end{enumerate}

\end{lemma}

\begin{proof}
  \begin{enumerate}[(i)]

  \item By symmetry, the positive and negative parts of $k_\delta$
    have the same integral, so   
    \begin{align*}
      \int_{-\infty}^\infty k_\delta(x)\,dx&=2\delta^{-4}\int_{-\delta}^0
      \left(3\delta(x+\delta)^2-2(x+\delta)^3\right)\,dx\\
      &=2\delta^{-4}\int_{0}^\delta
      (3\delta t^2-2t^3)\,dt\\
      &=2\delta^{-4}\left[\delta
        t^3-\frac{1}{2}t^4\right]_0^\delta\\
      &=2\delta^{-4}\left(\delta^4-\frac{1}{2}
      \delta ^4\right)\\
      &=1
    \end{align*}

\item We have $${k_\delta}''(x)=\delta^{-4}\left\{\begin{array}{ll}
6\delta-12(x+\delta)& \textrm{if }-\delta<x\leqslant 0\\
6\delta+12(x-\delta)& \textrm{if }0<x<\delta\\
0&\textrm{otherwise}
\end{array}\right.$$

  so
  \begin{align*}
  \psi(k_\delta)&=\int_{-\infty}^\infty ({k_\delta}''(x))^2\,dx\\
  &=\delta^{-8}\int_{-\delta}^\delta (6\delta-12|x|)^2\,dx\\
  &=36\delta^{-8}\int_{-\delta}^\delta\left(\delta^2-4\delta|x|+4x^2\right)\,dx\\
  &=36\delta^{-8}\left(2\delta^3-4\delta^3+\frac{8}{3}\delta^3\right)\\
  &=24\delta^{-5}
  \end{align*}

  \item This is immediate by evaluating $k_\delta(0)$.

  \end{enumerate}

\end{proof}

We can now define $f_n^\dagger(x)=\frac{1}{n}\sum_{i=1}^n
k_{\delta_n}(x-x_i+b)$ for some suitably chosen $\delta_n$. We obtain
the following properties of $f^\dagger$ for sufficiently large $n$.

\begin{lemma}\label{fdaggerbounds}
If $\lVert G_n-G\rVert_\infty\leqslant \sqrt{\frac{\log (n)}{n}}$,
and $n$ is sufficiently large, then
  
  \begin{enumerate}[(i)]

  \item $\int_{-\infty}^\infty f^\dagger(x)\,dx=1$.

  \item  $\psi(f^\dagger)\leqslant 48\left(\frac{5^4\psi(g)}{12}\right)^{\frac{1}{5}}\delta^{-4}$

  \item For all $i=1,\ldots,n$, $g^\dagger(x_i)\geqslant \frac{\epsilon}{4n}$
  \end{enumerate}

\end{lemma}

\begin{proof}
  \begin{enumerate}[(i)]

  \item This is obvious, since $f^\dagger$ is a mixture of $k_\delta$.
    
  \item

    \begin{align*}
      \psi(f^\dagger)&=\langle {f^\dagger}'',{f^\dagger}''\rangle\\
&=\left\langle \frac{1}{n}\sum k_{\delta,x_i-b}'',\frac{1}{n}\sum k_{\delta,x_i-b}''\right\rangle\\
&=\frac{1}{n^2}\sum_{i=1}^n\sum_{j=1}^n\left\langle  k_{\delta,x_i-b}'', k_{\delta,x_j-b}''\right\rangle\\
    \end{align*}
    For all $i,j$, $\left\langle  k_{\delta,x_i-b}'',
    k_{\delta,x_j-b}''\right\rangle\leqslant
    \psi(k_{\delta})=24\delta^{-5}$. On the other hand, if
    $|x_i-x_j|\geqslant 2\delta$, then $k_{\delta,x_i-b}''$ and
    $k_{\delta,x_j-b}''$ have disjoint supports, so 
    $\langle  k_{\delta,x_i-b}'', k_{\delta,x_j-b}''\rangle=0$.
    Thus 
    \begin{align*}
    \psi(f^\dagger)&\leqslant
    \frac{24}{n^2\delta^5}|\{(i,j)||x_i-x_j|<2\delta\}|\\
    &=\frac{24}{n\delta^5}\sum_{i=1}^n
    \left(G_n(x_i+2\delta)-G_n(x_i-2\delta)\right)\\
    &\leqslant\frac{24}{n\delta^5}\sum_{i=1}^n
    \left(G(x_i+2\delta)-G(x_i-2\delta)+2\lVert G_n-G\rVert_\infty\right)\\
    &\leqslant\frac{24}{n\delta^5}\sum_{i=1}^n
   \left( 4\delta\lVert g\rVert_\infty+2\lVert G_n-G\rVert_\infty\right)\\
    &\leqslant\frac{24}{n\delta^5}\sum_{i=1}^n
    \left(\delta\left(\frac{5^4\psi(g)}{12}\right)^{\frac{1}{5}}+2\sqrt{\frac{\log(n)}{n}}\right)
    \end{align*}
For large enough $n$, $2\sqrt{\frac{\log(n)}{n}}\leqslant
\left(\frac{5^4\psi(g)}{12}\right)^{\frac{1}{5}}\delta$, so
$\psi(f^\dagger)\leqslant 48\left(\frac{5^4\psi(g)}{12}\right)^{\frac{1}{5}}\delta^{-4}$
    
\item
 For $x\in\left[x_i-b-\frac{\delta}{2},x_i-b+\frac{\delta}{2}\right]$, we
 have $f^{\dagger}(x)\geqslant \frac{1}{n}k_\delta(x-(x_i-b))\geqslant
 \frac{1}{n}k_\delta\left(\frac{\delta}{2}\right)=\frac{\delta^{-1}}{2n}$. Therefore
 $g^\dagger(x_i)\geqslant \epsilon\frac{\delta}{2}\frac{\delta^{-1}}{2n}=\frac{\epsilon}{4n}$.

  \end{enumerate}

\end{proof}

Now we define $f_n^*(x)=(1-\gamma_n)f(x)+\gamma_n f_n^\dagger(x)$, where
\begin{equation}
  \gamma_n=
  \frac{2M\sqrt{\log(n)}}{\sqrt{n}+2M\sqrt{\log(n)}}\label{gammavalue}
\end{equation}
and $\delta_n$ is the solution to  
\begin{equation}
  \gamma_n=\left(\sqrt{2}-1\right)\delta_n^2\sqrt{\frac{\psi(f)}{48\left(\frac{5^4\psi(g)}{12}\right)^{\frac{1}{5}}}}
\label{deltavalue}
\end{equation}

\begin{proposition}\label{fstarbounds}
For the above definition of $f_n^*(x)$, if $\lVert
G_n-G\rVert_\infty<\sqrt{\frac{\log(n)}{n}}$ and $n$ is sufficiently
large, we have

\begin{enumerate}[(i)]

\item  $\psi(f_n^*)\leqslant 2\psi(f)$.
\item For $i=1,\ldots,n$, $g_n^*(x_i)\geqslant
  (1-\gamma_n)\left(g(x_i)\lor r_n\right)$, where
  \begin{equation}
  r_n=\frac{\epsilon\gamma_n}{4n}\label{rvalue}  
  \end{equation}

\end{enumerate}
  
\end{proposition}

\begin{proof}
\begin{enumerate}[(i)]

\item
$$\psi(f_n^*)\leqslant
  (1-\gamma_n)^2\psi(f)+2\gamma_n(1-\gamma_n)\sqrt{\psi(f)\psi(f^\dagger)}+\gamma_n^2\psi(f^\dagger)\leqslant
  \left(\gamma_n\sqrt{\psi(f^\dagger)}+\sqrt{\psi(f)}\right)^2$$
Substituting
$\gamma_n=\left(\sqrt{2}-1\right)\delta_n^2\sqrt{\frac{\psi(f)}{48\left(\frac{5^4\psi(g)}{12}\right)^{\frac{1}{5}}}} $ and
$\psi(f^\dagger)\leqslant 48\left(\frac{5^4\psi(g)}{12}\right)^{\frac{1}{5}}\delta^{-4}$
(from Lemma~\ref{fdaggerbounds}) gives
$$\gamma_n\sqrt{\psi(f^\dagger)}\leqslant(\sqrt{2}-1)\sqrt{\psi(f)}$$
so
$\psi(f_n^*)<2\psi(f)$.

\item

  \begin{align*}
  g_n^*(x_i)&= \gamma_n
g^\dagger(x_i)+
(1-\gamma_n)g(x_i)\\
&\geqslant\frac{\gamma_n\epsilon}{4n}\lor
(1-\gamma_n)g(x_i)\\
&=(1-\gamma_n)\left(g(x_i)\lor\frac{\gamma_n\epsilon}{4n(1-\gamma_n)}\right)\\
&\geqslant (1-\gamma_n)\left(g(x_i)\lor\frac{\gamma_n\epsilon}{4n}\right)
\end{align*}

\end{enumerate}
\end{proof}

This means that
$$\frac{1}{n}\sum_{i=1}^n\log(g_n^*(x_i))\geqslant 
\log(1-\gamma_n)+\frac{1}{n}\sum_{i=1}^n\log\left(b(x_i)\right) $$
where
$b_n(x)=g(x)\lor r_n$
where
$r_n=\frac{\gamma_n\epsilon}{4n}$.

\begin{proposition}\label{gStarLogLikelihood}
  Whenever, $\lVert G_n-G\rVert<\sqrt{\frac{\log(n)}{n}}$,
and $n$ is sufficiently large
$$\frac{1}{n}\sum_{i=1}^n \log(g_n^*(x_i))> -H(g)-
3M\log(n)^{\frac{3}{2}}n^{-\frac{1}{2}}$$
where
$H(g)$ is the entropy $-\int_{-\infty}^\infty g(x)\log(g(x))$
\end{proposition}

\begin{proof}

We set $b_n(x)=g(x)\lor r_n$, so that for all $i=1,\ldots,n$,
$g_n^*(x_i)\geqslant (1-\gamma_n)b_n(x_i)$.

\begin{align*}
\frac{1}{n}\sum_{i=1}^n \log(b_n(x_i))&=\log(r_n)-\int_{-\infty}^\infty
G_n(x)\left(\frac{d}{dx}\log(b_n(x))\right)\,dx\\
&=\log(r_n)-\int_{-\infty}^\infty G_n(x)\left(\frac{b_n'(x)}{b_n(x)}\right)\,dx\\
&=\log(r_n)-\int_{-\infty}^\infty G(x)\left(\frac{b_n'(x)}{b_n(x)}\right)\,dx+\int_{-\infty}^\infty (G(x)-G_n(x))\left(\frac{b_n'(x)}{b_n(x)}\right)\,dx\\
&=\int_{-\infty}^\infty g(x)\log(b_n(x))\,dx+\int_{-\infty}^\infty (G(x)-G_n(x))\left(\frac{b_n'(x)}{b_n(x)}\right)\,dx\\
&\geqslant -H(g)-\lVert G(x)-G_n(x)\rVert_\infty\int_{-\infty}^\infty \left|\frac{b_n'(x)}{b_n(x)}\right|\,dx\\
\frac{1}{n}\sum_{i=1}^n \log(g^*_n(x_i))&\geqslant
-H(g)+\log(1-\gamma_n)-\lVert G(x)-G_n(x)\rVert_\infty 
{\frac{2M}{5}}\log\left(\frac{5^4\psi(g)}{3\times
  2^{12}{r_n}^5}\right)\\
&\geqslant
-H(g)+\log(1-\gamma_n)- 
{\frac{2M}{5}}\log\left(\frac{5^4\psi(g)}{3\times 2^{12}{r_n}^5}\right)\sqrt{\frac{\log(n)}{n}}
\end{align*}
where the last line is by Proposition~\ref{logLowerBoundIntegral}.
Substituting $\gamma_n=\frac{c_n}{1+c_n}$ where
$c_n=2M\sqrt{\frac{\log(n)}{n}}$, and letting
$K_n=\left(\frac{5^4\psi(g)}{3\times 2^{12}}\right)^{\frac{1}{5}}{r_n}^{-1}$,
this inequality becomes
$$\frac{1}{n}\sum_{i=1}^n \log(g)_n^*(x_i))\geqslant
-H(g)-c_n\log(K_n)-\log(c_n+1)$$
We have $\log(c_n+1)\leqslant c_n$, and 
for large enough $n$, we have $n^3>e^2{K_n}^2$, so $\log(K_n)+1<\frac{3\log(n)}{2}$
Thus, the inequality becomes
$$\frac{1}{n}\sum_{i=1}^n \log(g_n^*(x_i))\geqslant
-H(g)-\frac{3c_n\log(n)}{2}
$$
\end{proof}

\subsection{Proving Consistency}

In this section, we use the $\tilde{g}_n$ and $f^*_n$ defined in the
previous sections to show that provided
$\frac{|S_n|\log(n)^{\frac{1}{4}}}{n^{\frac{1}{4}}}\rightarrow 0$,
when we set
$\lambda_n=C_1n^{\frac{7}{8}}\log(n)^{\frac{1}{8}}\sqrt{|S_n|}$, where $C_1=
\frac{2^{\frac{31}{20}}3^{-\frac{1}{10}}5^{\frac{3}{20}}}{\psi(f_x)^{\frac{4}{5}}\psi(g)^{\frac{1}{10}}}$, our
method is consistent.

\begin{lemma}\label{KLlor}
If $g$ and $h$ are density functions with support $S$, and $a>0$ then
$\int_S g(x)\log(h(x)\lor a)\,dx\leqslant -H(g)+a|S|$.
\end{lemma}

\begin{proof}
  \begin{align*}
    \int_S g(x)\log(h(x)\lor a)\,dx+H(g)&=\int_S
    g(x)\log\left(\frac{h(x)\lor a}{g(x)}\right)\,dx\\
    &\leqslant \int_S g(x)\left(\frac{h(x)\lor a-g(x)}{g(x)}\right)\,dx\\
    &\leqslant \int_S(h(x)+ a-g(x))\,dx\\
    &=1+a|S|-1=a|S|
  \end{align*}

\end{proof}

\begin{proposition}\label{HatPenalty}
  If $\lVert G_n-G\rVert\leqslant\sqrt{\frac{\log(n)}{n}}$, and $n$ is
  sufficiently large, then $\psi(\hat{f}_n)\leqslant 3\psi(f)$.
\end{proposition}

\begin{proof}
By Propositions~\ref{gStarLogLikelihood} and~\ref{fstarbounds}(i), the
penalised likelihood of $f_n^*$ is at least
$$-n\left(H(g)+3M\log(n)^{\frac{3}{2}}n^{-\frac{1}{2}}\right)-2\lambda_n\psi(f)$$
Since $\hat{f}_n$ maximises the penalised likelihood, we have that
\begin{equation}
\sum_{i=1}^n \log(\hat{g}_n(x_i))-\lambda_n\psi(\hat{f}_n)\geqslant
-n\left(H(g)+3M\log(n)^{\frac{3}{2}}n^{-\frac{1}{2}}\right)-2\lambda_n\psi(f)\label{pMLE}
\end{equation}
By Lemmas~\ref{psiBoundConvolution} and~\ref{psiBoundLipschitz},
$\hat{g}_n$ is Lipschitz with constant
$L=\left(\frac{125}{144}\right)^{\frac{1}{5}}\psi(\hat{f}_n)^{\frac{2}{5}}$.
By Proposition~\ref{Condition1b}, we have that for any $0<a<1$,
\begin{align}
\sum_{i=1}^n \log(\hat{g}_n(x_i))&\leqslant \sum_{i=1}^n
\log(\hat{g}_n(x_i)\lor a)\nonumber\\
&\leqslant n\int_{S_n}
\tilde{g}_n(x)\log(\hat{g}_n(x)\lor a)\,dx+\frac{n^{\frac{3}{4}}\log(n)^{\frac{1}{4}}}{C_3\psi(g)^{\frac{1}{5}}a}L\label{1chat}
\end{align}
Meanwhile, we have
\begin{adjustwidth}{-4cm}{-4cm}
\begin{align*}
\int_{S_n} \tilde{g}_n(x)\log(\hat{g}_n(x)\lor a)\,dx&=
\int_{S_n} (g(x)+\tilde{g}_n(x)-g(x))\log(\hat{g}_n(x)\lor a)\,dx\\
&\leqslant \int_{S_n} g(x)\log(\hat{g}_n(x)\lor a)\,dx+|S_n|\lVert
\tilde{g}_n-g\rVert_\infty\sup(|\log(\hat{g}_n(x)\lor a)|)\\
&\leqslant -H(g)+a|S_n|+|S_n|\lVert
\tilde{g}_n-g\rVert_\infty\left(\log(\sup(\hat{g}_n(x)))\lor (-\log(a)))\right)
\end{align*}
\end{adjustwidth}
Since $\hat{g}_n$ is Lipschitz with constant $L$, the equation
$\int_{-\infty}^\infty \hat{g}_n(x)\,dx=1$ gives
$\sup\hat{g}_n(x)\leqslant \sqrt{L}$, and by
Proposition~\ref{Condition1a}, $\lVert
\tilde{g}_n-g\rVert_\infty\leqslant
C_3\psi(g)^{\frac{1}{5}}\left(\frac{\log(n)}{n}\right)^{\frac{1}{4}}$
so $$\log(\sup(\hat{g}_n(x)))\lor (-\log(a))\leqslant
\log\left(\sqrt{L}\lor a^{-1}\right)=\frac{1}{2}\log\left(L\lor a^{-2}\right)$$ Therefore,
$$\int_{S_n} \tilde{g}_n(x)\log(\hat{g}_n(x)\lor a)\,dx\leqslant
-H(g)+a|S_n|+\frac{C_3}{2}\psi(g)^{\frac{1}{5}}|S_n|\left(\frac{\log(n)}{n}\right)^{\frac{1}{4}}\log\left(L\lor a^{-2}\right)
$$
Substituting this into~\eqref{1chat}, and substituting the result
into~\eqref{pMLE} gives
\begin{adjustwidth}{-2cm}{-2cm}
\begin{align}
-n\left(H(g)+3Mn^{-\frac{1}{2}}\log(n)^{\frac{3}{2}}\right)-2\lambda_n\psi(f)&\leqslant
n\left(-H(g)+a|S_n|+\frac{C_3}{2}\psi(g)^{\frac{1}{5}}|S_n|\left(\frac{\log(n)}{n}\right)^{\frac{1}{4}}\log\left(L\lor a^{-2}\right)
\right)\nonumber\\
&\qquad\qquad+
\frac{n^{\frac{3}{4}}\log(n)^{\frac{1}{4}}}{C_3\psi(g)^{\frac{1}{5}}a}L
-\lambda_n\psi(\hat{f}_n)\nonumber
\\
-3Mn^{\frac{1}{2}}\log(n)^{\frac{3}{2}}-2\lambda_n\psi(f)
&\leqslant
\frac{n}{2}|S_n|\left(2a+C_3\psi(g)^{\frac{1}{5}}\left(\frac{\log(n)}{n}\right)^{\frac{1}{4}}\left(\log\left(L\lor
a^{-2}\right)\right)\right)\nonumber\\
&\qquad\qquad+
\frac{n^{\frac{3}{4}}\log(n)^{\frac{1}{4}}}{C_3\psi(g)^{\frac{1}{5}}a}L
-\lambda_n\psi(\hat{f}_n)\label{psifhatboundequation}
\end{align}
\end{adjustwidth}
Since $a$ is arbitrary here, we can set
$a=\left(\frac{\log(n)}{n}\right)^{\frac{1}{8}}\frac{\sqrt{L}}{\psi(g)^{\frac{1}{10}}\sqrt{C_3|S_n|}}$
so that $an|S_n|+\frac{n^{\frac{3}{4}}\log(n)^{\frac{1}{4}}}{C_3\psi(g)^{\frac{1}{5}}a}L
=C_4n^{\frac{7}{8}}\log(n)^{\frac{1}{8}}\sqrt{L}\sqrt{|S_n|}$, where
$C_4=2\psi(g)^{-\frac{1}{10}}{C_3}^{-\frac{1}{2}}$. Also, if
$\psi(\hat{f}_n)\leqslant 3\psi(f)$, then the proposition is
true. Otherwise,
$a^{-2}=\frac{n^{\frac{1}{4}}\psi(g)^{\frac{1}{5}}C_3|S_n|}{L\log(n)^{\frac{1}{4}}}\leqslant
\frac{n^{\frac{1}{4}}\psi(g)^{\frac{1}{5}}C_3|S_n|}{\left(\frac{125}{144}\right)^{\frac{1}{5}}(3\psi(f))^{\frac{2}{5}}\log(n)^{\frac{1}{4}}}$.
For large enough $n$, we have $a^{-2}>L$, so we 
rearrange~\eqref{psifhatboundequation} to get
\begin{adjustwidth}{-4cm}{-4cm}
\begin{align*}
  &\psi(\hat{f}_n)-2\psi(f)\\
  \leqslant&\frac{1}{\lambda_n}\left(\frac{C_3\psi(g)^{\frac{1}{5}}}{2}n^{\frac{3}{4}}|S_n|\log(n)^{\frac{1}{4}}\left(\log\left(\frac{C_3\psi(g)^{\frac{1}{5}}|S_n|n^{\frac{1}{4}}}{\left(\frac{125}{16}\right)^{\frac{1}{5}}\psi(f)^{\frac{2}{5}}\log(n)^{\frac{1}{4}}}\right)\right)+3Mn^{\frac{1}{2}}\log(n)^{\frac{3}{2}}+C_4n^{\frac{7}{8}}\log(n)^{\frac{1}{8}}\sqrt{L}\sqrt{|S_n|}\right)
  \end{align*}
\end{adjustwidth}
Substituting
$L=\left(\frac{125}{144}\right)^{\frac{1}{5}}\psi(\hat{f}_n)^{\frac{2}{5}}$
and rearranging gives

$$\psi(\hat{f}_n)-\alpha_n\psi(\hat{f}_n)^{\frac{1}{5}}\leqslant
\beta_n+2\psi(f)$$
where
\begin{align*}
  \alpha_n
  &= \frac{C_4n^{\frac{7}{8}}\log(n)^{\frac{1}{8}}\sqrt{|S_n|}}{\lambda_n}
\left(\frac{125}{144}\right)^{\frac{1}{10}}\\
  \beta_n&=\frac{C_3\psi(g)^{\frac{1}{5}}}{2\lambda_n}n^{\frac{3}{4}}|S_n|\log(n)^{\frac{1}{4}}\log\left(\frac{C_3\psi(g)^{\frac{1}{5}}|S_n|n^{\frac{1}{4}}}{\left(\frac{125}{16}\right)^{\frac{1}{5}}\psi(f)^{\frac{1}{5}}\log(n)^{\frac{1}{4}}}\right)+
  \frac{3Mn^{\frac{1}{2}}\log(n)^{\frac{3}{2}}}{\lambda_n}
\end{align*}
We have defined $\lambda_n$ to satisfy $\lambda_n= \frac{2C_4}{\psi(f)^{\frac{4}{5}}}n^{\frac{7}{8}}\log(n)^{\frac{1}{8}}\sqrt{|S_n|}
\left(\frac{125}{144}\right)^{\frac{1}{10}}$,
so that $\alpha_n=\frac{\psi(f)^{\frac{4}{5}}}{2}$ and
$\beta_n\rightarrow 0$, so for large enough $n$,
$\beta_n<\left(1-\frac{1}{2}\times3^{\frac{1}{5}}\right)\psi(f)$, so we
have
$\frac{\psi(\hat{f}_n)}{\psi(f)}-\frac{1}{2}\left(\frac{\psi(\hat{f}_n)}{\psi(f)}\right)^{\frac{1}{5}}<3-\frac{1}{2}\times
3^{\frac{1}{5}}$, which gives $\psi(\hat{f}_n)<3\psi(f)$.
\end{proof}

This proves that $\hat{f}_n$ is Lipschitz. Under this Lipschitz
condition, we can show that if $\hat{g}_n$ is too far from $g$, then its
penalised likelihood will be less than the penalised likelihood of
$g_n^*$, contradicting the maximality.  

\begin{proposition}\label{FarBound}
  If $\lVert g_L-g\rVert_\infty\geqslant\rho_n$, where $g$ and $g_L$ are both
  Lipschitz density functions with constant $L$, 
  and $\lVert
  G_n-G\rVert_\infty<\sqrt{\frac{\log(n)}{n}}$, then

$$\frac{1}{n}\sum_{i=1}^n\log(g_L(x_i))\leqslant
-H(g)-\frac{{\rho_n}^4}{1536L^2}+\left(\frac{\log(n)}{n}\right)^{\frac{1}{4}}|S_n|\left(\frac{1536L^3}{C_3\psi(g)^{\frac{1}{5}}{\rho_n}^4}+C_3\psi(g)^{\frac{1}{5}}\log\left(\frac{1536|S_n|L^{\frac{5}{2}}}{{\rho_n}^4}\right)\right)$$

\end{proposition}

\begin{proof}
By Proposition~\ref{Condition1b}, we have that for any $a>0$, 
\begin{align*}
  \frac{1}{n}\sum_{i=1}^n\log(g_L(x_i))&\leqslant
  \frac{1}{n}\sum_{i=1}^n\log(g_L(x_i)\lor a)\\
&\leqslant
  \int_{S_n}
  \tilde{g}_n(x)\log(g_L(x)\lor
  a)\,dx+
\frac{\log(n)^{\frac{1}{4}}L}{C_3\psi(g)^{\frac{1}{5}}n^{\frac{1}{4}}a}
\end{align*}
We use Proposition~\ref{Condition1a} to show
  \begin{align*}
  \int_{S_n}
  \tilde{g}_n(x)\log(g_L(x)\lor
  a)\,dx &= \int_{S_n} \left(g(x)+
  \tilde{g}_n(x)-g(x)\right)\log(g_L(x)\lor a)\,dx\\
 &\leqslant \int_{S_n}
  g(x)\log(g_L(x)\lor a)\,dx+|S_n|\lVert\tilde{g}_n-g\rVert_\infty\log\left(\frac{\sup_{x\in
    S_n}(g_L(x))\lor1}{a}\right)\\
 &\leqslant \int_{S_n}
  g(x)\log(g_L(x)\lor a)\,dx\\
  &\qquad\qquad+C_3\psi(g)^{\frac{1}{5}}\left(\frac{\log(n)}{n}\right)^{\frac{1}{4}}|S_n|
  \log\left(\frac{\sup_{x\in
    S_n}(g_L(x))\lor 1}{a}\right)
\end{align*}
Let $\int_{S_n} g_L(x)\lor a\,dx=c$. Clearly $1\leqslant c\leqslant
1+a|S_n|$. Let $b(x)=\frac{g_L(x)\lor a}{c}$. By definition, $b(x)$
is a density function, and is Lipschitz with constant $L$, and $\lVert
b-g\rVert_\infty\geqslant\rho_n-\left((c-1)\lVert
g_L\rVert_\infty\lor a\right)$, since for any
$x$,
$$|b(x)-g_L(x)|=\left|\frac{g_L(x)\lor
  a}{c}-g_L(x)\right|=\left\{\begin{array}{ll}
\left(1-\frac{1}{c}\right)g_L(x) & \textrm{if }g_L(x)>a\\
\frac{a}{c}-g_L(x)& \textrm{if }g_L(x)\leqslant a\end{array}\right.
\leqslant
\left(c-1\right)g_L(x)\lor a
$$ Since $\lVert g_L-g\rVert_\infty\geqslant\rho_n$, there is some $x$ for which
$|g_L(x)-g(x)|\geqslant\rho_n$. For this $x$, $|b(x)-g(x)|\geqslant
|g_L(x)-g(x)|-|g_L(x)-b(x)|\geqslant \rho_n-\left((c-1)\lVert
g_L\rVert_\infty\lor a\right)$. Now
by Corollary~\ref{LinftyBound},
\begin{align*}
  \int_{S_n}
  g(x)\log(g_L(x)\lor a)\,dx&=  \int_{S_n}
  g(x)\log\left(cb(x)\right)\,dx\\
  &=\int_{S_n}
  g(x)(\log(b(x))+\log(c))\,dx\\
  &\leqslant -H(g)-\frac{\left(\rho_n-\left((c-1)\lVert
    g_L\rVert_\infty\right)\lor a\right)^4}{48L^2}+\log(c)\\
\end{align*}
Thus, using the fact that $(c-1)\lVert g_L\rVert_\infty\leqslant
a|S_n|\lVert g_L\rVert_\infty$, and $|S_n|\lVert
g_L\rVert_\infty\geqslant \int_{S_n}g_L(x)\,dx=1$, we get
\begin{adjustwidth}{-4cm}{-4cm}
  \begin{align*}
    \frac{1}{n}\sum_{i=1}^n\log(g_L(x_i))&\leqslant
    -H(g)-\frac{(\rho_n-a(|S_n|\lVert g_L\rVert_\infty))^4}{48L^2}+\log(1+a|S_n|)\\
    &\qquad\qquad
+C_3\psi(g)^{\frac{1}{5}}\left(\frac{\log(n)}{n}\right)^{\frac{1}{4}}|S_n|
  \log\left(\frac{\sup_{x\in
    S_n}(g_L(x))\lor1}{a}\right)\\
  &\qquad\qquad\qquad\qquad+\frac{\log(n)^{\frac{1}{4}}L}{C_3\psi(g)^{\frac{1}{5}}n^{\frac{1}{4}}a}\end{align*}
\end{adjustwidth}
for any $a>0$. In particular, we choose $a=
\frac{{\rho_n}^4}{1536L^2|S_n|}$. Since
$$\rho_n\leqslant\lVert g_L-g\rVert_\infty\leqslant\lVert
g_L\rVert_\infty+\lVert g\rVert_\infty\leqslant 2\sqrt{L}<768^{\frac{1}{3}}\sqrt{L}\leqslant \left(\frac{768L^2}{\lVert
  g_L\rVert_\infty}\right)^{\frac{1}{3}}$$
(because $\lVert
g_L\rVert_\infty\leqslant\sqrt{L}$, by the Lipschitz condition), it
follows that $a\leqslant\frac{\rho_n}{2|S_n|\lVert
  g_L\rVert_\infty}$, so we have $\frac{(\rho_n-a|S_n|\lVert
  g_L\rVert_\infty)^4}{48L^2}\geqslant \frac{{\rho_n}^4}{768L^2}$
and $\log(1+a|S_n|)\leqslant \frac{{\rho_n}^4}{1536L^2}$, so that
\begin{align*}
\frac{1}{n}\sum_{i=1}^n\log(g_L(x_i))&\leqslant
-H(g)-\frac{{\rho_n}^4}{1536L^2}+\frac{1536\log(n)^{\frac{1}{4}}|S_n|L^3}{C_3\psi(g)^{\frac{1}{5}}n^{\frac{1}{4}}{\rho_n}^4}\\
&\qquad\qquad+C_3\psi(g)^{\frac{1}{5}}\left(\frac{\log(n)}{n}\right)^{\frac{1}{4}}|S_n|\log\left(\frac{1536|S_n|L^2}{{\rho_n}^4}\left(\sup_{x\in
  S_n}g_L(x)\lor 1\right)\right)
\end{align*}

By
the Lipschitz condition, $\lVert g_L\rVert_\infty \leqslant\sqrt{L}$,
so for $L\geqslant 1$,

\begin{adjustwidth}{-4cm}{-4cm}
\begin{align*}
\frac{1}{n}\sum_{i=1}^n\log(g_L(x_i))&\leqslant
-H(g)-\frac{{\rho_n}^4}{1536L^2}+\left(\frac{\log(n)}{n}\right)^{\frac{1}{4}}|S_n|\left(\frac{1536L^3}{C_3\psi(g)^{\frac{1}{5}}{\rho_n}^4}+C_3\psi(g)^{\frac{1}{5}}\log\left(\frac{1536|S_n|L^{\frac{5}{2}}}{{\rho_n}^4}\right)\right)
\end{align*}

\end{adjustwidth}

\end{proof}

This allows us to show

\begin{theorem}
If $\lVert G_n-G\rVert_\infty<\sqrt{\frac{\log(n)}{n}}$
and
$\lambda_n=C_1n^{\frac{7}{8}}\log(n)^{\frac{1}{8}}\sqrt{|S_n|}$, where $C_1=\frac{2C_4}{\psi(f)^{\frac{4}{5}}}\left(\frac{125}{144}\right)^{\frac{1}{10}}=\frac{2^{\frac{31}{20}}3^{-\frac{1}{10}}5^{\frac{3}{20}}}{\psi(f_x)^{\frac{4}{5}}\psi(g)^{\frac{1}{10}}}$,
then for sufficiently large $n$, $\lVert
\hat{g}_n-g\rVert_\infty<C_2n^{-\frac{1}{32}}|S_n|^{\frac{1}{8}}\log(n)^{\frac{1}{32}}$
for some constant $C_2$ (depending on $\psi(f)$ and $\psi(g)$).
\end{theorem}

\begin{proof}
By Proposition~\ref{HatPenalty} and Lemma~\ref{psiBoundLipschitz}, $g$
and $\hat{g}_n$ are both Lipschitz with constant
$L=\left(\frac{125}{16}\right)^{\frac{1}{5}}\times
\psi(f)^{\frac{2}{5}}$. Let $\rho_n=\lVert
\hat{g}_n-g\rVert_\infty$. By Proposition~\ref{FarBound},
\begin{adjustwidth}{-4cm}{-4cm}
    $$\frac{1}{n}\sum_{i=1}^n\log(\hat{g}_n(x_i))\leqslant
  -H(g)-\frac{{\rho_n}^4}{1536L^2}+\left(\frac{\log(n)}{n}\right)^{\frac{1}{4}}|S_n|\left(\frac{1536L^3}{C_3\psi(g)^{\frac{1}{5}}{\rho_n}^4}+C_3\psi(g)^{\frac{1}{5}}\log\left(\frac{1536|S_n|L^{\frac{5}{2}}}{{\rho_n}^4}\right)\right)$$

\end{adjustwidth}
If $\rho_n>n^{-1}\left(1536|S_n|L^{\frac{5}{2}}\right)^{\frac{1}{4}}$, we have
$\log\left(\frac{1536|S_n|L^{\frac{5}{2}}}{{\rho_n}^4}\right)<4\log(n)$, so

\begin{adjustwidth}{-4cm}{-4cm}
\begin{align*}
\frac{1}{n}\sum_{i=1}^n\log(\hat{g}_n(x_i))&\leqslant
-H(g)-\frac{{\rho_n}^4}{1536L^2}+\frac{1536 \log(n)^{\frac{1}{4}}|S_n|L^3}{C_3\psi(g)^{\frac{1}{5}}n^{\frac{1}{4}}{\rho_n}^4}+4C_3\psi(g)^{\frac{1}{5}}\frac{\log(n)^{\frac{5}{4}}}{n^{\frac{1}{4}}}|S_n|
\end{align*}
\end{adjustwidth}

On the other hand, by  Proposition~\ref{gStarLogLikelihood}, 
$$\frac{1}{n}\sum_{i=1}^n\log(\hat g_n(x_i))-\frac{\lambda_n}{n}\psi(\hat{f}_n)\geqslant
  \frac{1}{n}\sum_{i=1}^n\log(g^*_n(x_i))-\frac{\lambda_n}{n}\psi(f^*_n)\geqslant -H(g)-3M\log(n)^{\frac{3}{2}}n^{-\frac{1}{2}}-\frac{2\lambda_n}{n}\psi(f)$$
Thus we have 

\begin{adjustwidth}{-4cm}{-4cm}
$$
-\frac{{\rho_n}^4}{1536L^2}+\frac{1536\log(n)^{\frac{1}{4}}|S_n|L^3}{C_3\psi(g)^{\frac{1}{5}}n^{\frac{1}{4}}{\rho_n}^4}+4C_3\psi(g)^{\frac{1}{5}}\frac{\log(n)^{\frac{5}{4}}}{n^{\frac{1}{4}}}|S_n|-\frac{\lambda_n}{n}\psi(\hat{f}_n)\geqslant
- 3M\log(n)^{\frac{3}{2}}n^{-\frac{1}{2}}-\frac{2\lambda_n}{n}\psi(f)$$
\end{adjustwidth}

$$
\frac{{\rho_n}^4}{1536L^2}-\frac{1536\log(n)^{\frac{1}{4}}|S_n|L^3}{C_3\psi(g)^{\frac{1}{5}}n^{\frac{1}{4}}{\rho_n}^4}\leqslant
4C_3\psi(g)^{\frac{1}{5}}\frac{\log(n)^{\frac{5}{4}}}{n^{\frac{1}{4}}}|S_n|+3M\log(n)^{\frac{3}{2}}n^{-\frac{1}{2}}+\frac{\lambda_n}{n}(2\psi(f)-\psi(\hat{f}_n))$$

For large enough $n$,
$n^{-\frac{1}{4}}\log(n)^{\frac{1}{4}}|S_n|^{-1}<\frac{C_3\psi(g)^{\frac{1}{5}}}{3M}$,
so 
$$
\frac{{\rho_n}^4}{1536L^2}-\frac{1536\log(n)^{\frac{1}{4}}|S_n|L^3}{C_3\psi(g)^{\frac{1}{5}}n^{\frac{1}{4}}{\rho_n}^4}\leqslant
5C_3\psi(g)^{\frac{1}{5}}\log(n)^{\frac{5}{4}}n^{-\frac{1}{4}}|S_n|
+\frac{\lambda_n}{n}(2\psi(f)-\psi(\hat{f}_n))$$

With
$$\lambda_n=\frac{2C_4}{\psi(f)^{\frac{4}{5}}}n^{\frac{7}{8}}\log(n)^{\frac{1}{8}}\sqrt{|S_n|}
\left(\frac{125}{144}\right)^{\frac{1}{10}}$$

for large enough $n$, and since $\psi(\hat{f}_n)>0$, we deduce
$$\frac{\lambda_n}{n}(2\psi(f)-\psi(\hat{f}_n))\leqslant \frac{2\lambda_n\psi(f)}{n}=2C_4\psi(f)^{\frac{1}{5}}n^{-\frac{1}{8}}\log(n)^{\frac{1}{8}}\sqrt{|S_n|}
\left(\frac{125}{144}\right)^{\frac{1}{10}}$$
so

$$
\frac{{\rho_n}^4}{1536L^2}-\frac{1536\log(n)^{\frac{1}{4}}|S_n|L^3}{C_3\psi(g)^{\frac{1}{5}}n^{\frac{1}{4}}{\rho_n}^4}\leqslant
5C_3\psi(g)^{\frac{1}{5}}\log(n)^{\frac{5}{4}}n^{-\frac{1}{4}}|S_n|
+C_5n^{-\frac{1}{8}}|S_n|^{\frac{1}{2}}\log(n)^{\frac{1}{8}}$$

  where $C_5=2C_4\psi(f)^{\frac{1}{5}}\left(\frac{125}{144}\right)^{\frac{1}{10}}$.
For sufficiently large $n$, the first term on the right-hand side is
much smaller than the second, so we get 

$$
\frac{{\rho_n}^4}{1536L^2}-\frac{1536\log(n)^{\frac{1}{4}}|S_n|L^3}{C_3\psi(g)^{\frac{1}{5}}n^{\frac{1}{4}}{\rho_n}^4}\leqslant
2C_5n^{-\frac{1}{8}}|S_n|^{\frac{1}{2}}\log(n)^{\frac{1}{8}}$$

Rewriting this as 
$a_n(\rho_n)^8-b_n(\rho_n)^4-c_n\leqslant 0$
where

\begin{align*}
  a_n&=\frac{1}{1536L^2}\\
  b_n&=2C_5n^{-\frac{1}{8}}|S_n|^{\frac{1}{2}}\log(n)^{\frac{1}{8}}\\
  c_n&=\frac{1536\log(n)^{\frac{1}{4}}|S_n|L^3}{C_3\psi(g)^{\frac{1}{5}}n^{\frac{1}{4}}}
\end{align*}
we deduce

\begin{adjustwidth}{-4cm}{-4cm}
\begin{align*}
  (\rho_n)^4&\leqslant \frac{b_n+\sqrt{{b_n}^2+4a_nc_n}}{2a_n}\\
  &=768L^2\left(2C_5n^{-\frac{1}{8}}|S_n|^{\frac{1}{2}}\log(n)^{\frac{1}{8}}
+\sqrt{\left(2C_5n^{-\frac{1}{8}}|S_n|^{\frac{1}{2}}\log(n)^{\frac{1}{8}}\right)^2+\frac{4\log(n)^{\frac{1}{4}}|S_n|L}{C_3\psi(g)^{\frac{1}{5}}n^{\frac{1}{4}}}}\right)\\
  &=768L^2n^{-\frac{1}{8}}|S_n|^{\frac{1}{2}}\log(n)^{\frac{1}{8}}\left(2C_5
+\sqrt{4{C_5}^2+\frac{4L}{C_3\psi(g)^{\frac{1}{5}}}}\right)
\end{align*}
\end{adjustwidth}

Thus
$$  \rho_n\leqslant C_2 \log(n)^{\frac{1}{32}}|S_n|^{\frac{1}{8}}n^{-\frac{1}{32}}$$

where, plugging in all the constants, 
$${C_2}=\left(
  1+\sqrt{1+\frac{3^{\frac{2}{5}}}{16}}\right)^{\frac{1}{4}}2^{\frac{179}{80}}3^{\frac{9}{40}}
  {5^{\frac{27}{80}}}\psi(f)^{\frac{1}{4}}\psi(g)^{-\frac{1}{40}}$$

\end{proof}

From this, and the Dvoretzky-Kiefer-Wolfowitz inequality,
Theorem~\ref{MainTheorem} follows immediately.

\section{Additional Simulation Results}

Figures~\ref{f1}--\ref{f9} show box plots for all simulation scenarios
not shown in the main paper. Like the scenarios shown in the main
paper, we see that P-MLE is consistently better than the other methods
in the vast majority of cases, rather than the differences in MISE
being due to a small number of outliers in MISE.

\newcommand{\simulationresultsfigurecaption}[2]{
  \caption{ Sample distribution of ISE for sample size #1 and SNR #2.}
  \parbox{\textwidth}{
    Each columnn corresponds to one of the seven true distributions in the
    simulation. Rows correspond to the error distribution.
    The \texttt{decon} package only allows a limited selection of
    error families, so could not be compared for simulations with a
    beta error. Some outliers where \texttt{decon} produced a large
    ISE are truncated from these plots. No P-MLE results have been
    truncated.
\\
}}

\begin{figure}
\centering
\includegraphics[width=14cm]{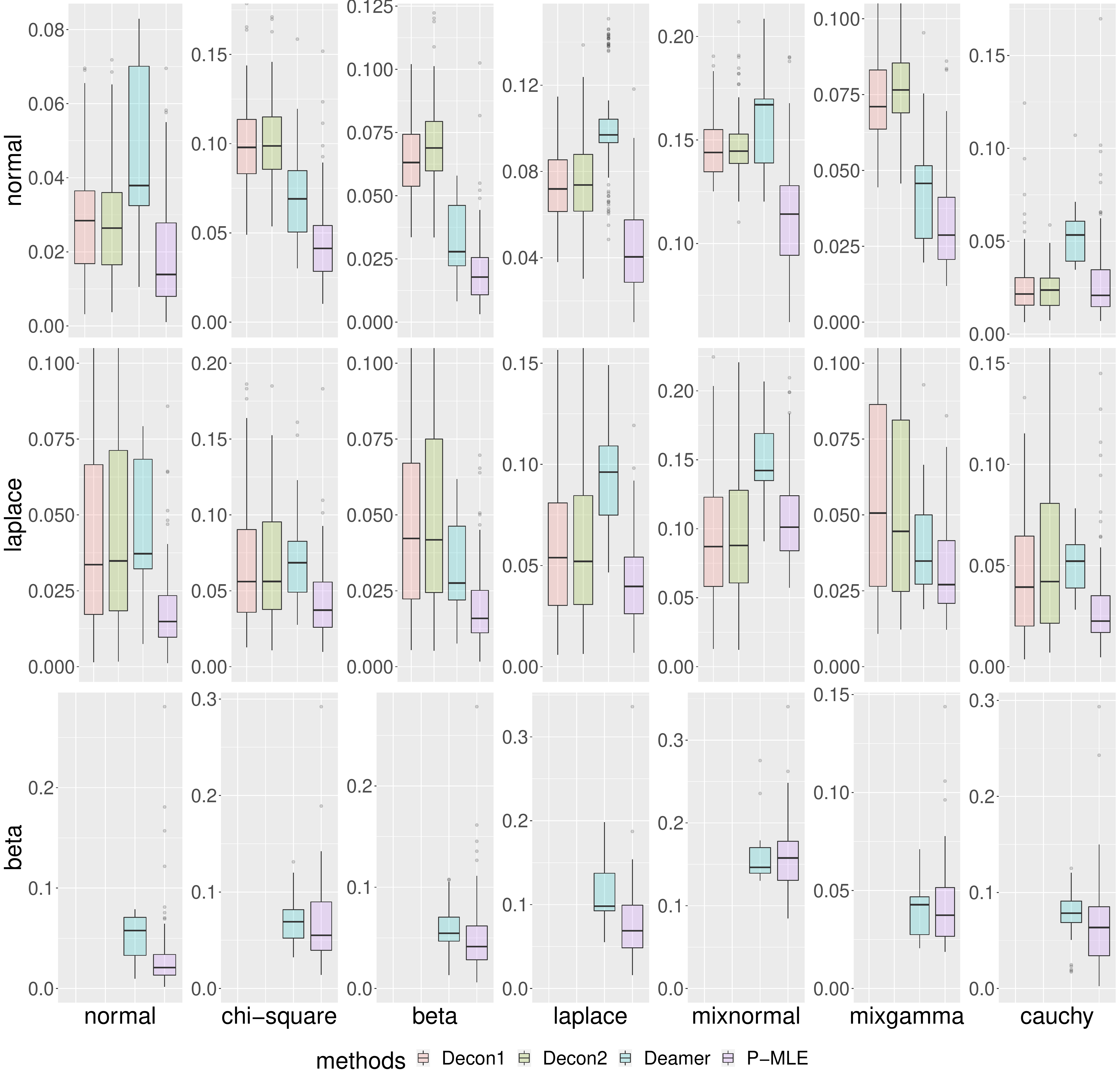}
\simulationresultsfigurecaption{30}{4}
\label{f1}
\end{figure}

\begin{figure}
\centering
\includegraphics[width=14cm]{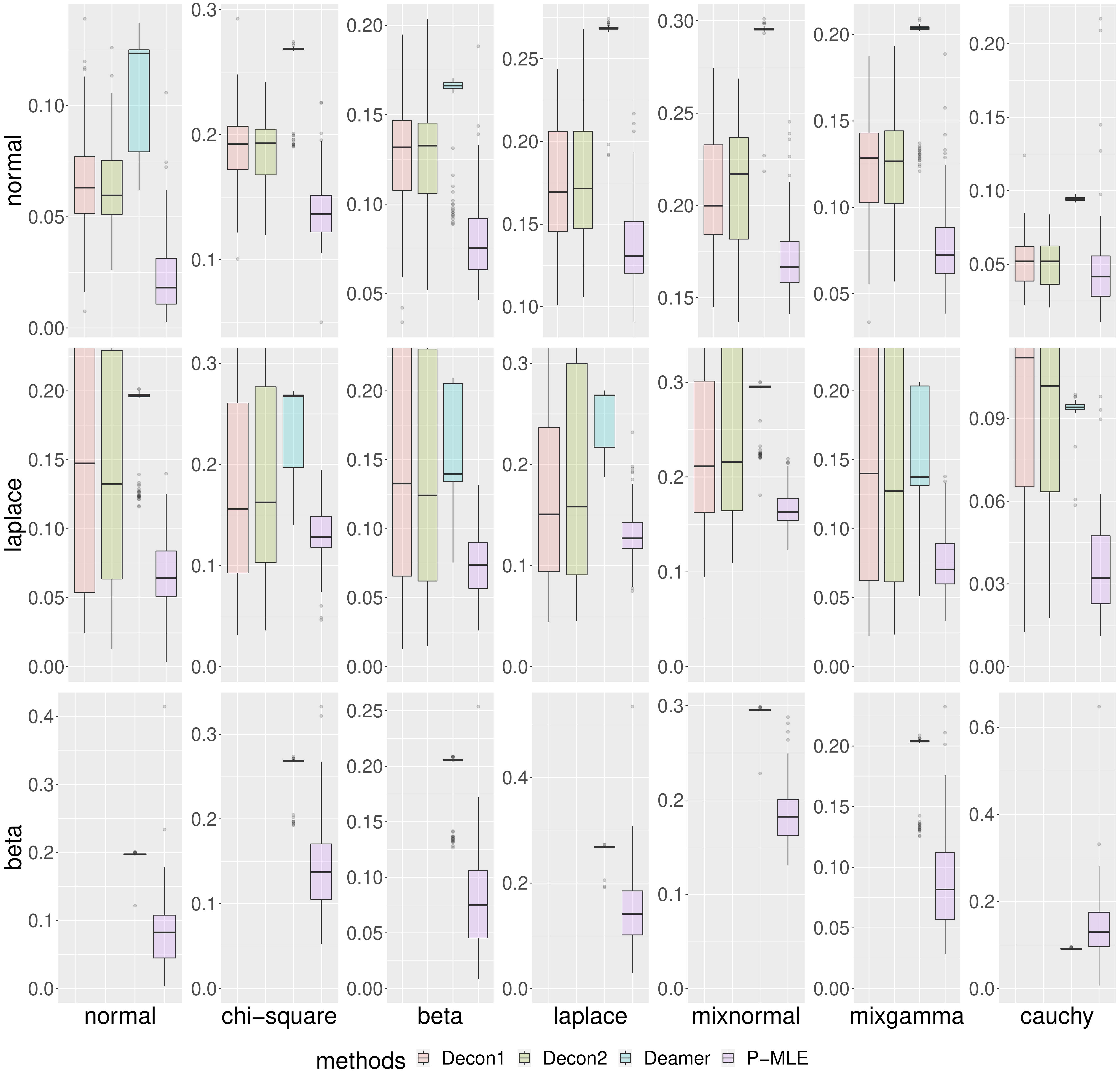}
\simulationresultsfigurecaption{30}{0.25}
\label{f3}
\end{figure}

\begin{figure}
\centering
\includegraphics[width=14cm]{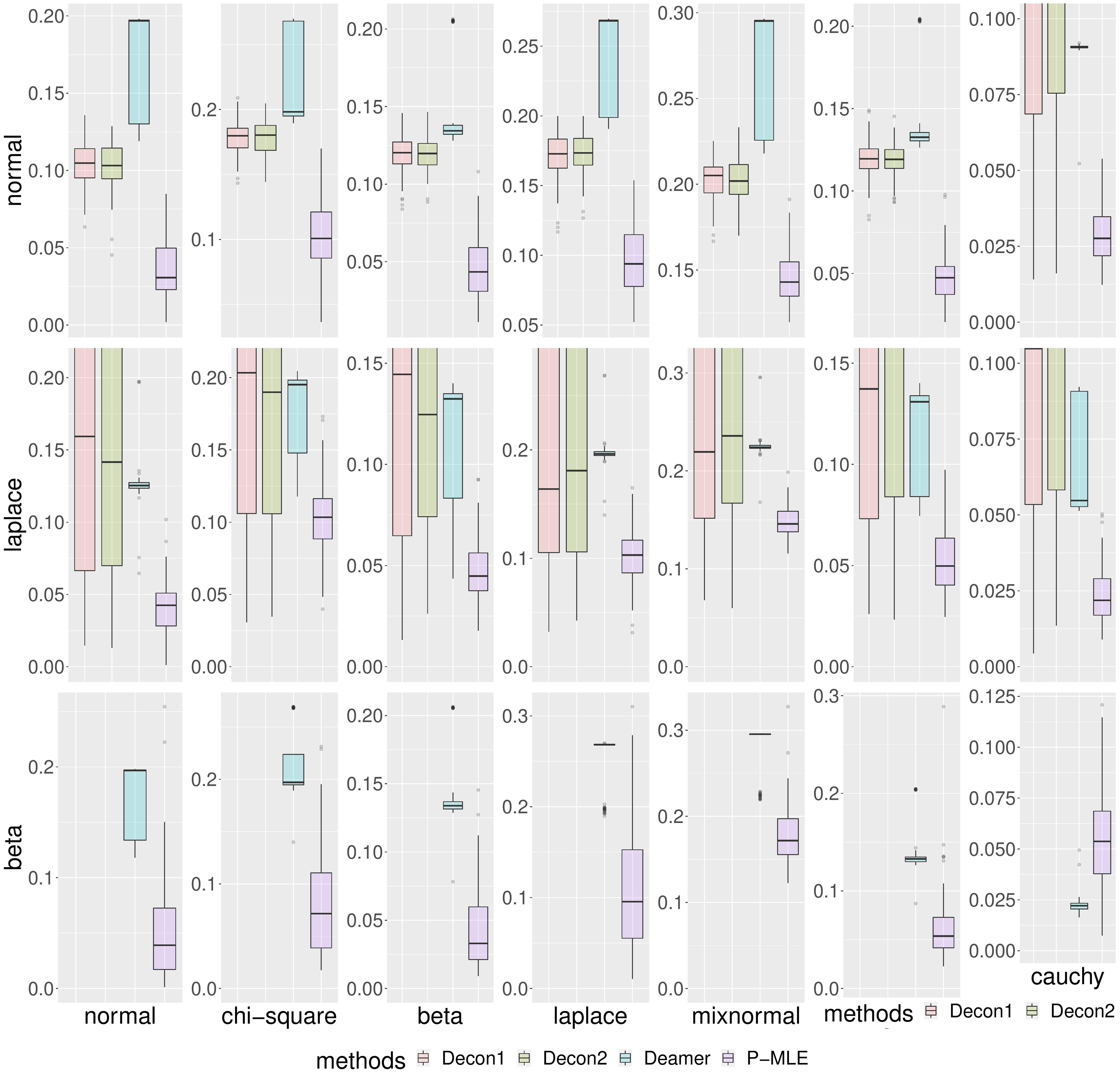}
\simulationresultsfigurecaption{100}{4}
\label{f4}
\end{figure}

\begin{figure}
\centering
\includegraphics[width=14cm]{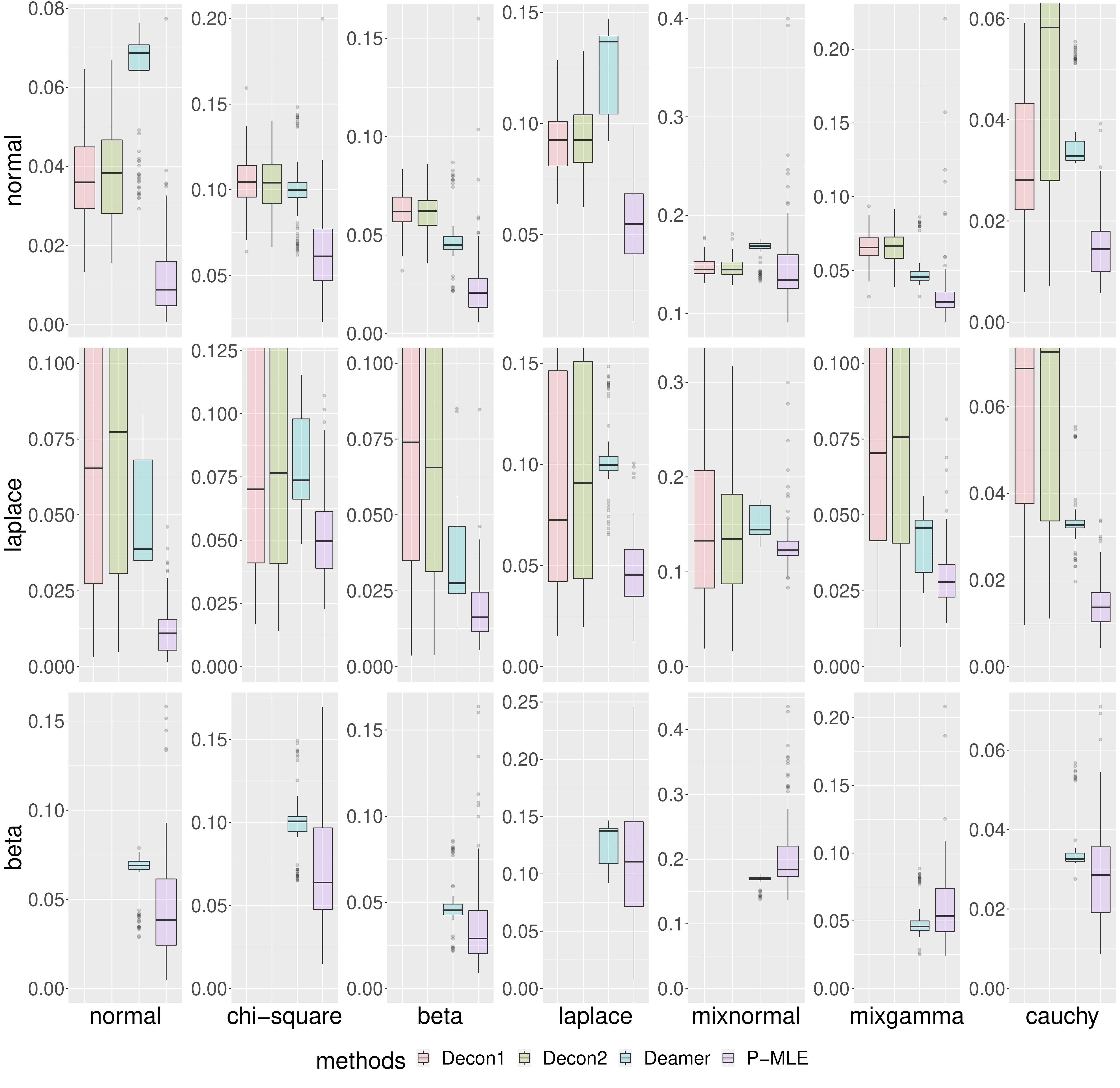}
\simulationresultsfigurecaption{100}{1}
\label{f5}
\end{figure}

\begin{figure}
\centering
\includegraphics[width=14cm]{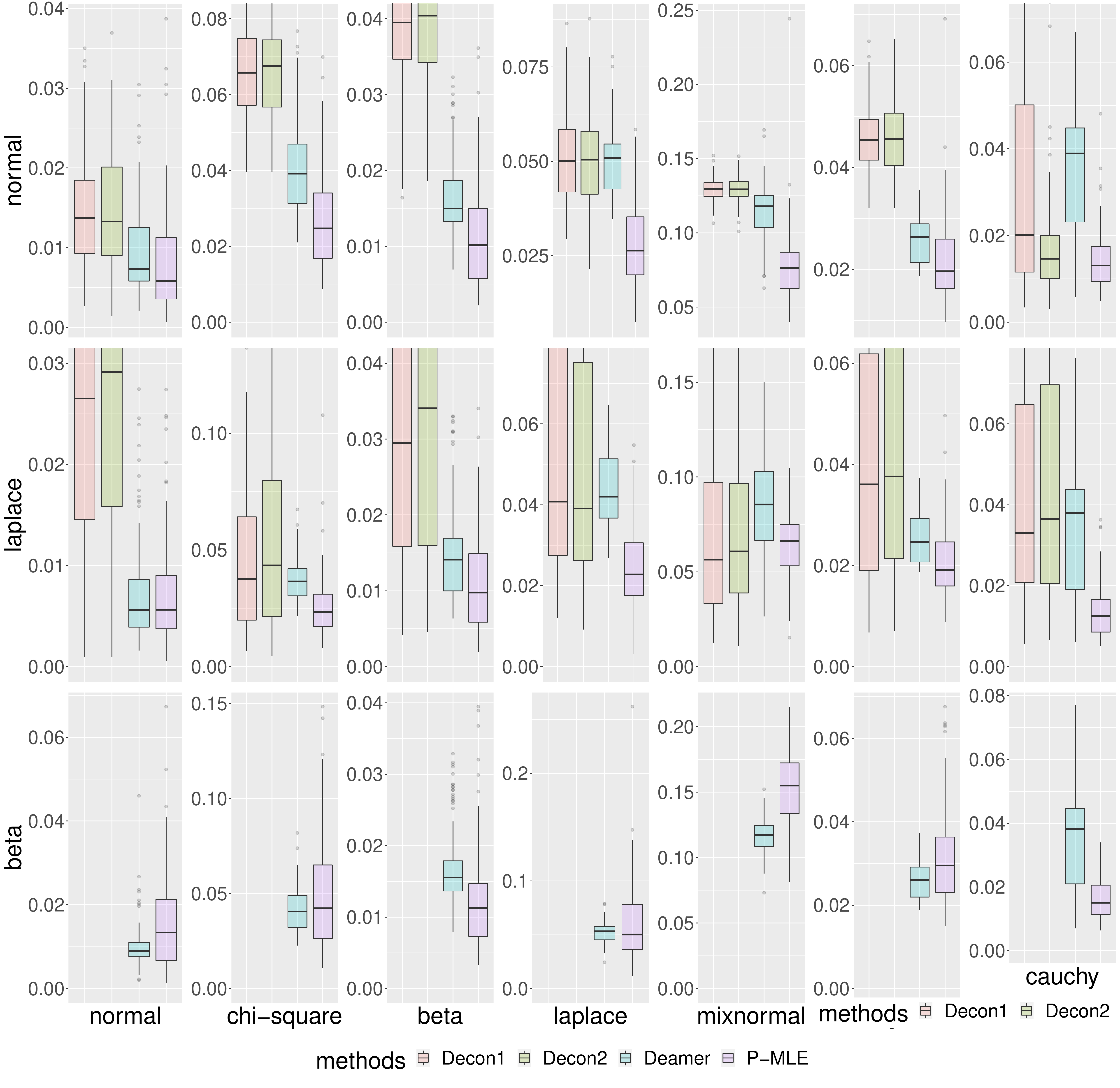}
\simulationresultsfigurecaption{100}{0.25}
\label{f6}
\end{figure}

\begin{figure}
\centering
\includegraphics[width=14cm]{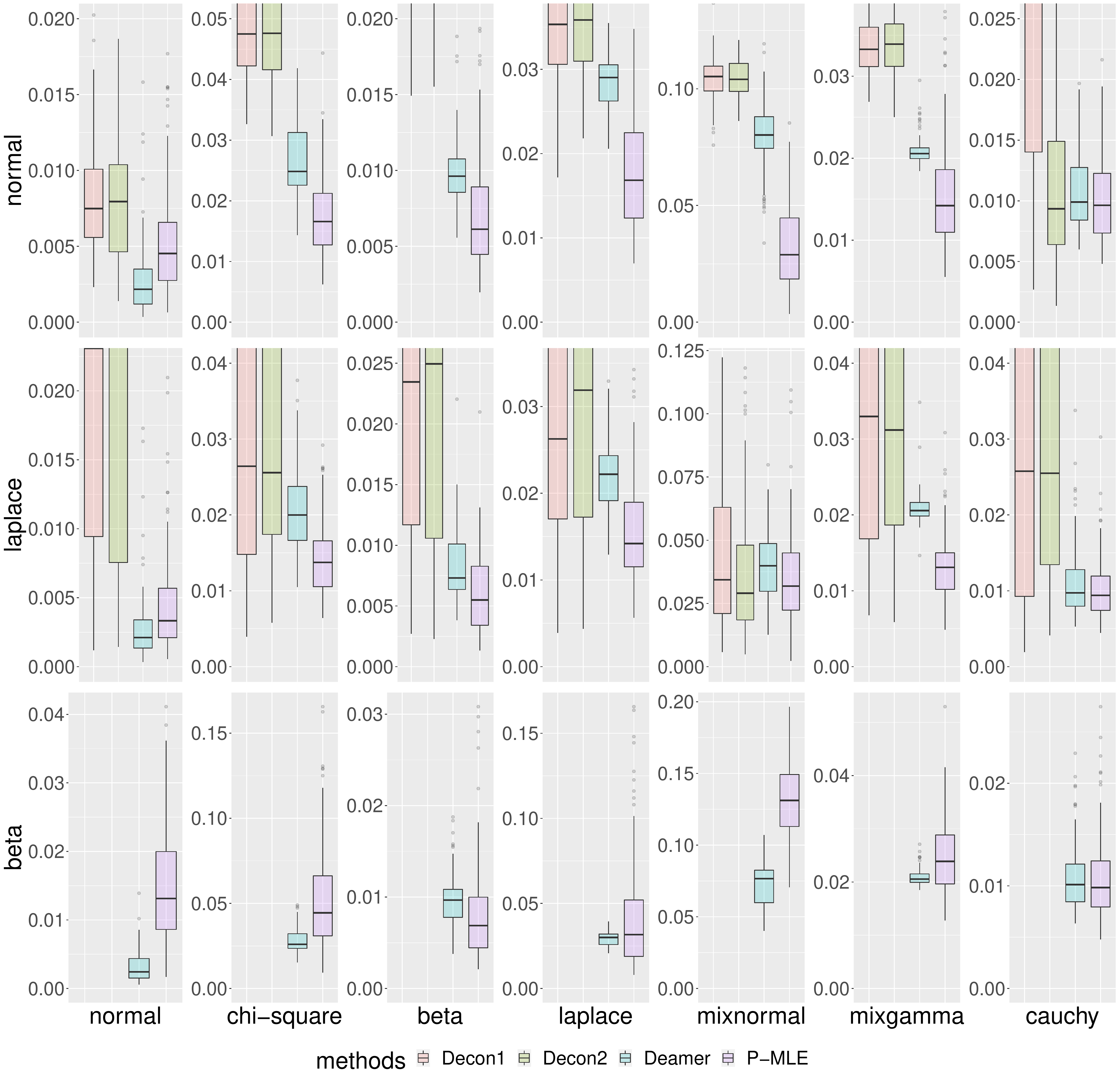}
\simulationresultsfigurecaption{300}{4}
\label{f7}
\end{figure}

\begin{figure}
\centering
\includegraphics[width=14cm]{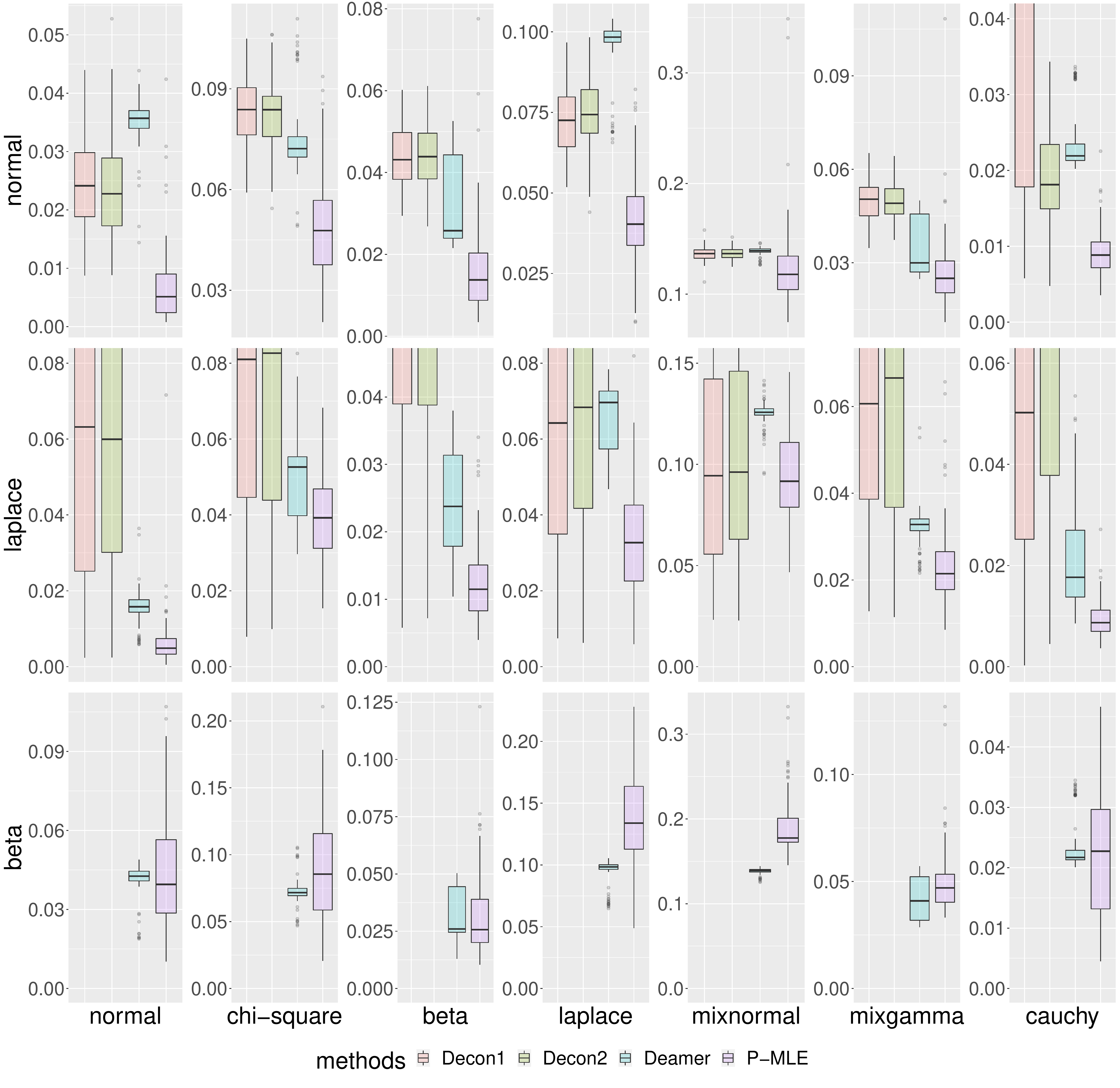}
\simulationresultsfigurecaption{300}{1}
\label{f8}
\end{figure}

\begin{figure}
\centering
\includegraphics[width=14cm]{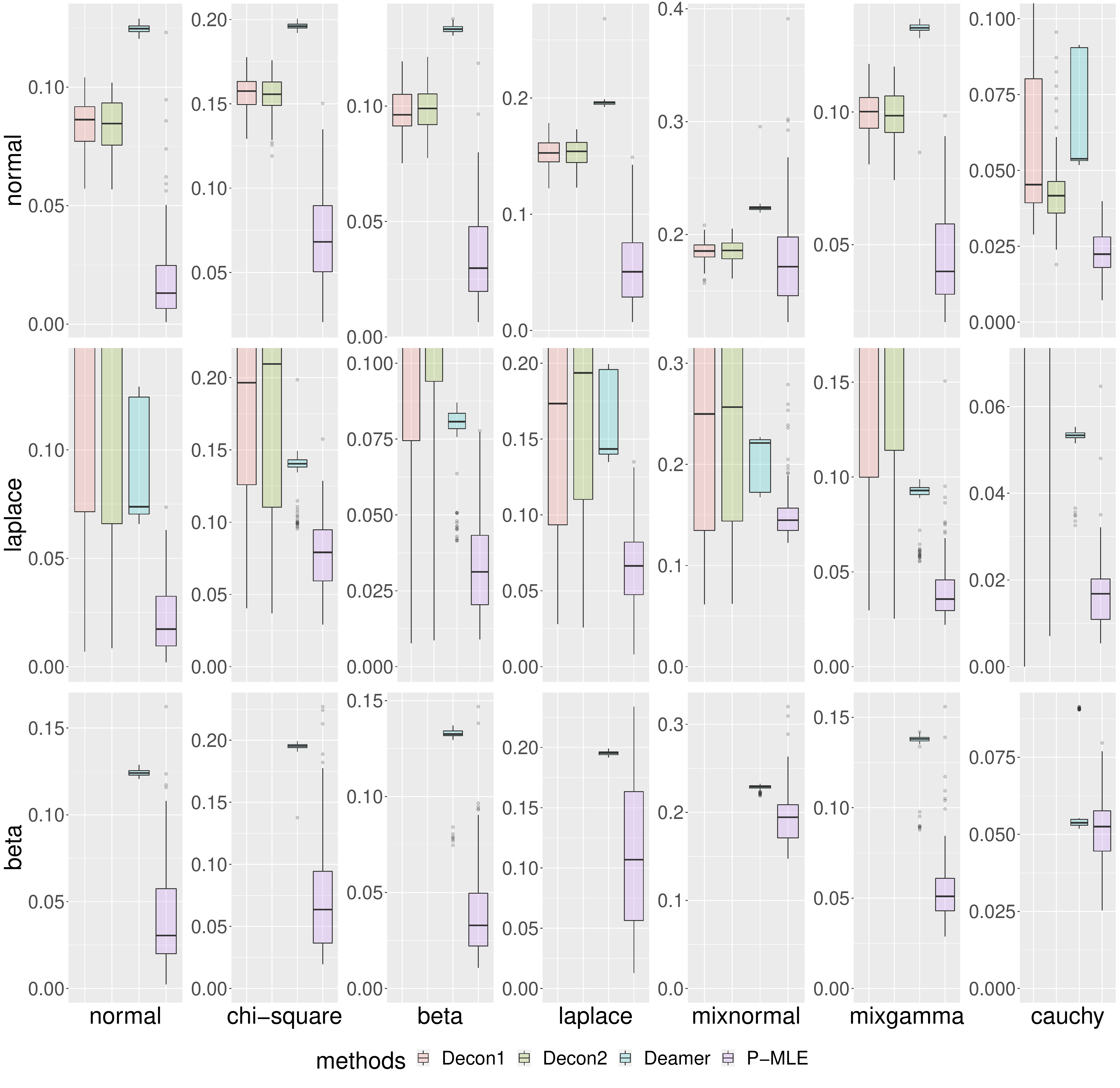}
\simulationresultsfigurecaption{300}{0.25}
\label{f9}
\end{figure}

\section{Effect of $\lambda_n$}

For the simulations, we wanted to know how much the results could be
improved with a better choice of the smoothness penalty $\lambda_n$.  We
reran the scenarios where P-MLE did not outperform the other methods
using different ratios for the heuristic method of setting
$\lambda_n$. We ran each scenario using values 1000, 10000, 100000,
1000000, and 10000000 for the ratio in the heuristic for setting
$\lambda_n$. We then looked at the best results in each scenario. In 9
of the scenarios, there was a difference from the outcome reported in
Table~\ref{SimulationSummaryTable} in the main paper. This is
selecting the value of $\lambda_n$ which performs best on the data, so
is not a completely reliable estimate of performance. However, it
shows that there is potential to improve performance with a better
choice of $\lambda_n$. Furthermore, these results are using the same
heuristic for $\lambda_n$ for all simulations in each scenario, rather
than tuning the value of $\lambda_n$ for each simulation. It is
therefore possible that more careful tuning could improve the
performance of P-MLE still further.

\begin{adjustwidth}{-4cm}{-4cm}
  
\begin{table}
  \caption[Results with different $\lambda_n$]{Simulation Results for some Scenarios with different
    heuristic values for $\lambda_n$.\label{lambdaEffect}

\vspace{\baselineskip}    
\parbox{\textwidth}{Significantly better results are highlighted in yellow. Results that
are not significantly different are highlighted in orange.}}
  
\hspace{-1cm}  \begin{tabular}{llrlrllll}
    \hline
\multicolumn{4}{c}{Scenario} & Best & Old  & New  & Deamer & Decon \\
\cline{1-4}
Truth & Error & $n$ & $C$ & Heuristic $\lambda_n$ & MISE (SE) &MISE (SE)
&MISE & MISE\\
\hline
cauchy&normal&30&0.5 & 1,000 &	0.0380(0.0080) &
\cellcolor{yellow}0.0294(0.0012) & & 0.0368(0.0010) \\
mixnormal&beta&100&1 & 10,000,000 & 0.2095(0.0063) &
\cellcolor{orange}0.1670(0.0054) & \cellcolor{orange}0.1679(0.0054) \\
mixgamma&beta&100&1 &	1,000,000 & 0.0614(0.0029) &
\cellcolor{orange}0.0506(0.0032) & \cellcolor{orange}0.0505(0.0014) \\
normal&laplace&300&2 & 100,000 & 0.0046(0.0004) &
\cellcolor{orange} 0.0028(0.0002) & \cellcolor{orange}0.0029(0.0003) \\
chi-squared&beta&300&1 & 10,000,000 & 0.0895(0.0038) &
\cellcolor{orange}0.0800(0.0048) & \cellcolor{orange}0.0724(0.0010) \\
mixgamma&beta&300&1 & 10,000,000 & 0.0503(0.0016) & \cellcolor{orange}0.0410(0.0015) & \cellcolor{orange}0.0420(0.0010) \\
laplace&beta&300&1 & 100,000 & 0.1361(0.0037) &
\cellcolor{orange}0.1010(0.0034) & \cellcolor{orange}0.0946(0.0010) \\
normal&beta&300&1 & 10,000,000 & 0.0431(0.0021) &
\cellcolor{yellow}0.0345(0.0012) & 0.0411(0.0006) \\
beta&beta&300&1 & 10,000,000 & 0.0316(0.0018) & \cellcolor{yellow}0.0250(0.0019) & 0.0311(0.0010) \\
\hline
  \end{tabular}

\end{table}
  \end{adjustwidth}
  
\newpage

\bibliographystyle{rss}
\bibliography{bmc_article}